\pdfoutput=1
\documentclass[12pt]{article}
\usepackage{amssymb,amsmath,mathtools,mathrsfs,slashed}
\usepackage{epsfig}
\usepackage{color,graphicx}
\usepackage{cite}
\usepackage{amssymb,amsmath,bm,bbm}
\usepackage{multirow}
\usepackage{array}
\usepackage{hyperref}
\usepackage{comment}
\usepackage{epstopdf}
\usepackage{cancel}

\usepackage{tikz}
\usepackage{tikz-3dplot}   
\tdplotsetmaincoords{60}{110}
\pgfmathsetmacro{\rvec}{.67}
\pgfmathsetmacro{\thetavec}{30}
\pgfmathsetmacro{\phivec}{60}

\usepackage{tabularx}
\newcolumntype{Y}{>{\centering\arraybackslash}X}
\usepackage{booktabs}
\AtBeginDocument{
    \heavyrulewidth=.08em
        \lightrulewidth=.05em
        \cmidrulewidth=.03em
        \belowrulesep=.65ex
        \belowbottomsep=0pt
        \aboverulesep=.4ex
        \abovetopsep=0pt
        \cmidrulesep=\doublerulesep
        \cmidrulekern=.5em
        \defaultaddspace=.5em
}

\newcommand{\be}{\begin{equation}}
\newcommand{\ee}{\end{equation}}
\newcommand{\bea}{\begin{eqnarray}}
\newcommand{\eea}{\end{eqnarray}}

\def\cN{{\ensuremath{\cal N}}}

\usepackage[top=2.5cm, bottom=2.5cm, left=2.5cm, right=2.5cm]{geometry}	
\numberwithin{equation}{section}

\begin{document}
 
\begin{flushright}
HIP-2021-16/TH
\end{flushright}

\begin{center}

\centering{\Large {\bf Risking your NEC}}

\vspace{6mm}

\renewcommand\thefootnote{\mbox{$\fnsymbol{footnote}$}}
Carlos Hoyos,${}^{1,2}$\footnote{hoyoscarlos@uniovi.es}
Niko Jokela,${}^{3,4}$\footnote{niko.jokela@helsinki.fi} \\
Jos\'e Manuel Pen\'\i n,${}^{3,4,5}$\footnote{jmanpen@gmail.com}
Alfonso V. Ramallo,${}^{6,7}$\footnote{alfonso@fpaxp1.usc.es} and
Javier Tarr\'io${}^{3,4}$\footnote{javier.tarrio@helsinki.fi}

\vspace{4mm}
${}^1${\small \sl Department of Physics}  and\\
${}^2$ {\small \sl Instituto de Ciencias y Tecnolog\'{\i}as Espaciales de Asturias (ICTEA)} \\
{\small \sl Universidad de Oviedo, c/Federico Garc\'ia Lorca 18} \\
{\small \sl E-33007/33004, Oviedo, Spain}

\vspace{4mm}
${}^3${\small \sl Department of Physics} and ${}^4${\small \sl Helsinki Institute of Physics} \\
{\small \sl P.O.Box 64} \\
{\small \sl FIN-00014 University of Helsinki, Finland} 

\vspace{2mm}

${}^5${\small \sl Mathematical Sciences and STAG Research Center,} \\ 
{\small \sl University of Southampton, University Road, Southampton SO17 1BJ, UK}

\vspace{2mm}

${}^6${\small \sl Departamento de  F\'\i sica de Part\'\i  culas} 
{\small \sl and} \\
${}^7${\small \sl Instituto Galego de F\'\i sica de Altas Enerx\'\i as (IGFAE)} \\
{\small \sl Universidade de Santiago de Compostela} \\
{\small \sl R\'ua de Xoaqu\'in D\'iaz de R\'abago s/n}  \\
{\small \sl E-15782 Santiago de Compostela, Spain} 

\vskip 0.2cm

\end{center}

\vspace{2mm}

\setcounter{footnote}{0}
\renewcommand\thefootnote{\mbox{\arabic{footnote}}}

\begin{abstract}
\noindent 
Energy conditions, especially the null energy condition (NEC), are generically imposed on solutions to retain a physically sensible classical field theory and they also play an important role in the AdS/CFT duality. Using this duality, we study non-trivially deformed strongly coupled quantum field theories at large-$N_c$. The corresponding dual classical gravity constructions entail the use of radially non-monotonic D-brane distributions. The distributions are phenomenological in the sense that they do not correspond to the smearing of known probe D-brane embeddings. The gravity backgrounds are supersymmetric and hence perturbatively stable, and do not possess curvature singularities. There are no short-cuts through the bulk spacetime for signal propagation which assures that the field theory duals are causal. Nevertheless, some of our solutions violate the NEC in the gravity dual. In these cases the non-monotonicity of the D-brane distributions is reflected in the properties of the renormalization group flow: none of the $c$-functions proposed in the literature are monotonic. This further suggests that the non-monotonic behavior of the $c$-functions within previously known anisotropic backgrounds does not originate from the breaking of Lorentz invariance. We surmise that NEC violations induced by quantum corrections also need to be considered in holographic duals, but can be studied already at the classical level.

\end{abstract}

\newpage
{\footnotesize{\tableofcontents}}
\newpage

\section{Introduction and summary}\label{sec:intro}

Energy conditions considered in the context of general relativity are crucial ingredients in several general theorems, such as Penrose's singularity theorem and no-go theorems for closed timelike curves or wormholes \cite{Kontou:2020bta}. Among the possible conditions, one of the weakest ones is the Null Energy Condition (NEC) that posits that for any null vector $n^\mu$, the matter energy-momentum tensor $T_{\mu\nu}$ satisfies the positivity condition
\be\label{eq:NEC}
 T_{\mu\nu} n^\mu n^\nu\geq 0 \ .
\ee  
In the context of gauge/gravity duality theories that violate the NEC in the gravity dual are considered to be problematic in some way. For instance, in Lifshitz geometries the bulk violation of (\ref{eq:NEC}) can be associated to violations of causality in the dual field theory \cite{Hoyos:2010at}. There is also an interesting relation between the properties of the RG flow in field theory and the NEC in the gravity dual. Under certain conditions, such as underlying Lorentz invariance and unitarity, it is possible to find quantities that measure the number of degrees of freedom and are monotonically decreasing along the RG flow as one goes from higher to lower energy scales \cite{Zamolodchikov:1986gt,Komargodski:2011vj,Casini:2004bw,Casini:2012ei}. Taking the two-dimensional case as a reference we will call all such quantities collectively as $c$-functions. In theories with a gravity dual it is possible to find holographic $c$-functions that are monotonic along the holographic radial direction, corresponding to different energy scales in the field theory. The monotonicity of holographic $c$-functions in most cases depend on whether the NEC is satisfied in the gravity dual \cite{Freedman:1999gp,Myers:2010xs,Myers:2010tj}. Interestingly, the NEC can be satisfied and still the $c$-function can be non-monotonic if it has an infinite jump, as is the case in \cite{Bigazzi:2009gu}. This is reminiscent of field theory examples with multivalued $c$-functions \cite{Bernard:2001sc,LeClair:2003hj,Leclair:2003xj}.

However, local energy conditions, including NEC, are generically violated in quantum field theory (QFT), even though they can be satisfied in an averaged sense (see, \emph{e.g.}, \cite{Kontou:2020bta,Fewster:2012yh,Curiel:2014zba} for reviews). This raises the question of whether the NEC is a condition that should be satisfied in any classical supergravity theory with a holographic dual, or whether a weaker constraint (if any) is sufficient. This question becomes notably pertinent once quantum corrections are to be taken into account. In reverse order, the NEC violations abound to gravitational energy-momentum tensor would then in principle be translated in the RG flow of the dual quantum field theory, see, {\emph{e.g.}}, \cite{Nakayama:2012jv}. We note that violations of the NEC in holography have been considered in search for traversable wormholes (see, {\emph{e.g.},~\cite{Gao:2016bin}), though their existence in $AdS$ spacetimes do not seem to necessitate NEC-violating matter content \cite{Anabalon:2020loe}.

Recently, it has been shown that there are anisotropic supergravity solutions interpolating between (isotropic) $AdS$ geometries of the same radius \cite{Donos:2017sba,Hoyos:2020zeg}. The field theory dual interpretation is an RG flow between two CFTs with the same value of the $c$-function. These flows have been dubbed ``boomerang flows'' \cite{Donos:2017sba} as apparently one returns to the same fixed point. Although this does not contradict the existing $c$-theorems \cite{Zamolodchikov:1986gt,Komargodski:2011vj,Casini:2004bw,Casini:2012ei}, which all assume Lorentz invariance,\footnote{For non-monotonic field theory examples with broken Lorentz invariance, see \cite{Swingle:2013zla,Daguerre:2020pte}.} it seems in tension with the interpretation of $c$-functions as counting the number of degrees of freedom with the intuition that this number should decrease along the RG flow hitherto suggesting possible violations of the NEC  in the gravity dual. It should be stressed that in anisotropic solutions the monotonicity of the usual holographic $c$-functions is not guaranteed by the NEC but there are several concurrent proposals generalizing the holographic $c$-functions assuming particular matter content \cite{Giataganas:2017koz,Chu:2019uoh,Ghasemi:2019xrl,Cremonini:2020rdx}.

In \cite{Hoyos:2020zeg} we explored several holographic $c$-functions in anisotropic solutions sourced by smeared D5-branes in a geometry asymptotically dual to D3-branes and found them to be non-monotonic, in boomerang flows and also in a family of flows possessing anisotropic scaling in the IR limit. The non-monotonic behavior appears naturally because the distribution of smeared D5-branes is not monotonic along the holographic radial direction, it decays both at the asymptotic boundary and close to the origin, with a peak in an intermediate region. Using the intuition given by flavor branes, this could translate into having more degrees of freedom in the intermediate energy scales than close to the UV or IR fixed points. In this work we construct also isotropic solutions for other types of smeared D-branes with non-monotonic distributions and check if holographic $c$-functions are monotonic. Specifically, we will study configurations for smeared D6-branes in geometries that asymptotically approach the duals of ABJM and D2-branes as well as smeared D7-branes in geometries that asymptote to the dual of D3-branes. We furthermore test the monotonicity of alternative proposals for holographic $c$-functions in the anisotropic D3-D5 case. We in particular set out to search for violations of the NEC in all these cases. The monotonicity properties of the $c$-functions and the validity of energy conditions turn out to depend on the type of D-brane intersection we consider. It should be noted that the distributions we use do not correspond to the smearing of known probe D-branes, so it is possible that the solutions we find are not part of string theory. In this sense the solutions we present are ``bottom-up''.

We observe non-monotonic behavior of holographic $c$-functions and violations of NEC in most cases, except for the D2-D6 setup if the density of D6-branes is not too large.  However, there is no obvious sickness in the solutions we study: they are all supersymmetric, void of curvature singularities, and there are no apparent violations of causality in the dual field theory. To our knowledge these are the first examples of this kind that have been reported in the literature. Previous works in the context of cosmology \cite{Rubakov:2006pn,Creminelli:2006xe,Rubakov:2014jja} found non-supersymmetric models of scalars with higher derivative terms that are apparently free of issues despite violating the NEC.\footnote{In addition, there is a trivial example of a free scalar non-minimally coupled to gravity.} This suggests that the NEC in the gravity dual could indeed be relaxed under certain conditions, although there might be more subtle issues that do not emerge in the observables we have studied.

The rest of the paper is organized as follows. In Section~\ref{sec:setup} we briefly acknowledge the common features shared by all the theories mentioned above and studied in this work. We list all utile proposals for holographic $c$-functions that we are aware of and can hence straightforwardly test. The subsequent sections will then flesh out salient details of the dual gravity theories that are needed for $c$-functions. We unveil the results with a number of analytic arguments as well as with figures where seen appropriate. We will conclude in Section~\ref{sec:discussion} with a summary of our findings and discuss the repercussions in the larger AdS/CFT context.  Finally, in the appendices we will provide many technical details that lead to the results and conclusions presented in the main text. We will start with ABJMf in Appendix~\ref{app:ABJM}, followed by an analysis of D2-D6 in Appendix~\ref{app:D2D6}, D3-D7 case in Appendix~\ref{app:D3D7}, and finally D3-D5 case in Appendix~\ref{app:D3D5}.

\section{Setup}\label{sec:setup}

Theories with a gravity dual usually contain matter in adjoint or bifundamental representations of the gauge group, {\emph{i.e.}}, with ${\cal O}(N_c^2)$ degrees of freedom, where $N_c$ is roughly speaking the rank of the group. Quenched matter in the fundamental representation can be incorporated by introducing probe branes in the dual geometry when the number of flavors is small $N_f\ll N_c$ \cite{Karch:2002sh}. However, if the number of flavors is comparable to the rank of the group $N_f\sim N_c$ (unquenched), the probe approximation will not be valid and one has to take into account the backreaction of flavor branes on the geometry. This is quite a hard problem, but it can be significantly simplified by considering smooth distributions of branes (``smearing'') instead of localized brane sources that bring in delta functions resulting in partial differential equations. Techniques to construct backreacted geometries with smeared brane distributions have been developed and applied successfully to describe  gravity duals of gauge theories with flavors (see \cite{Nunez:2010sf} for a review). 

A remarkable feature of geometries with smeared branes is that it is possible to construct families of supersymmetric solutions depending on an arbitrary brane distribution. These families include distributions corresponding to the backreaction of branes with the profile corresponding to probe flavor branes but also (infinitely) many others. Although most of those solutions might not correspond to any limit of a configuration of sources existing in string theory, it is still interesting to study them as we expect them to enjoy the perks associated to supersymmetry, in particular they should be stable within the classical supergravity approximation. 

We will study the following cases:
\begin{itemize}
\item ABJMf: Gravity dual to ABJM with smeared D$6$-branes.
\item D2-D6: Gravity dual to D$2$-branes with smeared D$6$-branes.
\item D3-D7: Gravity dual to D$3$-branes with smeared D$7$-branes.
\item D3-D5: Gravity dual to D$3$-branes with smeared D$5$-branes.
\end{itemize}
The first two are dual to ($2+1$)-dimensional theories, while the last two are dual to ($3+1$)-dimensional theories. The last case is anisotropic along one of the spatial directions of the field theory, while the rest are isotropic. The ten-dimensional geometry splits into a ($3+1$)- or ($4+1$)-dimensional part involving the field theory directions and the holographic radial coordinate $r$, plus the internal space directions. The smeared D$p$-branes act as a source of magnetic flux for the $F_{8-p}$ Ramond-Ramond forms: $F_2$ (D$6$-branes), $F_1$ (D$7$-branes), or $F_3$ (D$5$-branes). In the family of D2-D6 backgrounds we found the D6-branes are smeared over a compact generic six-dimensional  nearly-K\"ahler manifold, whereas in the D3-D7 and D3-D5 cases the internal space is an arbitrary Sasaki-Einstein space.

The flux in supersymmetric solutions can depend on an arbitrary function $p(r)$ that determines the brane distribution
\be
 F_{8-p}\propto Q_f p(r) \ ,
\ee
where $Q_f$ is chosen such that $p(r)=1$ corresponds to the solutions with massless flavors. The first order BPS equations of ten-dimensional supergravity needed to preserve some amount of supersymmetry allow the introduction of arbitrary profile functions $p(r)$. The backgrounds with $p(r\to 0)=0$ and $p(r\to\infty)=1$ correspond to massive flavors. In those cases one can use kappa symmetry to find the supersymmetric embeddings of the brane and obtain a unique profile function $p(r)$ for every quark mass. 
An interesting choice of distribution is one where $p(r)$ vanishes at the asymptotic boundary $p(r)\to 0$ as $r\to \infty$. In some cases (see, {\emph{e.g.}}, \cite{Hoyos:2020zeg}) this can be obtained in such a way that the leading asymptotic decay of the supergravity fields is the same as the theory without flavors, so that the configuration with smeared branes would correspond to a different state of the unflavored theory. For the sake of concreteness we will fix $p(r)$ to be of the form\footnote{In fact, for the D$3$-D$5$ case $p(r)$ is not exactly the distribution but there is an additional factor \cite{Hoyos:2020zeg}.}
\be\label{eq:pdist}
 p=\frac{r^n}{(1+r^m)^{\frac{n+d}{m}}} \ , \ \ n\geq 0,\ m\geq 1 \ .
\ee
The exponent $d$ is chosen in such a way that the contribution of the branes to the energy density remains positive as $r\to \infty$. Expressions for the energy density in each case are derived in the Appendices. We find that in this limit, the positive energy condition becomes
\be
 T_{00}^{Dp} \propto Q_f(d p+r p')\geq 0 \ ,
\ee
where $d=2$ (ABJMf), $d=3$ (D3-D5), or $d=4$ (D2-D6 and D3-D7).

With the choice in \eqref{eq:pdist}, the brane distribution also vanishes at the origin of space (except for $n=0$), so it is not a monotonic function of the radial coordinate. The non-monotonicity of the distribution is propagated to the supergravity fields, in particular the metric. Given the relation between the radial coordinate and energy or length scales in the dual field theory, many quantities in the dual field theory will also be non-monotonic along the RG flow. This raises two possible issues in the field theory side
\begin{enumerate}
\item The Averaged Null Energy Condition in the field theory is not satisfied, this translates into acausality. 
\item There is no $c$-function monotonically decreasing along the RG flow, signalling a possible issue with unitarity of the dual field theory. 
\end{enumerate}
Let us review how each of these are determined using the gravity dual description.

\subsection{Causality and boundary Averaged Null Energy Condition}

As mentioned above, energy conditions may be violated locally but satisfied on average. In particular, the averaged null energy condition (ANEC), defined by integrating the NEC along the full null geodesic, should be satisfied in flat spacetime if the theory is causal. For theories with holographic duals, causality in field theory requires that signals travelling through the bulk of the gravity dual between two points at the boundary take a longer time than a light ray localized at the boundary\footnote{However, a more refined analysis might allow to relax this condition, see~\cite{Hernandez-Cuenca:2021eyf}.}. As shown in \cite{Kelly:2014mra}, this condition is equivalent to demanding that the energy-momentum tensor of the dual field theory satisfies the ANEC, at least for asymptotically $AdS$ spacetimes. For D$p$-brane geometries (with $p<5$) the same argument is expected to hold, since one can choose a frame where D$p$-brane geometries are obtained by the compactification of an $AdS$ spacetime \cite{Kanitscheider:2008kd}. Morally, a signal propagating also through internal directions will take a longer time to return to the boundary than a signal propagating only in the field theory directions, as the distance is longer. It is hence sufficient to focus on propagation in the field theory and radial directions. 

The family of metrics we will encounter are diagonal
\be
 ds^2=g_{rr} dr^2+g_{\mu\nu}dx^\mu dx^\nu \ , \ g_{\mu\nu}=0\; \text{if}\; \mu\neq \nu \ .
\ee
The local speed of light at a fixed constant-$r$ slice of the metric is given by the ratio of two metric components. We can define a local refraction index ${\bm n}_x$ as follows
\be
 {\bm n}_x=\left(\frac{g_{xx}}{|g_{00}|}\right)^{1/2} \ .
\ee 
The local speed of light will be larger closer to the asymptotic boundary if the refraction index is monotonically decreasing with $r$:
\be\label{eq:drnx}
 \frac{d}{dr}{\bm n}_x \leq 0 \ .
\ee
If the condition (\ref{eq:drnx}) is satisfied, a signal propagating at the speed of light through the bulk would take a longer time to travel between two boundary points than a signal bound at the boundary. In this case the dual field theory energy-momentum tensor will satisfy the ANEC and no acausality is expected. 

If the geometry is Poincar\'e invariant the refraction index is trivially unity ${\bm n}_x=1$ and we do not expect violations of causality. However, if the geometry is anisotropic in either time or space directions, \eqref{eq:drnx} is a non-trivial condition. Then, the only model in our study where the ANEC of the dual field theory is at stake is D3-D5. However, in \cite{Hoyos:2020zeg} it was shown that ${\bm n}_x$ has the right monotonicity, so this type of unphysical behavior is absent.

\subsection{Holographic $c$-functions and bulk Null Energy Condition}

Let us now gear our attention to various measures of degrees of freedom and their possible relation to bulk NEC. In the Wilsonian view of the renormalization group flow, heavy degrees of freedom are integrated out above some energy scale yielding an effective theory for the light degrees of freedom. The effective theory is then expected to contain fewer degrees of freedom than the original microscopic theory. If this picture is correct, then there might be some quantity which gives a measure of the number of degrees of freedom and which is decreasing along the RG flow. This quantity is usually dubbed a ``$c$-function''. The first instance of a $c$-function is the one found by Zamolodchikov for two-dimensional theories \cite{Zamolodchikov:1986gt}. Zamolodchikov's $c$-function takes the value of the coefficient of the conformal anomaly when the flow is at a fixed point. The proof \cite{Zamolodchikov:1986gt} implies that this coefficient is smaller at an IR fixed point than at a UV fixed point, for a flow interpolating between the two. More recently, $c$-functions for three and four dimensional theories have also been found \cite{Casini:2004bw,Casini:2012ei,Komargodski:2011vj}. In the case of the four-dimensional theory the $c$-function at a fixed point also coincides with the coefficient of the conformal anomaly (in particular the type A term, proportional to the Euler density), while in three dimensions is given by the value of the partition function on a sphere.

A theory can admit multiple $c$-functions, however. In a theory with a holographic dual it is possible to find quantities in the gravity dual that are monotonic functions of the radial coordinate \cite{Freedman:1999gp,Sahakian:1999bd,Alvarez:1998wr}. In the standard Poincar\'e invariant geometries, monotonicity requires that the bulk energy-momentum tensor satisfies the Null Energy Condition $T_{MN}n^M n^N\geq 0$, $n_M n^M=0$. Through Einstein's equations, the NEC is equivalent to (the term containing the Ricci scalar vanishes as the metric is contracted with null vectors)
\be
 R_{MN}n^M n^N \geq 0 \ ,
\ee
where $R_{MN}$ is the Ricci tensor of the dual $(D+1)$-dimensional metric. Thus, the existence of a monotonic $c$-function translates into a purely geometrical statement. However, when the theory is not Poincar\'e invariant, as for instance will happen if there is an anisotropy in the time or some of the spatial directions, it is not straightforward to find a holographic $c$-function. Several proposals exist, some if not all assume NEC \cite{Giataganas:2017koz,Chu:2019uoh,Ghasemi:2019xrl,Cremonini:2020rdx}.

In the Poincar\'e invariant cases (ABJMf, D2-D6, and D3-D7) a first check is then to see if the NEC is satisfied when the brane distribution is not monotonic in the radial direction. There are actually {\em two different NECs} that we can check, one involving the Ricci tensor of the ten-dimensional theory, and the other with the Ricci tensor of the reduced four- or five-dimensional gravity. It should be noted that holographic $c$-theorems always invoke the latter, and it is not immediately obvious whether the ten-dimensional theory has to satisfy its own NEC for a monotonic holographic $c$-function to exist. 

If the geometry is Poincar\'e invariant, in the reduced four- or five-dimensional theory there is a choice of coordinates that puts the metric in domain wall form
\be\label{eq:metdw}
 ds^2_{D+1}=du^2+e^{2A(u)}\eta_{\mu\nu}dx^\mu dx^\nu \ .
\ee
If the $(D+1)$-dimensional NEC is satisfied, a holographic $c$-function is proportional to \cite{Freedman:1999gp}
\be\label{eq:cdw}
 c_{dw}(u)=\frac{1}{(A'(u))^{D-1}} \ .
\ee
The monotonicity condition is
\be\label{eq:nec0}
 \frac{d}{du} c_{dw}(u)\geq 0\ \ \Rightarrow \ \  A''(u)\leq 0 \ \Leftrightarrow \ \text{NEC} \ .
\ee
There are other proposals that avoid selecting a particular coordinate system. A related holographic $c$-function in the reduced theory is given by the expansion of null congruences in the geometry \cite{Sahakian:1999bd,Bousso:1999xy,Alvarez:1998wr}.  The null vector field generating the congruence is of the form
\be
 {\bm k}=F(r)\left( \frac{1}{\sqrt{|g_{00}|}}\partial_0-\frac{1}{\sqrt{g_{rr}}}\partial_r\right) \ ,
\ee
where the function $F(r)$ is such that the vector field satisfies the affine condition
\be
 k^M \nabla_M k^N=0 \ .
\ee
The expansion parameter of the congruence is $\theta=\nabla_M k^M$. The holographic $c$-function is defined to be proportional to 
\be
 c_{nc}(r)=\frac{1}{\sqrt{H}\theta^{D-1}} \ ,
\ee
where $H$ is the determinant of the induced metric on surfaces of constant $x^0$ and $r$.  In the domain wall metric
\be
 {\bm k}=-e^{-2A}\left(\partial_0-e^A\partial_u\right) \ ,
\ee
from where it follows that $\theta=(D-1)A'(u) e^{-A}$, and
\be\label{eq:cnc}
 c_{nc}=\frac{1}{((D-1)A'(u))^{D-1}} \ .
\ee
The monotonicity condition is then the same as for the domain wall $c$-function.

Another similar quantity that is usually monotonic is the refraction index in the radial direction \cite{Gubser:2009gp}
\be
 {\bm n}_r=\left(\frac{g_{rr}}{|g_{00}|}\right)^{1/2} \ .
\ee 
In the domain wall geometry it is simply ${\bm n  }_r=e^{-A}$ and the monotonicity condition is
\be
 \frac{d}{du}{\bm n}_r\leq 0\ \Rightarrow \ A'(u)\geq 0 \ .
\ee
This implies that the warp factor $A$ is a monotonic function of the radial coordinate.

A different set of candidates for holographic $c$-functions that have a more direct interpretation in the field theory dual are obtained from the entanglement entropy computed using the Ryu-Takayanagi (RT) prescription \cite{Ryu:2006bv}, which makes it equal to a codimension two spatial extremal surface in the gravity dual. If one computes the entanglement entropy $S_{EE}$ for a strip of width $\ell$, the holographic $c$-function is proportional to
\be\label{eq:cEE}
 c_{EE}(\ell)=\ell^{D-1}\frac{\partial S_{EE}}{\partial \ell} \ .
\ee

The expression in (\ref{eq:cEE}) is a monotonically decreasing function of $\ell$ when the NEC is satisfied in the gravity dual and there is Poincar\'e invariance \cite{Myers:2010xs,Myers:2010tj,Myers:2012ed}. The dependence in $\ell$ can be traded for a dependence on the lowest point $u_*$ of the RT surface associated to the strip \cite{Myers:2012ed,Jokela:2020auu}. 
For a Poincar\'e invariant system that is dual to a $(D+1)$-dimensional geometry, written in the domain wall coordinates (\ref{eq:metdw}) in Einstein frame, the EE derivative becomes
\be\label{eq:ceenec}
 \frac{d}{du_*} c_{EE} \propto -\int_0^\ell dx \frac{A''}{(A')^2} \ .
\ee
Then the $c$-function is monotonic for $A''\leq 0$:
\be
 \frac{d}{d\ell}c_{EE} \leq 0 \ ,
\ee
provided $d\ell/du_*\leq 0$. In all known cases the condition on $\ell(u_*)$ holds for the extremal surfaces that determine the entanglement entropy, although other extremal surfaces that do not satisfy the condition but are not global minima of the area also exist. Note that the monotonicity condition involves an integral along the surface, so it only requires an averaged version of the NEC, which is a weaker constraint than the one required for the domain wall or null congruence $c$-functions.

When Lorentz invariance is broken, the $c$-functions defined above are generically non-monotonic \cite{Hoyos:2020zeg}. In addition, we may consider different NECs taking null vectors pointing in different directions, but they do not seem to lead to new monotonicity conditions. Nevertheless, the domain wall $c$-function admits generalizations that are monotonic in at least a larger family of geometries with broken Lorentz invariance. One example is that of \cite{Cremonini:2020rdx}. If the metric is put in domain wall form
\be\label{eq:anismet}
 ds_{D+1}=du^2+e^{2A(u)}\left(-e^{2b(u)}dt^2+d\vec{x}^2\right) \ ,
\ee
up to constant factors the candidate holographic $c$-function is
\be\label{eq:cdw1}
 c_{dw,an1}(u)=-\left( A'+\frac{1}{D}b'\right) \ .
\ee
In principle, a similar function could be defined when the anisotropic direction is spacelike instead of timelike. Another proposal is that of \cite{Giataganas:2017koz}.  In this case the anisotropic direction is spatial and the metric is
\be\label{eq:metdwanis}
 ds_{D+1}=du^2+e^{2A(u)}\left(-dt^2+d\vec{x}_\perp^2+e^{2b(u)}dx_\parallel^2\right) \ .
\ee
The proposed holographic $c$-function is
\be\label{eq:cdw2}
 c_{dw,an2}(u)=-\left(A'+\frac{1}{D-1}b'\right) e^{b/(D-1)} \ .
\ee
The proposal (\ref{eq:cdw2}) was only given for $D=4$, but we extrapolated it to arbitrary dimensions. To show that this quantity is monotonic does not require using the NEC, but it may depend on the field content of the gravity dual.

In this case with a single anisotropic direction we may consider a null vector pointing partially in the radial and field theory directions, as depicted in Fig.~\ref{fig:nullvector}, 
\be
 n^0=e^{-A}\ , \ n^u=\cos \alpha \ , \ n_\perp^i=e^{-A}\hat{n}^i\sin\alpha\cos\beta \ , \ n^\parallel=e^{-A-b} \sin\alpha\sin\beta \ ,
\ee
where $\hat{n}$ is a unit vector in the $D-2$ isotropic spatial directions. The general NEC in this case is of the form
\be\label{eq:complexcond}
 R_{MN} n^M n^N=-\cos^2\alpha {\cal R}_u-\sin^2\alpha \sin^2\beta {\cal R}_\parallel \geq 0 \ .
\ee
Here ${\cal R}_u$ is the expression for $n^M$ pointing purely in the radial direction $(\alpha=0)$ and ${\cal R}_\parallel$ is an additional contribution that vanishes in the isotropic case
\be\label{eq:anisoNEC}
 {\cal R}_u=(D-1) A''+b''+(b')^2+A'b' \ , \ {\cal R}_\parallel=b''+(b')^2+DA'b' \ .
\ee
If $b$ is constant or $\beta=0$ then there are no additional conditions from imposing the NEC in other directions. In the anisotropic case when $b$ is not constant, there can be additional conditions obtained from imposing the NEC for non-zero $\beta$. In particular, the condition (\ref{eq:complexcond}) would impose ${\cal R}_\parallel\leq 0$ for $\alpha=\pi/2$. If both ${\cal R}_u\leq 0$ and ${\cal R}_\parallel\leq 0$ then clearly the NEC is satisfied for null vectors $n^M$ pointing in any direction. So it is necessary and sufficient to impose these two conditions, although the connection of ${\cal R}_\parallel$ to the RG flow remains unclear. 

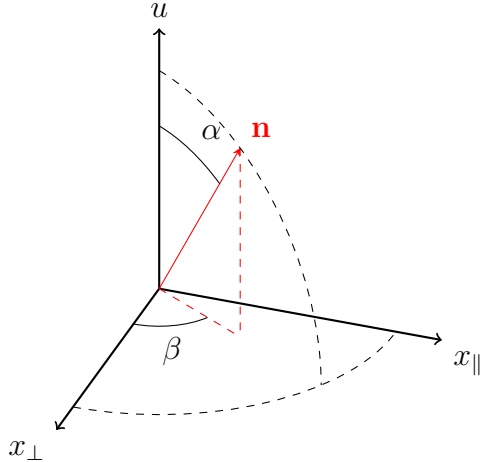
\begin{figure}\label{fig:nullvector}
\begin{center}
\begin{tikzpicture}[scale=5,tdplot_main_coords]
\coordinate (O) at (0,0,0);
\tdplotsetcoord{P}{\rvec}{\thetavec}{\phivec}
\tdplotsetcoord{X}{1.}{50.}{90.}
\draw[thick,->] (0,0,0) -- (0.8,0,0) node[anchor=north east]{$x_\perp$};
\draw[thick,->] (0,0,0) -- (0,0.8,0) node[anchor=north west]{$x_\parallel$};
\draw[thick,->] (0,0,0) -- (0,0,0.8) node[anchor=south]{$u$};
\draw[-stealth,color=red] (O) -- (P) node[above right] {${\bf n}$};
\draw[dashed, color=red] (O) -- (Pxy);
\draw[dashed, color=red] (P) -- (Pxy);
\tdplotdrawarc{(O)}{0.2}{0}{\phivec}{anchor=north}{$\beta$}
\tdplotsetthetaplanecoords{\phivec}
\tdplotdrawarc[tdplot_rotated_coords]{(0,0,0)}{0.5}{0}{\thetavec}{anchor=south west}{$\alpha$}
\draw[dashed,tdplot_rotated_coords] (\rvec,0,0) arc (0:90:\rvec);
\draw[dashed] (\rvec,0,0) arc (0:90:\rvec);
\tdplotsetrotatedcoords{\phivec}{\thetavec}{0}
\tdplotsetrotatedcoordsorigin{(P)}
\end{tikzpicture}
\end{center}\caption{\small The null vector $n^M$ is pointing in general to an arbitrary direction, which we parametrize by two angles.}
\end{figure}

The holographic $c$-function based on entanglement entropy of a strip has also been generalized for the anisotropic case \cite{Chu:2019uoh}. They consider metrics with anisotropic scaling symmetry
\be
 t\to \lambda^{n_t} t \ , \ x_i\to \lambda^{n_1} x_i \ , \ y_j\to \lambda^{n_2} y_j \ ,
\ee
with $D_1$ directions $x_i$ and $D_2$ directions $y_j$. The holographic $c$-functions are adapted to this scaling and depend on the orientation of the strip 
\be
 c_{EE,an\, a}(\ell_a)=\ell^{d_a} \frac{\partial S_{EE\,a}}{\partial \ell_a} \ ,  \ a=x,y \ ,
\ee 
where $a=x$ corresponds to a strip of width $\ell_x$ with sides separated along one of the $x$ directions, and similarly for $a=y$. The effective dimensions are
\be
 d_x=D_1+D_2\frac{n_2}{n_1} \ , \ d_y=D_2+D_1\frac{n_1}{n_2} \ .
\ee

In the next sections we will study the NEC and monotonicity of holographic $c$-functions in backgrounds with non-monotonic smeared $D$-brane distributions. From the examples we study, only the D3-D5 background was constructed previously in \cite{Hoyos:2020zeg}, while the rest are new solutions. The metric Ansatz and BPS equations were all derived previously in other constructions with smeared brane solutions, the main new ingredient here is the type of D-brane distribution we use to construct the solutions and the analysis of the NEC and holographic $c$-theorems.

We will distinguish between the NEC in ten dimensions and the NEC in the reduced theory that is linked to the holographic $c$-theorems. They are in general different so ten-dimensional NEC violations may be admissible from the point of view of the $c$-theorem. We will restrict to the NEC along a null radial direction, which is the one relevant for the holographic $c$-functions, although NEC along other directions can also be studied in the anisotropic case (see, \emph{e.g.}, \cite{Dong:2012se}). The ten-dimensional NEC is computed directly by projecting the components of the ten-dimensional energy-momentum tensor with a null vector pointing in the radial direction. In the dimensionally reduced theory the NEC can be more easily determined by putting the metric in domain wall coordinates \eqref{eq:metdw} and computing the derivative of the warp factor as in \eqref{eq:nec0}. Other quantities like the refraction indices or anisotropic $c$-functions will be computed following their definition. We will use the BPS equations that the fields of the background geometry satisfy to trade radial derivatives for combinations of those functions. In some cases this allows to identify some positive definite quantities, but in general we have to resort to numerical calculations.

 \section{ABJMf}\label{sec:ABJM}

The ABJM theory \cite{Aharony:2008ug} is a $U(N_c)\times U(N_c)$ Chern-Simons theory with level $(k,-k)$ and $\cN=6$ supersymmetry. In the large-$N_c$, large-$k$, limit with $N_c/k\gg 1$ fixed, the holographic dual is an $AdS_4\times \mathbb{CP}^3$ geometry in type IIA supergravity. Matter in the $(N_c,1)\oplus (1,N_c)$ fundamental representations and preserving $\cN=3$ supersymmetry can be added by introducing localized D$6$-branes extended in the $AdS_4$ directions and wrapping $\mathbb{RP}^3\subset \mathbb{CP}^3$ \cite{Hohenegger:2009as,Gaiotto:2009tk,Jensen:2010vx,Zafrir:2012yg}. For a large number of flavors $N_f\sim N_c$, the smeared brane configurations, preserving ${\cal N}=1$ supersymmetry, corresponding to massless and massive flavors were constructed in \cite{Conde:2011sw,Bea:2013jxa}. Generalizations to finite temperature or to noncommutative geometry are found in \cite{Jokela:2012dw,Bea:2014yda,Bea:2017iqt}.

\subsection{Solutions}

Let us discuss the ten-dimensional background geometry with smeared D6-brane sources. The 10d metric in type IIA supergravity in Einstein frame takes the following form:
\be\label{eq:ABJMmet10d}
 ds^2_{10}=e^{-\frac{\phi}{2}}\left(  h^{-\frac{1}{2}} (dx^2_{1,2}) + h^{\frac{1}{2}} \left[  \frac{e^{2g}}{r^2}dr^2 + e^{2f} (\mathcal{S}_\xi^2+\sum_{a=1}^3 \mathcal{S}_a^2)+e^{2g}\sum_{i=1}^2E_i^2 \right] \right) \ , 
\ee
where the functions $h$, $f$, and $g$, together with the dilaton $\phi$, depend on the radial coordinate $r\in [0,\infty)$.\footnote{Notice that we denote the radial coordinate as $r_\text{here}=x_\text{there}$ \cite{Bea:2013jxa}.} The one-forms $E_i$, $i=1,2$, and $\mathcal{S}_a$, $a=1,2,3,\xi$ are appropriate for the description of the internal space as a fibration of a $SU(2)$ instanton over a $S^4$ base. Explicit expressions for them will not be needed in the following, but can be found in Appendix~\ref{app:ABJMprel}. The metric is supported by the fluxes
\begin{eqnarray}
 F_2 & = &\frac{k}{2} \left[ E_1 \wedge E_2-\eta(r) (\mathcal{S}_\xi \wedge \mathcal{S}_3+\mathcal{S}_1 \wedge \mathcal{S}_2) \right] \\
 F_4 & = & G(r)\frac{e^g}{r} dt \wedge dx^1 \wedge dx^2 \wedge dr \ ,
\end{eqnarray}
where $G=3\pi^2 N_c h^{-2} e^{-4f-2g}$. In the absence of flavors $\eta=1$ and $f=g$, so that the internal space becomes $\mathbb{CP}^3$. The functions appearing in the metric satisfy the following first order BPS equations:
\begin{eqnarray}
 \phi' & = & \frac{e^g}{r} \big[-\frac{3k}{8}e^\phi h^{-\frac{1}{4}} (e^{-2g}-2\eta e^{-2f}) -\frac{1}{4}h^{\frac{3}{4}} G e^{\phi}\big]  \nonumber \\
 g' & = & \frac{e^g}{r} \big[ \frac{k}{2}h^{-\frac{1}{4}}e^{\phi-2f} \eta+e^{-g}-e^{-2f+g} \big] \nonumber \\
 f' & = & \frac{e^g}{r}  \big[ \frac{k}{2}h^{-\frac{1}{4}}e^\phi(\eta e^{-2f}-e^{-2g})+e^{-2f+g} \big] \nonumber \\
 h' & = & \frac{e^g}{r} \big[ \frac{k}{2}h^{\frac{3}{4}}e^\phi (e^{-2g}-2 \eta e^{-2f}) -h^{\frac{7}{4}}G e^{\phi} \big] \label{eq:ABJMBPSeq} \ .
\end{eqnarray} 
The reduction to four dimensions detailed in Appendix~\ref{app:ABJMred} produces the following metric
\be\label{eq:ABJMmet4d}
 ds_4^2 = he^{4f+2g-2\phi} \left(\eta_{\mu\nu}dx^\mu dx^\nu+h e^{2g}\frac{dr^2}{r^2}\right) \ .
\ee
To get solutions representing flavor, one should look for solutions in which $\eta$ goes to a constant in the UV, $r \rightarrow \infty$. When $\eta=1$ there is a solution to the equations which gives asymptotically $AdS_4$. We will be interested now in solutions in which the profile goes to zero in the UV. We will require $T_{00}>0$. This constrains $\eta$ to decay not faster than $r^{-2}$ as $r\to\infty$. To solve the BPS equations, we will consider small deviations from $\eta=1$, and solve perturbatively in the would be 'flavor' deformation $\hat\epsilon\sim Q_f/Q_c$.  Let us consider the following profile:
\be\label{eq:ABJMprofile}
 \eta(r)=1+\hat\epsilon p(r)=1+\hat\epsilon \frac{r^n}{(1+r^m)^{\frac{n+2}{m}}} \ .
\ee
The full solution can be obtained from a master function \cite{Bea:2013jxa} that we expand in powers of $\hat\epsilon$:
\be
 W(r)=W_0(r) +\hat\epsilon W_1(r)+\hat\epsilon^2W_2(r)+\hat\epsilon^3W_3(r)+\ldots \ ,
\ee
where $W_0=2r$. Each of the metric functions and the dilaton have a similar expansion, which, at leading order in $\hat{\epsilon}$ reads
\bea
 e^g & \approx & \frac{r}{2^{\frac{1}{3}}}(1+\hat\epsilon g_1) \\
 e^f & \approx & \frac{r}{2^{\frac{1}{3}}}(1+\hat\epsilon f_1)  \\
 h & \approx & \frac{4\pi^2N_c 2^{\frac{1}{3}}}{kr^4}(1+\hat\epsilon h_1) \\
 e^\phi & \approx & \frac{2^\frac{5}{4}N_c^\frac{1}{4}}{k^{\frac{5}{4}}}\sqrt{\pi}(1+\hat\epsilon \phi_1) \ .
\eea
In these equations $f_1,g_1,h_1,\phi_1$ are functions of $r$ as well as of the parameters $m$ and $n$ through the profile (\ref{eq:ABJMprofile}). Their explicit expressions are written in (\ref{eq:ABJMg1})-(\ref{eq:ABJMphi1}) of Appendix~\ref{app:ABJMper} where we detail the perturbative expansion in $\hat\epsilon$.

 \subsection{Null Energy Condition in ten and four dimensions}

The ten-dimensional NEC is straightforward to obtain from the energy-momentum tensor of SUGRA fields and branes. Projecting on a null vector pointing in the radial direction $n^M$,
\be
 T_{{}_M{} _N}^{\rm 10d} n^M n^N = \frac{e^{-8f-4g+\frac{3}{2}\phi}}{128h^3}[9e^{\phi}(2N_c\pi^2+e^{4f}kh-2e^{2f+2g}kh\eta)^2+64 e^{6f+3g}kr h^{\frac{9}{4}}\eta'] \ . 
\ee
One can check, expanding for small $\hat\epsilon$, that
\be
T_{{}_M{} _N}^{\rm 10d} n^M n^N\simeq \hat\epsilon \frac{2^{\frac{1}{8}} }{k^{\frac{1}{8}}N_c^{\frac{3}{8}}\pi^{\frac{3}{4}}}r p' = \hat\epsilon \frac{2^{\frac{1}{8}} }{k^{\frac{1}{8}}N_c^{\frac{3}{8}}\pi^{\frac{3}{4}}}\frac{r^n(n-2r^m)}{(1+r^m)^{\frac{n+2+m}{m}}} \ .
\ee  
  
We will now expand for large and small values of the radial coordinate, using the results in Appendix~\ref{app:ABJMexp}. Using the UV expansions of the solutions, for $r\to\infty$, 
\be
 T_{{}_M{} _N}^{\rm 10d} n^M n^N\simeq  -2\hat\epsilon \frac{2^{\frac{1}{8}} }{k^{\frac{1}{8}}N_c^{\frac{3}{8}}\pi^{\frac{3}{4}}}\frac{1}{r^2} \leq 0 \ .
\ee 
On the other hand, in the IR $r\to 0$, if $n>0$,
\be
 T_{{}_M{} _N}^{\rm 10d} n^M n^N\simeq n\hat\epsilon \frac{2^{\frac{1}{8}} }{k^{\frac{1}{8}}N_c^{\frac{3}{8}}\pi^{\frac{3}{4}}}r^n  \geq 0 \ .
\ee 
Therefore, the ten-dimensional NEC is always satisfied in the IR and violated in the UV.

Let us now check the NEC in four dimensions. To this end, let us transform the metric \eqref{eq:ABJMmet4d} in domain wall form \eqref{eq:metdw} through the change of radial coordinate
\be
 \frac{du}{dr}= \frac{he^{2f+2g-\phi}}{r}\equiv \partial_r u \ .
\ee
The warp factor is
\be
 A=\frac{1}{2}\log h+2 f+  g-\phi \ .
\ee
The NEC in four dimensions is equivalent to the condition \cite{Freedman:1999gp}
\be
 A''(u)\leq 0 \ .
\ee
The radial derivative of the warp factor is, after using the BPS equations \eqref{eq:ABJMBPSeq},
\bea
 A'(u) & = & \frac{1}{\partial_r u}\left[\frac{h'}{2h}+2 f'+g'-\phi'\right] \nonumber \\
       & = & \frac{1}{r\partial_r u}\left[1+e^{2g-2f}-\frac{1}{4} \frac{3\pi^2 N_c e^{-4 f-g+\phi}}{h^{5/4}}-\frac{k e^{-g+\phi}}{h^{1/4}}\left( \frac{3}{8}-\frac{1}{4}e^{-2f+2g}\eta\right) \right] \ .
\eea
In the following it will be convenient to use the functions
\be
F\equiv e^{2g-2f} \ ,\   P\equiv e^{-2f+2g}\eta \ ,\  H\equiv  \frac{3\pi^2 N_c e^{-4 f-g+\phi}}{h^{5/4}} \ ,\  K=\frac{k e^{-g+\phi}}{h^{1/4}} \ .
\ee
The second derivative gives
\bea
 A''(u) & = & -\frac{1}{r^2(\partial_r u)^2}\left[2+4F^2+\frac{7}{16}H^2-\frac{3}{2}H(1+F)-\frac{1}{4}K(5+3F)+\frac{9}{16}K\left(H-\frac{K}{4}\right)\right. \nonumber \\
& & \left. +KP\left(\frac{1}{2}+F-\frac{K}{16}-\frac{3 H}{8}-\frac{1}{4}\frac{x \eta'}{\eta} \right)-\frac{K^2P^2}{16}  \right] \ .\label{eq:NECABJM}
\eea
Expanding for a small number of flavors one finds
\be\label{eq:NECABJM2}
A''(u)\simeq -\frac{\hat\epsilon}{r^2(\partial_r u)^2}\left[ \frac{1}{2}\left(20 g_1-16 f_1-4\phi_1+h_1-r p'\right) \right] \ .
\ee
The term in parenthesis can be simplified to
\be\label{eq:necplot}
\frac{1}{2}\left(20 g_1-16 f_1-4\phi_1+h_1-r p'\right)=6 \frac{W_1'+4p}{W_0'+4}-\frac{W_1}{W_0}-\frac{r}{2}p' \ ,
\ee
where $W_0$ and $W_1$ are given in Appendix~\ref{app:ABJMper}. Using the UV expansions of the solutions we find
\be
A''(u)\simeq -\frac{\hat\epsilon}{r^2(\partial_r u)^2}\left[ \frac{2}{r^2} \right]\leq 0  \ , \ r\to \infty \ .
\ee 
Therefore, the NEC is always satisfied in the UV. On the other hand, in the IR:
\be
A''(u)\simeq -\frac{\hat\epsilon}{r^2(\partial_r u)^2}\left[ \frac{4-4n-n^2}{2(n+4)} r^n\right] \ , \ r\to 0 \ .
\ee 
For $n<2(\sqrt{2}-1)\approx 0.828$ the term in the parenthesis is positive and the NEC is satisfied. For any larger values $n>2(\sqrt{2}-1)$, the NEC is violated in the IR. When the NEC is satisfied both in the IR and UV, we find by numerical analysis that NEC is also satisfied in intermediate regions as shown in Fig.~\ref{fig:abjmnec}. Recall that we are in an approximation with a small number of flavors $\hat\epsilon\ll 1$, this conclusion therefore may change when the number of flavors is significant $\hat\epsilon\sim 1$.

\begin{figure}[h!]
\begin{center}
\begin{tabular}{cc}
\includegraphics[width=0.45\textwidth]{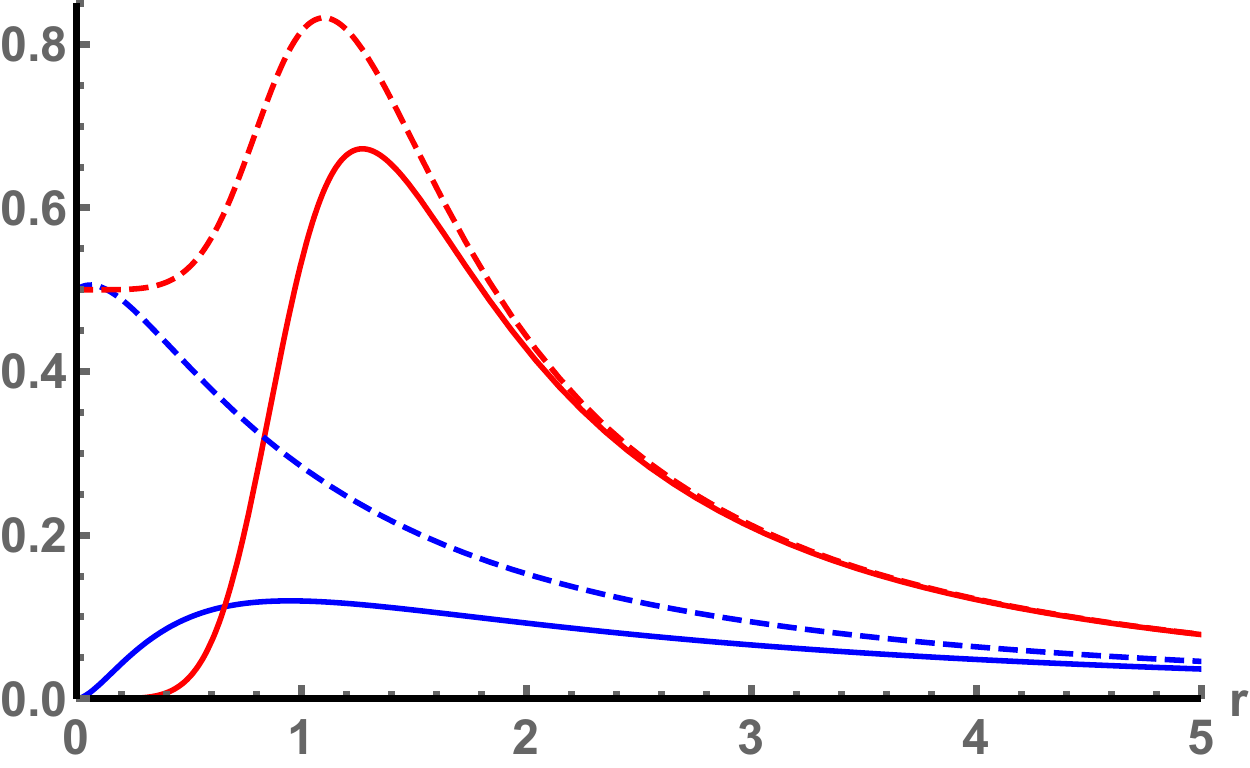} \includegraphics[width=0.45\textwidth]{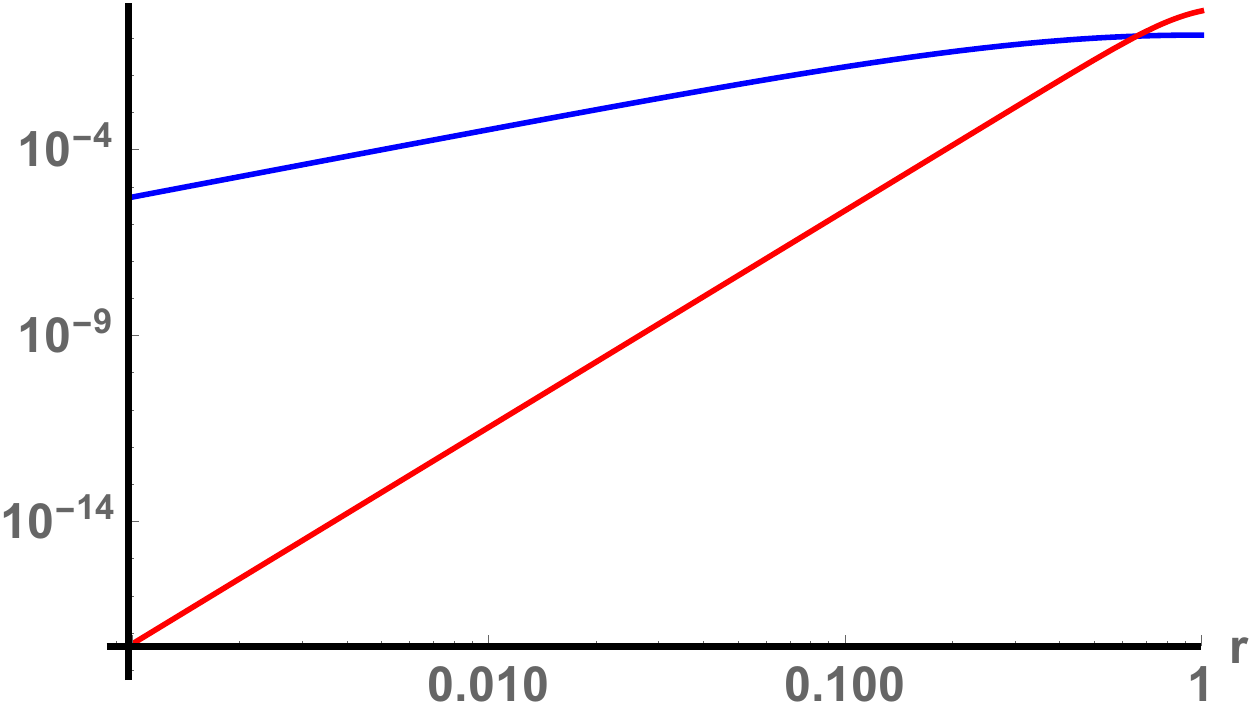}
\end{tabular}
\end{center}
\caption{\small  In these plots we depict the quantity in square parenthesis in \eqref{eq:NECABJM2}, given in \eqref{eq:necplot}. Both plots are for solutions with  $n=0$ (dashed), $n=2(\sqrt{2}-1)$ (solid) and $m=1$ (blue), $m=5$ (red). If the curves are positive then it means that the four-dimensional NEC holds.  We observe that the NEC is satisfied for the full range of $r$.}\label{fig:abjmnec}
\end{figure}

\subsection{Holographic $c$-functions}

\begin{itemize}

\item {\bf Domain wall $c$-function and null congruences.} In this case the NEC is satisfied in the UV, while in the IR is generically violated except for small values of $n$, $0\leq n\leq 2(\sqrt{2}-1)$. Therefore the $c$-functions will be monotonic for values of $n$ in this range, at least for small number of flavors considered in this work.

\item {\bf Refraction index in the radial direction.} From either \eqref{eq:ABJMmet10d} or \eqref{eq:ABJMmet4d} we find an explicit expression for the refraction index ${\bm n}_r$: 
\be
 {\bm n}_r=\frac{e^g h^{1/2}}{r} \ .
\ee
The derivative along the radial direction gives
\be
 \frac{d}{dr} {\bm n}_r={\bm n}_r\left[ g'-\frac{1}{r}+\frac{1}{2}\frac{h'}{h}\right] \ .
\ee
We are left with computing the combination inside the parenthesis. By using the BPS equations \eqref{eq:ABJMBPSeq} we find 
\be\label{eq:refindABJM}
 g'-\frac{1}{r}+\frac{1}{2}\frac{h'}{h}=-\frac{1}{r}  \left[F+\frac{H}{2}-\frac{K}{4}\right] = -\frac{2}{r}+{\cal O}(\hat\epsilon) \ .
\ee
Therefore, in the limit of a small number of flavors the refraction index is indeed decreasing for any radial coordinate and hence the dual quantum field theory is expected to be causal.

\item {\bf Entanglement entropy.} In the cases where the NEC is satisfied, $n<2(\sqrt{2}-1)$, the associated $c$-function is monotonic. Since the NEC is always satisfied in the UV, it is possible that it is also monotonic in the cases where there is some violation of the NEC in the IR, as \eqref{eq:ceenec} involves an average. If this were the case, this would be an example where the one-to-one correspondence between the NEC and the monotonicity of the RG flow is broken.

\end{itemize}

 \section{D2-D6}\label{sec:D2D6}

The D$2$-brane geometry was first considered as an example of a holographic dual of a non-conformal theory, $\cN=8$ supersymmetric Yang-Mills theory in $2+1$ dimensions \cite{Itzhaki:1998dd}. By putting the D$2$-branes on a $G_2$-holonomy cone supersymmetry is reduced to $\cN=1$ \cite{Acharya:1998db}. Flavors can be added by introducing D$6$-branes in the D$2$-brane geometry \cite{Arean:2006pk,Myers:2006qr,Itsios:2016ffv}. Backreacted geometries preserving ${\cal N}=1$ supersymmetry, produced by smeared D$6$-branes were constructed in \cite{Faedo:2015ula}.

\subsection{Solutions}

The ten-dimensional metric and forms in type IIA supergravity in Einstein frame are the following:
\bea
 ds_{10}^2 & = & e^{-\frac{\phi}{2}}[h^{-1/2}\eta_{\mu\nu}dx^\mu dx^\nu+h^{1/2}e^{2\chi}\left( dr^2+r^2ds_6^2\right)] \label{eq:d2d6met10d}\\
 F_2 & = & Q_f p\, J \\
 F_6 & = & \frac{Q_c}{6} \, J\wedge J \wedge J \\
 e^\phi & = & h^{1/4}e^{3\chi} \ ,
\eea
where $ds_6^2$ is the metric of a compact six-dimensional nearly-K\"ahler manifold. These type of manifolds are endowed with a fundamental two-form $J$, which is used to construct the RR two- and six-forms of our ansatz following \cite{Faedo:2015ula}.  The precise expression of the two-form $J$ along the internal directions  in a specific example can be found in Appendix~\ref{app:D2D6prel}. 
The functions $h$ and $\chi$ depend on the radial coordinate $r$ and are determined by BPS equations:
\bea
 \chi' & = & Q_f\frac{e^{2\chi}}{r^2}p  \nonumber \\
 h' &= & -\frac{Q_c}{r^6}e^{-2\chi}-3 Q_f \frac{e^{2\chi}}{r^2}p h \ . \label{eq:d2d6BPSeq}
\eea
Solutions to this system of equations with a distribution of the form \eqref{eq:pdist} can be found in Appendix~\ref{app:D2D6prel}. The dimensional reduction to four dimensions is detailed in Appendix~\ref{app:D2D6red}. For the purposes of discussing the null energy conditions and the $c$-functions in subsequent sections we need the reduced metric, which reads 
\be\label{eq:d2d6met4d}
 ds_4^2=h r^6 e^{-2\phi+6\chi}  \left( \eta_{\mu\nu}dx^\mu dx^\nu+h e^{2\chi}dr^2\right) \ .
\ee

 \subsection{Null Energy Condition in ten and four dimensions}
 
The ten-dimensional NEC is again straightforward to obtain from the energy-momentum tensor of SUGRA fields and branes. Projecting on a null vector pointing in the radial direction $n^M$,
\be\label{eq:NEC10D2D6}
 T_{{}_M{} _N}^{\rm 10d} n^M n^N=\frac{e^{-\frac{1}{2}\chi}}{32r^2h^{\frac{3}{8}}}\left[\frac{Q_c^2e^{-4\chi}}{r^{10}h^2}+3Q_f \frac{e^{2\chi} p}{r}\left(-6\frac{Q_c e^{-2\chi}}{r^5 h}+27Q_f\frac{e^{2\chi}p}{r}+16\frac{rp'}{ p}\right)\right] \ . 
\ee
  
The expansions of $h$ and $\chi$ for large and small values of the radial coordinate $r$ are written in Appendix~\ref{app:D2D6exp}. Using the UV expansions of the solutions,  
\be
 T_{{}_M{} _N}^{\rm 10d} n^M n^N\simeq  25\frac{e^{-\frac{1}{2}\chi}}{32r^2h^{\frac{3}{8}}}  \geq 0 \ , \ r\to\infty \ .
\ee 
On the other hand, in the IR, if $n>1$,
\be
 T_{{}_M{} _N}^{\rm 10d} n^M n^N\simeq 25\frac{e^{-\frac{1}{2}\chi}}{32r^2h^{\frac{3}{8}}}   \geq 0 \ , \ r\to 0
\ee 
while if $n<1$,
\be
 T_{{}_M{} _N}^{\rm 10d} n^M n^N\simeq n(n+24)\frac{e^{-\frac{1}{2}\chi}}{32r^2h^{\frac{3}{8}}}  \geq 0 \ , \ r\to 0 \ .
\ee 
Therefore, the ten-dimensional NEC is always satisfied in the IR and UV for any number of flavors $Q_f$.  

Question arises if the ten-dimensional NEC is satisfied in the intermediate energy scales. It turns out to depend on the parameters $n,m$ of the profile, especially if the number of flavors $Q_f$ is larger than some critical value. In Fig.~\ref{fig:d2d6nec10} we illustrate how the positivity of (\ref{eq:NEC10D2D6}) varies with these parameters.

\begin{figure}[h!]
\begin{center}
\begin{tabular}{cc}
\includegraphics[width=0.45\textwidth]{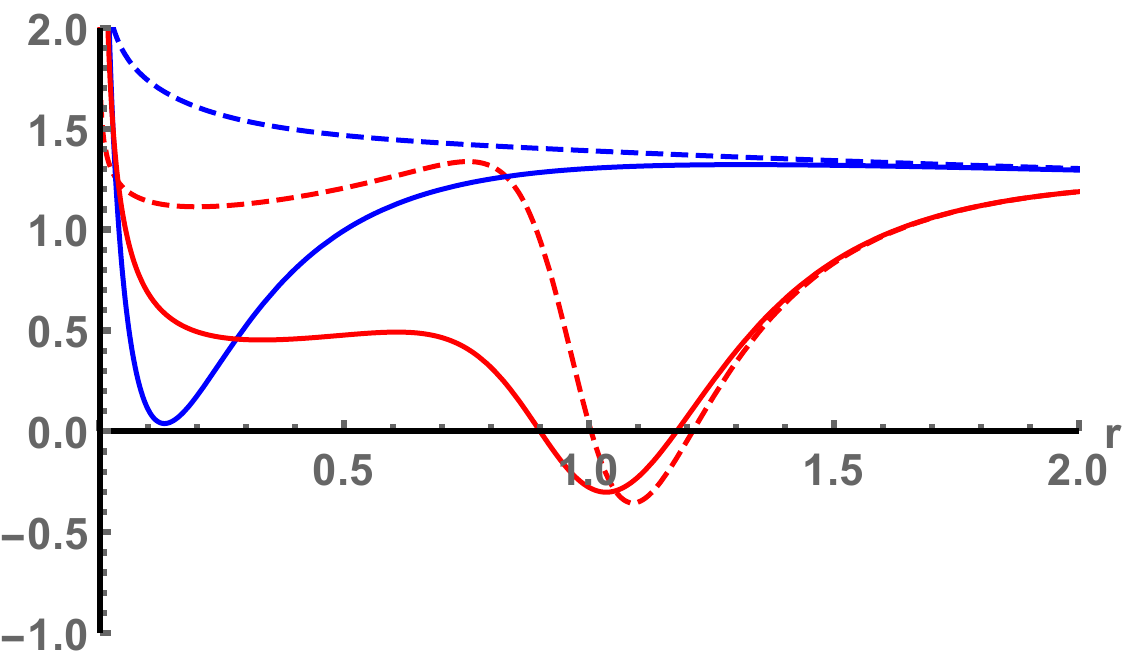} &\includegraphics[width=0.45\textwidth]{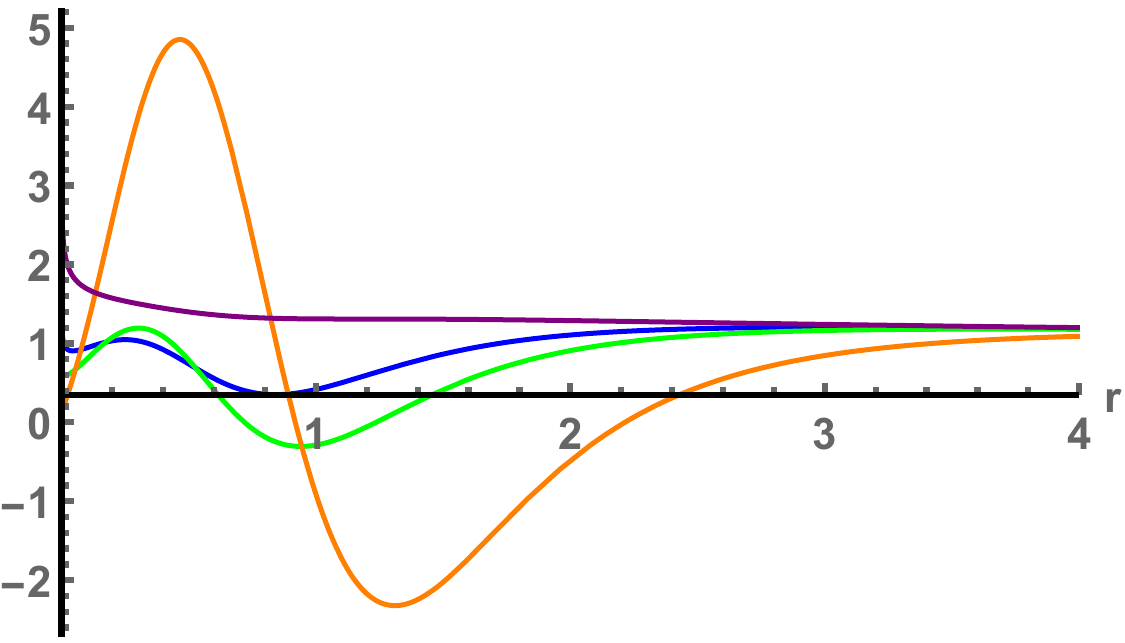}
\end{tabular}
\end{center}
\caption{\small In these plots we depict the quantity in square parenthesis in \eqref{eq:NEC10D2D6}. The ten-dimensional NEC is satisfied whenever this quantity is positive.  {\bf{Left}}: $Q_f=1/4$ and varying the values of $(n,m)=(1/2,1)$ (solid blue), $(1/2,10)$ (solid red), $(2,1)$ (dashed blue), and $(2,15)$ (dashed red). {\bf{Right}}: $(n,m)=(2,2)$ and varying $Q_f=1/10$ (purple), $1$ (blue), $2$ (green), $10$ (orange). We observe that the NEC${}_{10}$ is satisfied for low enough values of $m$ and $Q_f$ but is violated in some range of the radial coordinate as $m$ or $Q_f$ is increased above some critical value.}\label{fig:d2d6nec10}
\end{figure}

Let us now check the NEC in four dimensions. We can transform the metric \eqref{eq:d2d6met4d} in domain wall form \eqref{eq:metdw} through the change of radial coordinate
\be
 \frac{du}{dr}=h r^3 e^{-\phi+4\chi}\equiv \partial_r u \ .
\ee
The warp factor is
\be
 A=\frac{1}{2}\log h+3\log r-\phi+3\chi \ .
\ee
The NEC in four dimensions is equivalent to the condition \cite{Freedman:1999gp}
\be
 A''(u)\leq 0 \ .
\ee
The radial derivative of the warp factor is, after using the BPS equations \eqref{eq:d2d6BPSeq},
\be
 A'(u)=\frac{1}{\partial_r u}\left[\frac{h'}{2h}+\frac{3}{r}-\phi'+3 \chi'\right]=\frac{1}{4r\partial_r u}\left[12-\frac{Q_c e^{-2\chi}}{r^5 h}-3Q_f\frac{e^{2\chi}p}{r} \right] \ .
\ee
In the following it will be convenient to use the functions
\be
 P\equiv Q_f\frac{e^{2\chi}p}{r}  \ , \ H\equiv \frac{Q_c e^{-2\chi}}{r^5 h} \ .
\ee
The second derivative gives
\be\label{eq:NECD2D6}
 A''(u)=-\frac{1}{r^2(\partial_r u)^2}\left[12 +\frac{7}{16}H^2-\frac{9}{2}H+\frac{39}{16}P^2-\frac{9}{8} HP-\frac{3}{4}P\left(10-\frac{rp'}{p}\right) \right] \ .
\ee

Using the UV expansions of the solutions,
\be
 A''(u)\simeq -\frac{1}{r^2(\partial_r u)^2}\left[ \frac{7}{16}\right]\leq 0 \ , \ r\to\infty
\ee 
On the other hand, in the IR $r\to 0$, if $n>1$
\be
 A''(u)\simeq -\frac{1}{r^2(\partial_x u)^2}\left[ \frac{7}{16}\right]\leq 0 \ , \ r\to 0 \ ,
\ee 
while if $n<1$,
\be
 A''(u)\simeq -\frac{1}{r^2(\partial_x u)^2}\left[ \frac{n(n+6)}{16}\right]\leq 0 \ , \ r\to 0 \ .
\ee 
The NEC is therefore satisfied both in the IR and in the UV. By numerical analysis we also find that the four-dimensional NEC is also satisfied in intermediate regions as long as $Q_f$ is not too large; see Fig.~\ref{fig:d2d6nec}.

\begin{figure}[h!]
\begin{center}
\begin{tabular}{cc}
\includegraphics[width=0.45\textwidth]{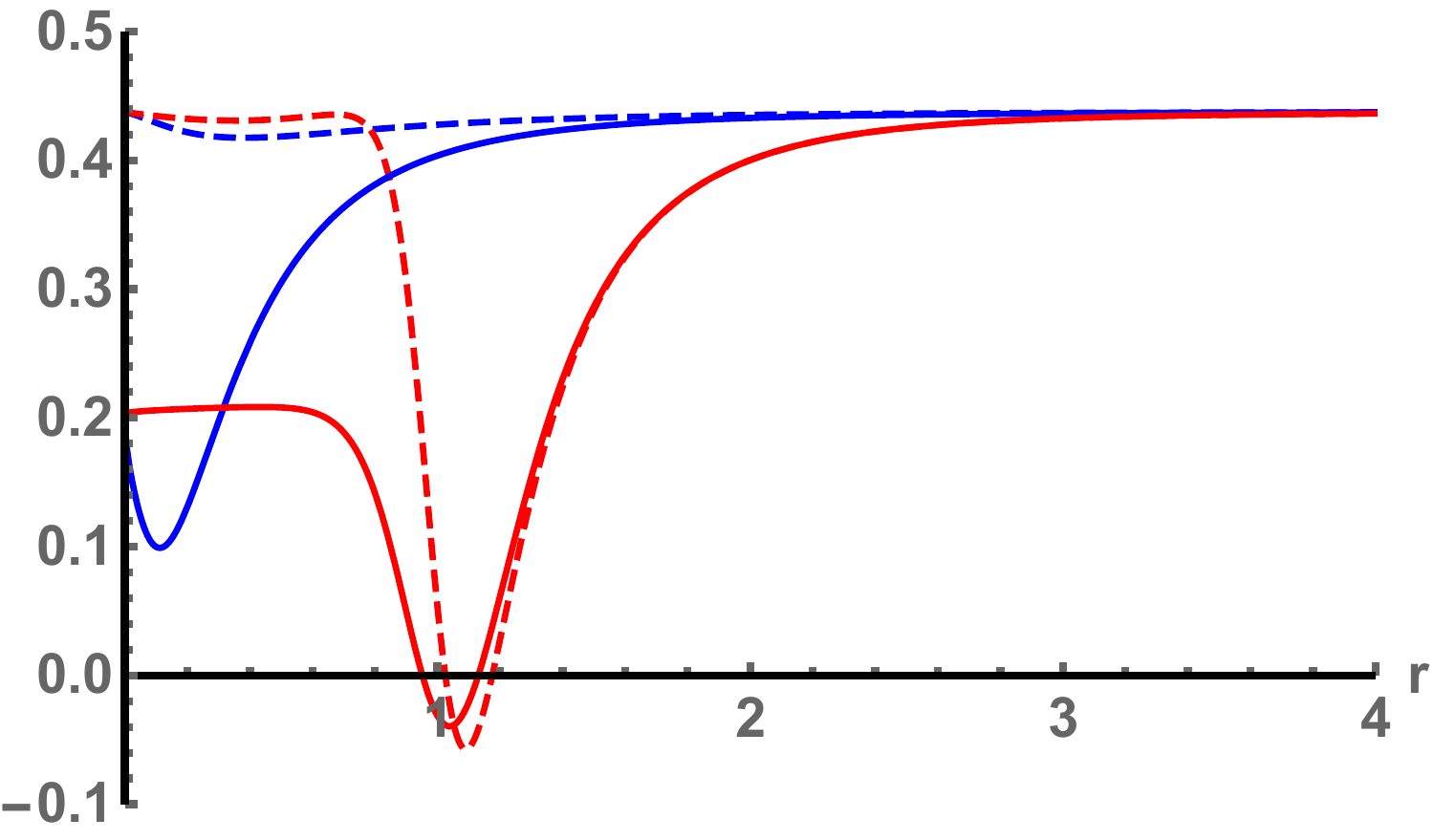} &\includegraphics[width=0.45\textwidth]{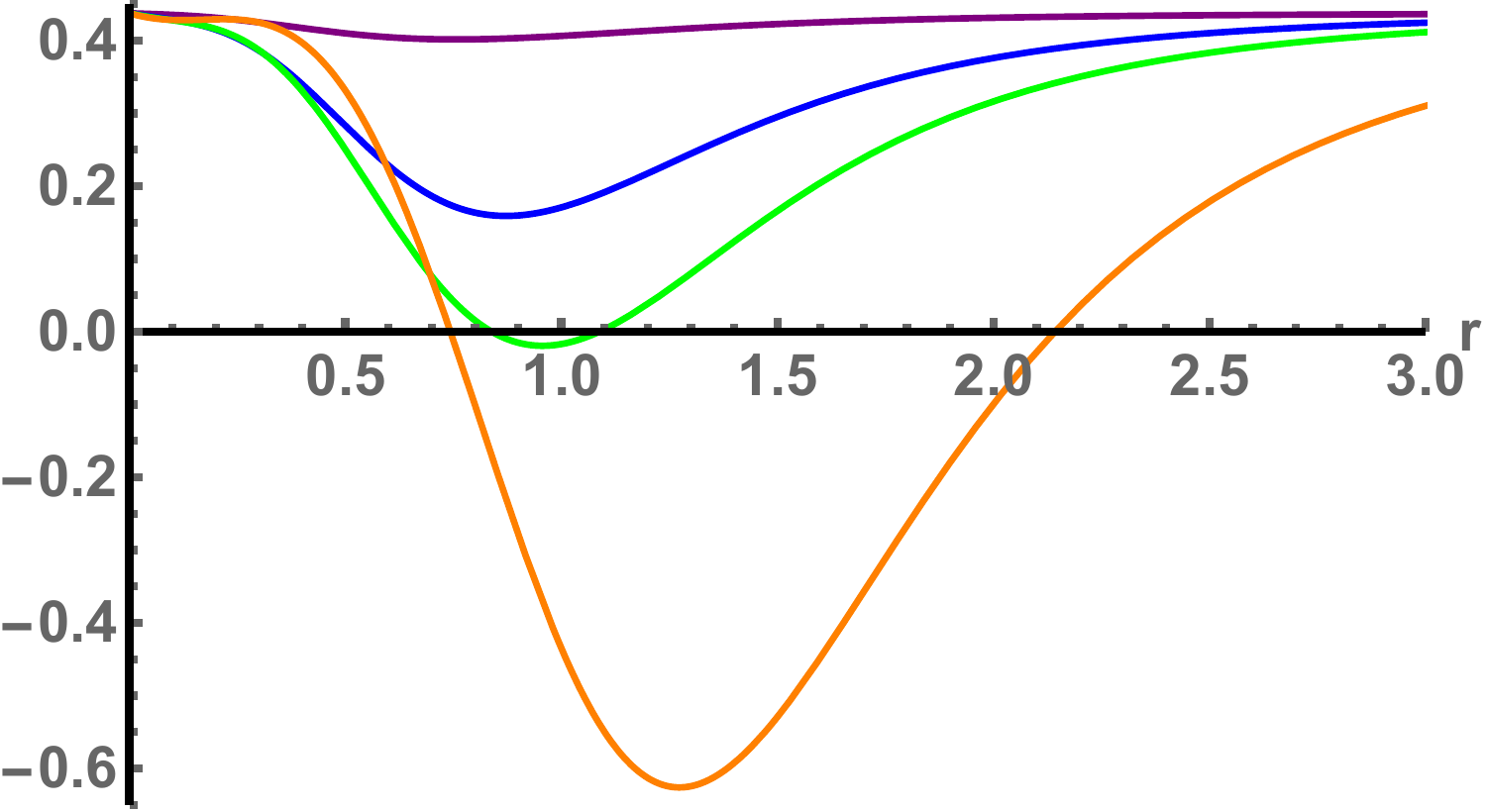}
\end{tabular}
\end{center}
\caption{\small In this plot we depict the quantity in square parenthesis in \eqref{eq:NECD2D6}. The four-dimensional NEC is violated if this quantity becomes negative. {\bf{Left}}: $Q_f=1/4$ and varying the values of $(n,m)=(1/2,1)$ (solid blue), $(1/2,10)$ (solid red), $(2,1)$ (dashed blue), and $(2,15)$ (dashed red). {\bf{Right}}: $(n,m)=(2,2)$ and varying $Q_f=1/10$ (purple), $1$ (blue), $2$ (green), and $10$ (orange). We observe that the NEC is satisfied for low enough values of $m$ and $Q_f$ but is violated in some range of the radial coordinate if $m$ or $Q_f$ is increased above some critical value.}\label{fig:d2d6nec}
\end{figure}

\subsection{Holographic $c$-functions}

\begin{itemize}

\item {\bf Domain wall $c$-function and null congruences.} In this case, if the number of flavors is not too large, the NEC is satisfied everywhere and the $c$-functions are monotonically increasing towards the UV. If $Q_f$ is large enough, there will be violations of the NEC at intermediate values of the radial coordinate and in those regions the $c$-functions will cease to be monotonic. Nevertheless, if the regions where this happens are small enough, a weak version of a $c$-theorem could still hold, in the sense that the UV value of the holographic $c$-function would be larger than the IR value even if the function is not monotonic along the full flow.

\item {\bf Refraction index in the radial direction.} We get an explicit expression for the refraction index from either \eqref{eq:d2d6met10d} or \eqref{eq:d2d6met4d} as follows 
\be
 {\bm n}_r=e^\chi h^{1/2} \ .
\ee
The derivative along the radial direction gives
\be
 \frac{d}{dx} {\bm n}_r={\bm n}_r\left[ \chi'+\frac{h'}{2h}\right] \ .
\ee
By using the BPS equations \eqref{eq:d2d6BPSeq} we find 
\be\label{eq:refindD2D6}
 \chi'+\frac{h'}{2h}=-\frac{1}{2r}  \left[H+P\right] \ .
\ee
The quantity in parenthesis is manifestly positive for any value of the radial coordinate, so the refraction index is monotonically decreasing throughout the geometry.

\item {\bf Entanglement entropy.} The associated $c$-function will be monotonic in the cases where the NEC is satisfied everywhere. When the NEC is violated in a finite region along the radial coordinate there is the possibility that it is still monotonic, as \eqref{eq:ceenec} involves an average. This could be another example breaking the one-to-one correspondence between the NEC and the monotonicity of the RG flow.

\end{itemize}

\section{D3-D7}\label{sec:D3D7}

This is the paradigmatic example of a holographic dual with flavor \cite{Karch:2002sh,Kruczenski:2003be}. The low energy theory on the D$3$-branes is $\cN=4$ super Yang-Mills, that has an $AdS_5\times S^5$ geometry in type IIB supergravity as holographic dual \cite{Maldacena:1997re}. This can be generalized to $\cN=1$ supersymmetric theories with $AdS_5\times X^5$ dual by placing the branes on a Calabi-Yau cone \cite{Klebanov:1998hh}. A small number of flavors can be introduced preserving $\cN=2$ (for the $\cN=4$ dual) or $\cN=1$ supersymmetry by adding in the dual probe D$7$-branes extended in $AdS_5$ directions \cite{Karch:2002sh,Ouyang:2003df}. For a large number of flavors, the backreacted geometries produced by smeared D$7$-branes were constructed in \cite{Benini:2006hh}. The smeared D7-branes preserve $\cN=1$ supersymmetry in all cases.
 
\subsection{Solutions}

The ten-dimensional metric is 
\bea
 ds_{10}^2 & = & h^{-1/2}\eta_{\mu\nu}dx^\mu dx^\nu+h^{1/2}\frac{e^{2f}}{r^2}dr^2+e^{2f}(d\tau+A)^2+e^{2g}ds^2_{KE} \label{eq:d3d7met10d}\\
 F_1 & = & Q_f p (d\tau+A) \\
 F_5 & = & Q_c \frac{e^{-4g}}{r h^2}(1+\star)( dx^0 \wedge dx^1 \wedge dx^2 \wedge dx^3 \wedge dr) \ ,
\eea
where $A$ is a one-form and the functions $h$, $f$, $g$, and $p$, together with the dilaton $\phi$, depend on the radial coordinate $r\in [0,\infty)$. The internal space is constructed by squashing a five-dimensional  Sasaki-Einstein manifold obtained as a $U(1)$ fibration over a K\"ahler-Einstein base. Our Ansatz above is valid for an arbitrary five-dimensional Sasaki-Einstein space. As a concrete example, in Appendix~\ref{app:D3D7prel}} we consider the case of the five-sphere ${\mathbb S}^5$, for which the  K\"ahler-Einstein base is  ${\mathbb C} {\mathbb P}^2$. In that example we give the expressions for the one-form $A$ and for the metric $ds^2_{KE}$ in a suitable system of coordinates.

The BPS equations are
\bea
 g' & = & \frac{1}{r}e^{2f-2g} \nonumber \\
 f' & = & \frac{3}{r}-\frac{2}{r} e^{2f-2g}-\frac{Q_f}{2}\frac{e^\phi}{r}\, p \nonumber \\
 \phi' & = & \frac{Q_f}{r}e^\phi\, p \nonumber \\
 h' & = & -\frac{Q_c}{r}e^{-4g} \ . \label{eq:d3d7BPSeq}
\eea
The explicit integration of this system of equations is performed in Appendix~\ref{app:D3D7prel}. The reduction to five dimensions is detailed in Appendix~\ref{app:D3D7red}. The resulting five-dimensional metric reads
\be\label{eq:d3d7met5d}
 ds_5^2=h^{4/3}e^{8(f+g)/3}  \left( \frac{e^{-2f }}{h} \eta_{\mu\nu}dx^\mu dx^\nu+\frac{dr^2}{r^2}\right) \ .
\ee

\subsection{Null Energy Condition in ten and five dimensions}
 
The ten-dimensional NEC is straightforward to obtain from the energy-momentum tensor of SUGRA fields and branes. Projecting on a null vector pointing in the radial direction $n^M$,
\be
 T_{{}_M{} _N}^{\rm 10d} n^M n^N=\frac{1}{2} h^{-1/2}e^{-2f}\, Q_f e^\phi p\, \left(Q_f e^\phi p +\frac{r p'}{p} \right) \ .
\ee
The expansion of the solutions for small and large values of the radial coordinate is collected in Appendix~\ref{app:D3D7exp}. Using the UV expansions of the solutions, 
\be
 T_{{}_M{} _N}^{\rm 10d} n^M n^N\simeq -\frac{1}{2} h^{-1/2}e^{-2f}\, \frac{4 Q_f}{r^4}\leq 0 \ , \ r\to\infty \ .
\ee 
On the other hand, in the IR
\be
 T_{{}_M{} _N}^{\rm 10d} n^M n^N\simeq +\frac{1}{2} h^{-1/2}e^{-2f}\,  n Q_f e^{\phi_0} r^n\geq 0 \ , \ r\to 0
\ee 
Therefore, the ten-dimensional NEC is always satisfied in the IR and violated in the UV for any nonzero $Q_f$. 

Let us now check the NEC in five dimensions. We can transform the metric \eqref{eq:d3d7met5d} in domain wall form \eqref{eq:metdw} through the change of radial coordinate
\be
 \frac{du}{dr}=\frac{h^{2/3} e^{4(f+g)/3}}{r}\equiv \partial_r u \ .
\ee
The warp factor is
\be
 A=\frac{1}{3}\log h+\frac{2}{3}(f+4g) \ .
\ee
The NEC in five dimensions is equivalent to the condition \cite{Freedman:1999gp}
\be
 A''(u)\leq 0 \ .
\ee
The radial derivative of the warp factor is, after using the BPS equations \eqref{eq:d3d7BPSeq},
\be
 A'(u)=\frac{1}{3\partial_r u}\left[\frac{h'}{h}+2f'+8 g'\right]=\frac{1}{3}h^{-2/3}e^{-4(f+g)/3}\left[6+4 e^{2f-2g}-\frac{Q_c e^{-4g}}{h}-Q_f e^\phi p \right] \ .
\ee
In the following it will be convenient to use the functions
\be
F\equiv e^{2f-2g} \ , \ P\equiv Q_f e^\phi p \ ,  \ H\equiv \frac{Q_c e^{-4g}}{h} \ .
\ee
The second derivative gives
\bea
 A''(u) & = & -\frac{8}{r^2(\partial_r u)^2}\left[1+\frac{2}{3}F\left(\frac{7}{6}F-1 \right)+\frac{5}{72}H^2-\frac{1}{9}H\left( 3+2 F\right)\right. \nonumber\\
 & & \left. +\frac{5}{72} P^2+\frac{1}{18}(H+2F)P+\frac{1}{24}P\left(\frac{r p'}{p}-8 \right) \right] \ .
\eea

Using the UV expansions of the solutions,
\be
 A''(u)\simeq +\frac{8}{r^2(\partial_r u)^2} \frac{Q_f}{6 r^4}\geq 0 \ , \ r\to\infty \ .
\ee 
On the other hand, in the IR
\be
 A''(u)\simeq -\frac{8}{r^2(\partial_r u)^2} \frac{n}{24} Q_f e^{\phi_0} r^n\leq 0 \ , \ r\to 0 \ .
\ee 
Therefore, the five-dimensional NEC behaves similarly to the ten-dimensional NEC, it is always satisfied in the IR and violated in the UV for any nonzero $Q_f$.

\subsection{Holographic $c$-functions}

\begin{itemize}

\item {\bf Domain wall $c$-function and null congruences.}  Since the NEC changes sign, the $c$-functions are non-monotonic. In addition, since  $A''\geq 0$ in the UV, so the $c$-function is decreasing towards the UV, instead of increasing.

\item {\bf Refraction index in the radial direction.} We find the refraction index from either \eqref{eq:d3d7met10d} or \eqref{eq:d3d7met5d}, resulting in
\be
 {\bm n}_r=\frac{e^f h^{1/2}}{r} \ .
\ee
The derivative along the radial direction gives
\be
 \frac{d}{dr} {\bm n}_r={\bm n}_r\left[ f'-\frac{1}{r}+\frac{1}{2}\frac{h'}{h}\right] \ .
\ee
By using the BPS equations \eqref{eq:d3d7BPSeq} we find 
\be\label{eq:refindD3D7}
 f'-\frac{1}{r}+\frac{1}{2}\frac{h'}{h}=-\frac{1}{2r}  \left[H+4F-4+P\right] \ .
\ee
Both in the UV and in the IR the dominant term is the first one inside the parenthesis $F\sim 1$, so the refraction index is decreasing in both limits. In order for it not to be monotonically decreasing it would be necessary that there are at least two critical points where $\frac{d}{dr}{\bm n}_r=0$ at intermediate values of the radial coordinate. This seems highly unnatural and indeed a (non-exhaustive, however) numerical analysis seems to confirm that the refraction index is monotonically decreasing, see Fig.~\ref{fig:d3d7nr}.

\begin{figure}[h!]
\begin{center}
\begin{tabular}{cc}
\includegraphics[width=0.45\textwidth]{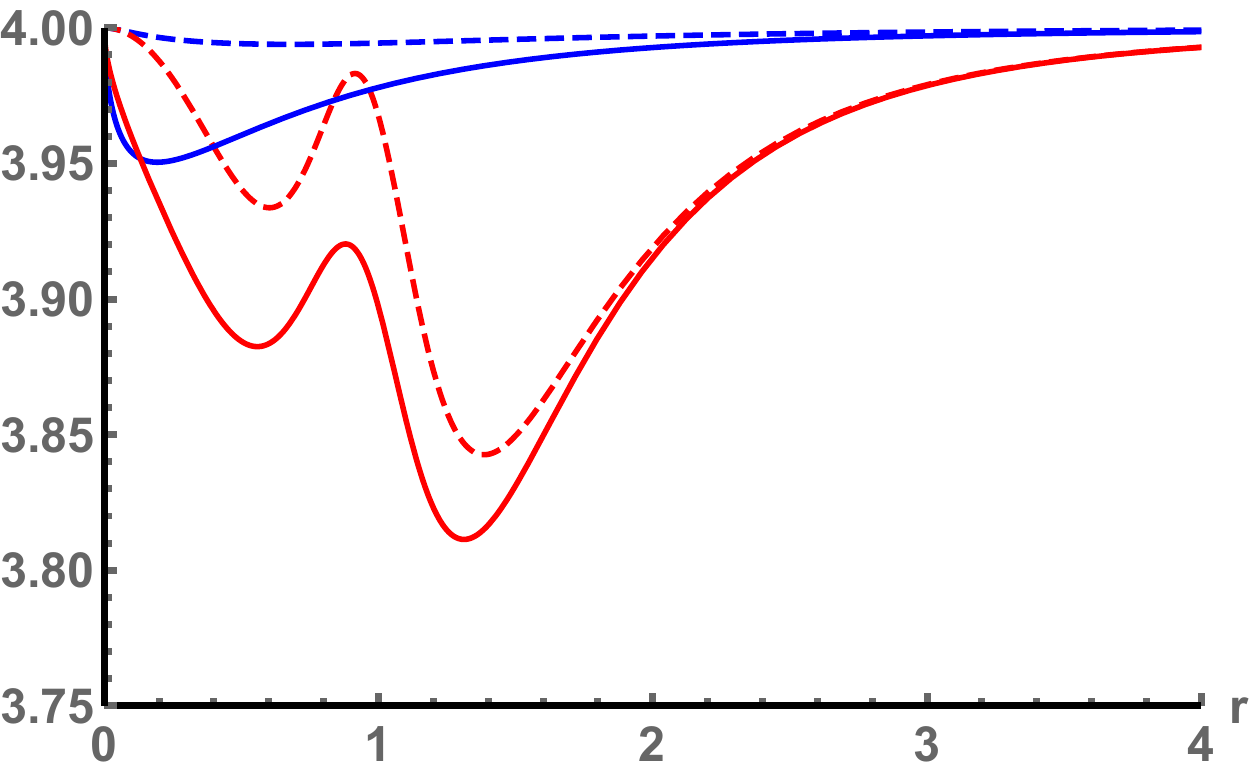}  & \includegraphics[width=0.45\textwidth]{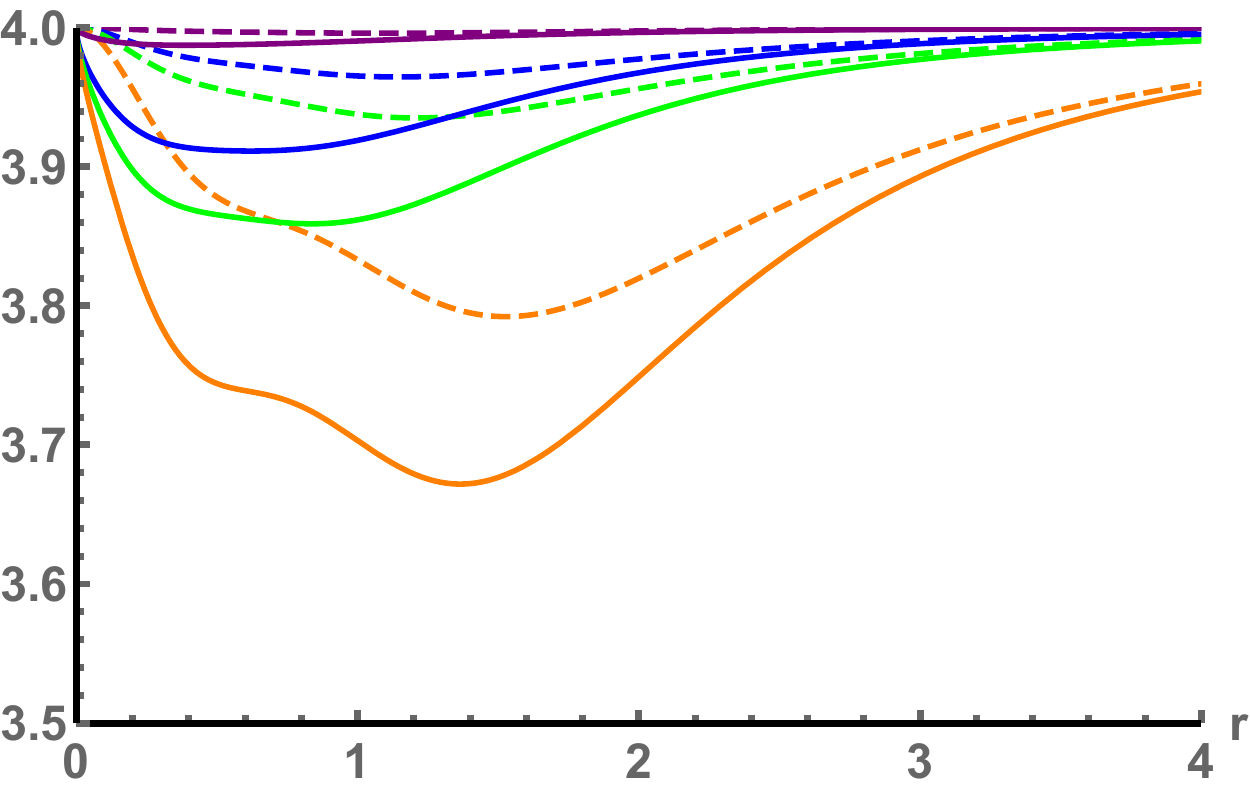}
\end{tabular}
\end{center}
\caption{\small In these plots we depict the quantity in square parenthesis in \eqref{eq:refindD3D7} as a function of the radial coordinate $r$. If the values of the curves stay positive, the refraction index is monotonically decreasing, corresponding to causal dual quantum field theory. {\bf{Left}}: Solutions with $Q_f=1$ and varying $n=1/2$ (solid), $2$ (dashed), $m=1$ (blue), $m=10$ (red). {\bf{Right}}: Solutions with $m=2$, $n=1/2$ (solid), $2$ (dashed) and varying $Q_f=1/10$ (purple), $1$ (blue), $2$ (green), and $10$ (orange).}\label{fig:d3d7nr}
\end{figure}

\item {\bf Entanglement entropy.} For strips of small enough width the integral along the RT surface in \eqref{eq:ceenec} will be performed for values where $A''\geq 0$ everywhere on the surface. Then the corresponding $c$-function will be decreasing towards the UV instead of increasing, {\emph{i.e.}}, having values larger in close proximity of the UV fixed point.

\end{itemize}

\section{D3-D5}\label{sec:D3D5}

In this example the dual to the D$3$-branes is the same as in the (probe) D$3$-D$7$ case. The intersection of D$3$- with D$5$-branes along $2+1$ dimensions corresponds to introducing a supersymmetric codimension one defect in the theory of D$3$-branes. If the number of D$5$-branes is small the holographic dual contains probe D$5$-branes \cite{Karch:2002sh,DeWolfe:2001pq,Erdmenger:2002ex,Arean:2006pk}. For a large number of D$5$-branes, the geometries obtained after smearing in all transverse directions result in anisotropic background solutions preserving two supercharges \cite{Conde:2016hbg,Penin:2017lqt,Jokela:2019tsb}.

 \subsection{Solutions}

The ten-dimensional metric is
\bea
 ds_{10}^2 & = & h^{-1/2}\left[\eta_{\mu\nu}dx^\mu dx^\nu+e^{-2\phi}(dx^3)^2\right]+h^{1/2}\left[e^{-2f}r^2dr^2+e^{2f}(d\tau+A)^2+r^2 ds^2_{KE}\right] \label{eq:d3d5met10d}\\
 F_3 & = & Q_f p dx^3\wedge {\rm Im}\hat{\Omega}_2 \\
 F_5 & = & \partial_r(e^{-\phi}h^{-1})(1+\star)( dx^0 \wedge dx^1 \wedge dx^2 \wedge dx^3 \wedge dr) \ ,
\eea
where $\mu,\nu=0,1,2$ and the functions $h$, $f$, and $p$, together with the dilaton $\phi$, depend on the radial coordinate $r$. As in the D$3$-D$7$ case, the internal space is a Sasaki-Einstein manifold obtained as a fibration over a K\"ahler-Einstein base, which is squashed by the backreaction of the smeared D5-branes. Details about the internal space as well as the precise definition of ${\rm Im}\hat{\Omega}_2$ can be found in the original references \cite{Conde:2016hbg,Hoyos:2020zeg}.

The BPS equations are
\bea
 f' & = &3r e^{-2f}-\frac{2}{r}+\frac{Q_f\, e^{\frac{3}{2}\phi}}{2r}e^{-f}p \nonumber\\
 \phi' & = & \frac{Q_f\, e^{\frac{3}{2}\phi}}{r}e^{-f}p  \label{eq:d3d5BPSeq} \\
 h' & = & -\frac{Q_c}{r^3}e^{-2f}-\frac{Q_f\,  e^{\frac{3}{2}\phi}}{r}e^{-f}h\, p \nonumber \ .
\eea
Solutions for the type of profile we study were constructed in \cite{Hoyos:2020zeg}, but for ease of reference we have collected them in Appendix~\ref{app:D3D5prel}. The reduction to five dimensions (also in \cite{Hoyos:2020zeg}) produces the following metric
\be\label{eq:d3d5met5d}
 ds_5^2=h^{4/3}r^{20/3}e^{-4f/3}  \left[ \frac{e^{2f }}{r^4 h}\left( \eta_{\mu\nu}dx^\mu dx^\nu+e^{-2\phi}(dx^3)^2\right)+\frac{dr^2}{r^2}\right] \ .
\ee

The geometry can be reduced further to four dimensions along the directions of the D3-D5 intersection. The details are gathered in Appendix~\ref{app:D3D5reductionto4d}. The reduced four-dimensional metric is
\be\label{eq:d3d5met4d}
 ds_4^2=h^{3/2}r^8 e^{-f-\phi}  \left[ \frac{e^{2f }}{r^4 h}\eta_{\mu\nu}dx^\mu dx^\nu+\frac{dr^2}{r^2}\right] \ .
\ee

 \subsection{Null Energy Condition in ten, five, and four dimensions}

The ten-dimensional NEC is straightforward to obtain from the energy-momentum tensor of SUGRA fields and branes. Projecting on a null vector pointing in the radial direction $n^M$,
\be
 T_{{}_M{} _N}^{\rm 10d} n^M n^N=\frac{e^{2f}}{2r^2h^{1/2}}Q_f e^{\frac{3}{2}\phi}e^{-f}p\left( Q_f e^{\frac{3}{2}\phi}e^{-f}p+\frac{2 r p'}{p}\right) \ .
\ee
The expansion for large and small values of the radial coordinate were worked out in \cite{Hoyos:2020zeg}, but we have also gathered them in Appendix~\ref{app:D3D5exp}.  Using the UV expansions of the solutions,
\be
 T_{{}_M{} _N}^{\rm 10d} n^M n^N\simeq -6 \frac{e^{2f}}{2r^2h^{1/2}}Q_f e^{\frac{3}{2}\phi}e^{-f}p\leq 0 \ , \ r\to\infty \ .
\ee 
On the other hand, in the IR,
\be
 T_{{}_M{} _N}^{\rm 10d} n^M n^N\simeq 2n \frac{e^{2f}}{2r^2h^{1/2}}Q_f e^{\frac{3}{2}\phi}e^{-f}p\geq 0 \ , \ r\to 0 \ .
\ee 
Therefore, the ten-dimensional NEC is always satisfied in the IR and violated in the UV for any nonzero $Q_f$. 

Let us now check the NEC in five dimensions. We can transform the metric \eqref{eq:d3d5met5d} in domain wall form \eqref{eq:metdwanis} through the change of radial coordinate
\be
 \frac{du}{dr}=h^{2/3}r^{7/3} e^{-2f/3}\equiv \partial_r u \ .
\ee
The warp factors in the anisotropic metric \eqref{eq:anismet} are
\be
 A=\frac{1}{6}\log h+\frac{1}{3}f+\frac{4}{3}\log r \ , \ b=-\phi \ .
\ee

The NEC in five dimensions is equivalent to the conditions

\be
{\cal R}_u \leq 0,\ \ {\cal R}_\parallel\leq 0 \ ,
\ee
where ${\cal R}_u$, ${\cal R}_\parallel$ are determined by \eqref{eq:anisoNEC} with $D=4$. Let us write these open in the current context.

The radial derivative of the warp factors are, after using the BPS equations \eqref{eq:d3d5BPSeq},
\bea
 A'(u) & = & \frac{1}{6\partial_r u}\left[\frac{h'}{h}+2f'+\frac{8}{r} \right]=\frac{1}{6}h^{-2/3}r^{-10/3} e^{2f/3}\left[6r^2e^{-2f}+4 -\frac{Q_c e^{-2f}}{r^2h} \right] \nonumber \\
 b'(u) & = & -\frac{1}{\partial_r u} \frac{Q_f\, e^{\frac{3}{2}\phi}}{r}e^{-f}p =-h^{-2/3}r^{-10/3} e^{2f/3}\frac{Q_f\, e^{\frac{3}{2}\phi}}{r}e^{-f}p \ .
\eea
In the following it will be convenient to define the functions
\be
 F\equiv\frac{e^{2f}}{r^2} \ ,\  P\equiv Q_f e^{\frac{3}{2}\phi}e^{-f} p \ ,  \ H\equiv \frac{Q_c }{r^4 h} \ .
\ee
Then,
\bea
 A'(u) & = & \frac{1}{r \partial_r u}\frac{6+4F-H}{6F} \nonumber\\
 b'(u) & = & -\frac{1}{r \partial_r u}P \ .
\eea
The second derivative of $A(u)$ gives
\bea
 A''(u) & = & -\frac{4}{F^2(r\partial_r u)^2}\left[1+\frac{2}{3}F\left(\frac{7}{6}F-1 \right)+\frac{5}{72}H^2 -\frac{1}{9}H\left(3+2F \right)\right. \nonumber\\
        &   & \left. +\frac{1}{24}(H-4F)FP\right] \ . \label{eq:NECD3D5}
\eea
The second derivative of $b(u)$ is
\be
 b''(u)  =  \frac{P}{F(r\partial_r u)^2}\left[1+\frac{8}{3}F-\frac{2}{3}H-2 F P-F\frac{r p'}{p}\right] \ . 
\ee
We therefore obtain
\bea
{\cal R}_u & = & - \frac{12}{F^2 (r \partial_r u)^2}\left[1+\frac{2}{3}F\left(\frac{7}{6}F-1 \right)+\frac{5}{72}H^2 -\frac{1}{9}H\left(3+2F \right)\right. \nonumber \\
        &  & \left. +\frac{F P}{12}\left(-4F+H+F P +F\frac{r p'}{p} \right)\right] \nonumber \\
{\cal R}_\parallel & = & - \frac{P}{F (r \partial_r u)^2}\left[3+F P+F \frac{r p'}{p}\right].  \label{eq:NECD3D5R}
\eea
Next we will discuss whether these quantities lead to violations of the NEC both asymptotically, an analysis that can be performed analytically, or at intermediate energy scales where we need to invoke numerical analysis.

Using the UV expansions of the solutions,
\be
 A''(u)\simeq -\frac{4r^2e^{-4f}}{(\partial_r u)^2} \left(\frac{3}{32} \frac{Q_f^2}{r^8} \right) \leq 0 \ , \ r\to\infty
\ee 

we find the radial and anisotropic contributions to the NEC:
\bea
 {\cal R}_u & \simeq & -\frac{12 e^{-2f}}{(\partial_r u)^2} \left( -\frac{Q_f}{4 r^4}\right) \geq 0 \nonumber\\
 {\cal R}_\parallel & \simeq & -\frac{Q_fe^{\frac{3}{2}\phi-3f}p}{(\partial_r u)^2} \left( \frac{n+3}{r^m}\right)\leq 0 \ .
\eea

On the other hand, in the IR $r\to 0$:
\begin{itemize}
\item Boomerang flows $n>1$ ($n\neq 5$)
\be
 A''(u)\simeq -\frac{4r^2e^{-4f}}{(\partial_r u)^2} \left(\frac{2n(n-1)}{(n^2-25)^2} \frac{Q_f^2}{w_{n,m}^2}r^{2(n-1)}  \right) \leq 0 \ .
\ee 

The expansion of the radial and anisotropic contributions to the NEC are
\bea
 {\cal R}_u & \simeq & -\frac{12 e^{-2f}}{(\partial_r u)^2} \left( \frac{n}{12}\frac{Q_f}{w_{n,m}}r^{n-1}\right) \leq 0 \nonumber\\
 {\cal R}_\parallel & \simeq & -\frac{Q_fe^{\frac{3}{2}\phi-3f}p}{(\partial_r u)^2} \left( n+3\right)\leq 0 \ .
\eea
\color{black}
\item Lifshitz flows $1>n\geq 1/3$
\be
A''(u)\simeq -\frac{4r^2e^{-4f}}{(\partial_r u)^2} \left(-\frac{(1-n)^4(3-n)}{8(n+1)(n+5)^2} \frac{w_{nm}}{Q_f}r^{1-n} \right) \geq 0 \ .
\ee

The expansion of the radial and anisotropic contributions to the NEC are
\bea
 {\cal R}_u & \simeq & -\frac{12 e^{-2f}}{(\partial_r u)^2} \left( \frac{3n(1-n)}{(n+5)^2} \right) \leq 0 \nonumber\\
 {\cal R}_\parallel & \simeq & -\frac{Q_fe^{\frac{3}{2}\phi-3f}p}{(\partial_r u)^2} \left( 6\frac{n+3}{n+5}\right)\leq 0 \ .
\eea

\end{itemize}

\begin{figure}[h!]
\begin{center}
\begin{tabular}{cc}
\includegraphics[width=0.45\textwidth]{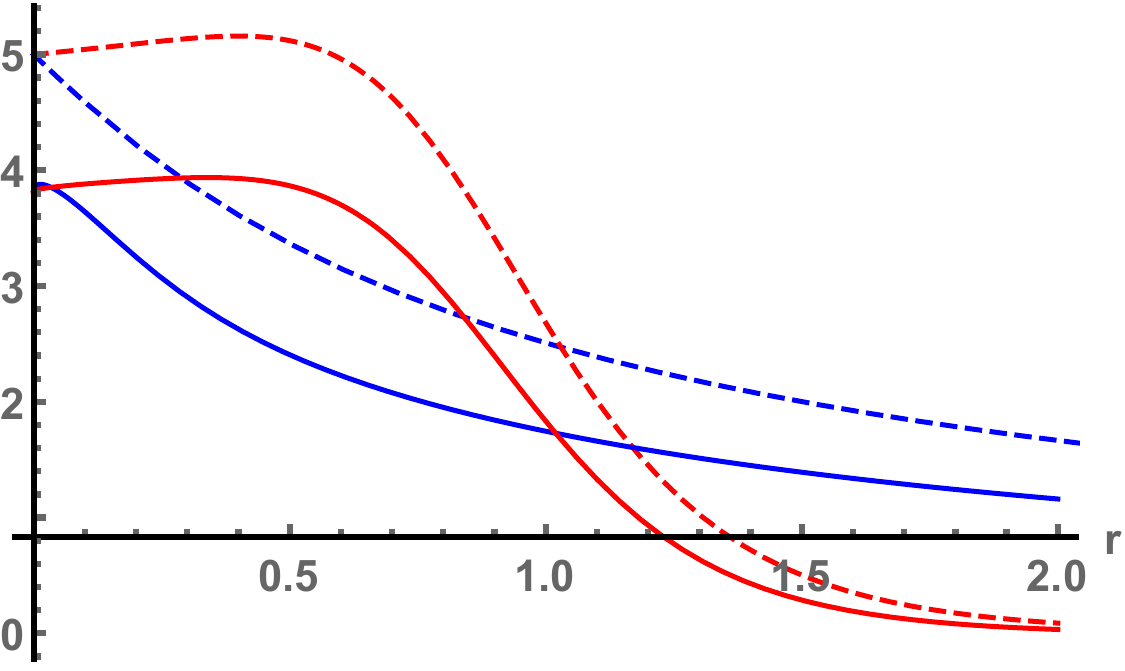} &\includegraphics[width=0.45\textwidth]{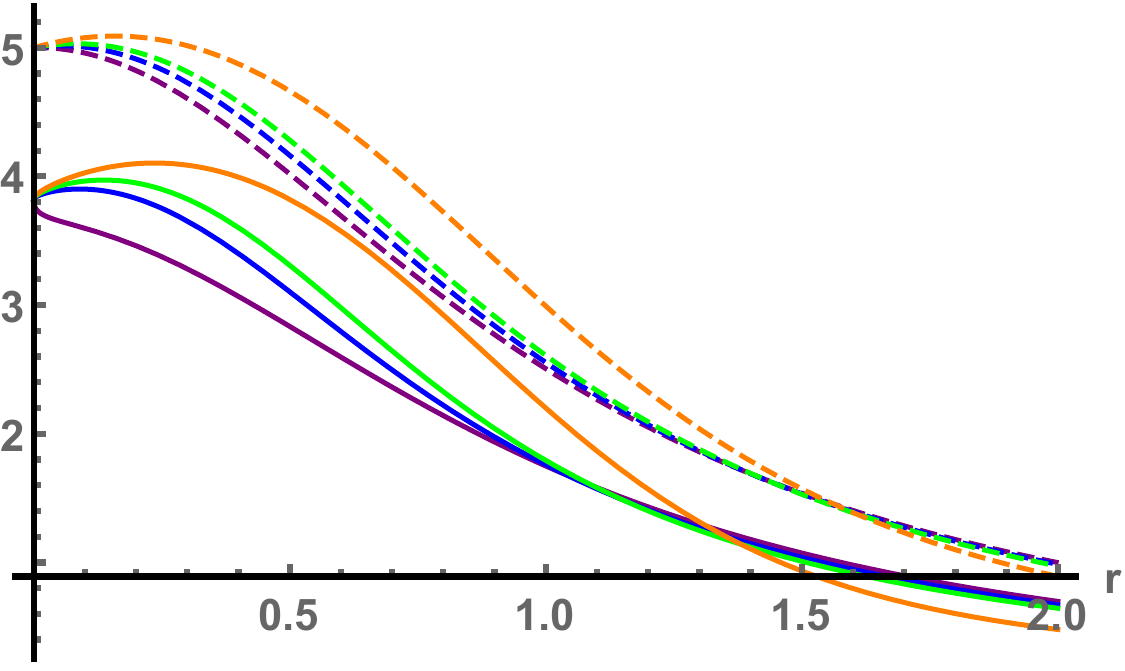}
\end{tabular}
\end{center}
\caption{\small In these plots we depict the quantity in square parenthesis in the formula for ${\cal R}_\parallel$ in the third line of \eqref{eq:NECD3D5R} as a function of the radial coordinate $r$. The NEC in the anisotropic direction is satisfied whenever this quantity is positive. {\bf{Left}}: Solutions with $Q_f=1$, $n=1/2$ (solid) or $n=2$ (dashed) and varying $m=1$ (blue), $m=5$ (red). {\bf{Right}}: Solutions with $n=1/2$ (solid) or $n=2$ (dashed), $m=2$ and varying $Q_f=1/10$ (purple), $1$ (blue), $2$ (green), and $10$ (orange).}\label{fig:d3d5necparallel}
\end{figure}

\begin{figure}[h!]
\begin{center}
\begin{tabular}{cc}
\includegraphics[width=0.45\textwidth]{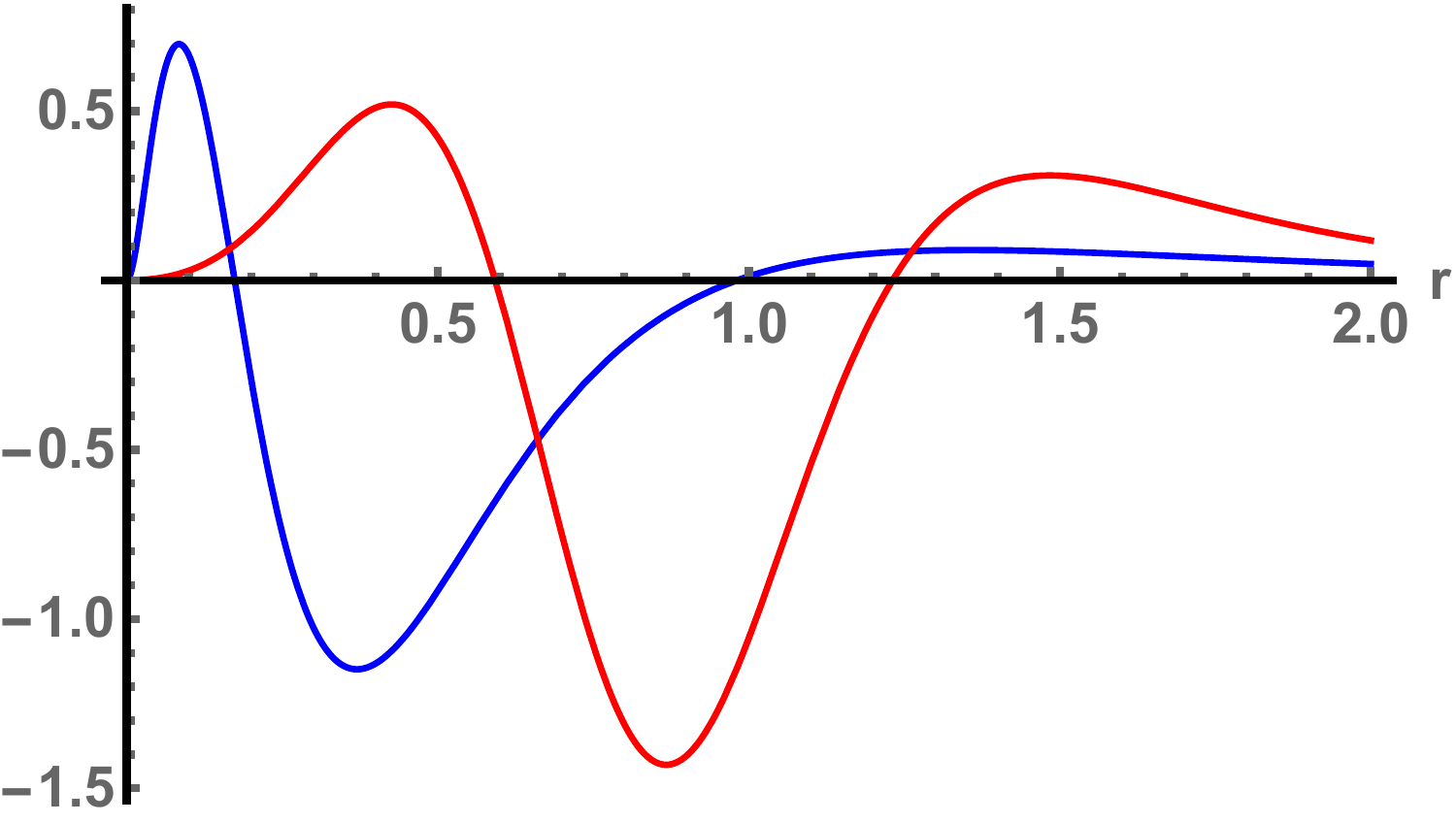} &\includegraphics[width=0.45\textwidth]{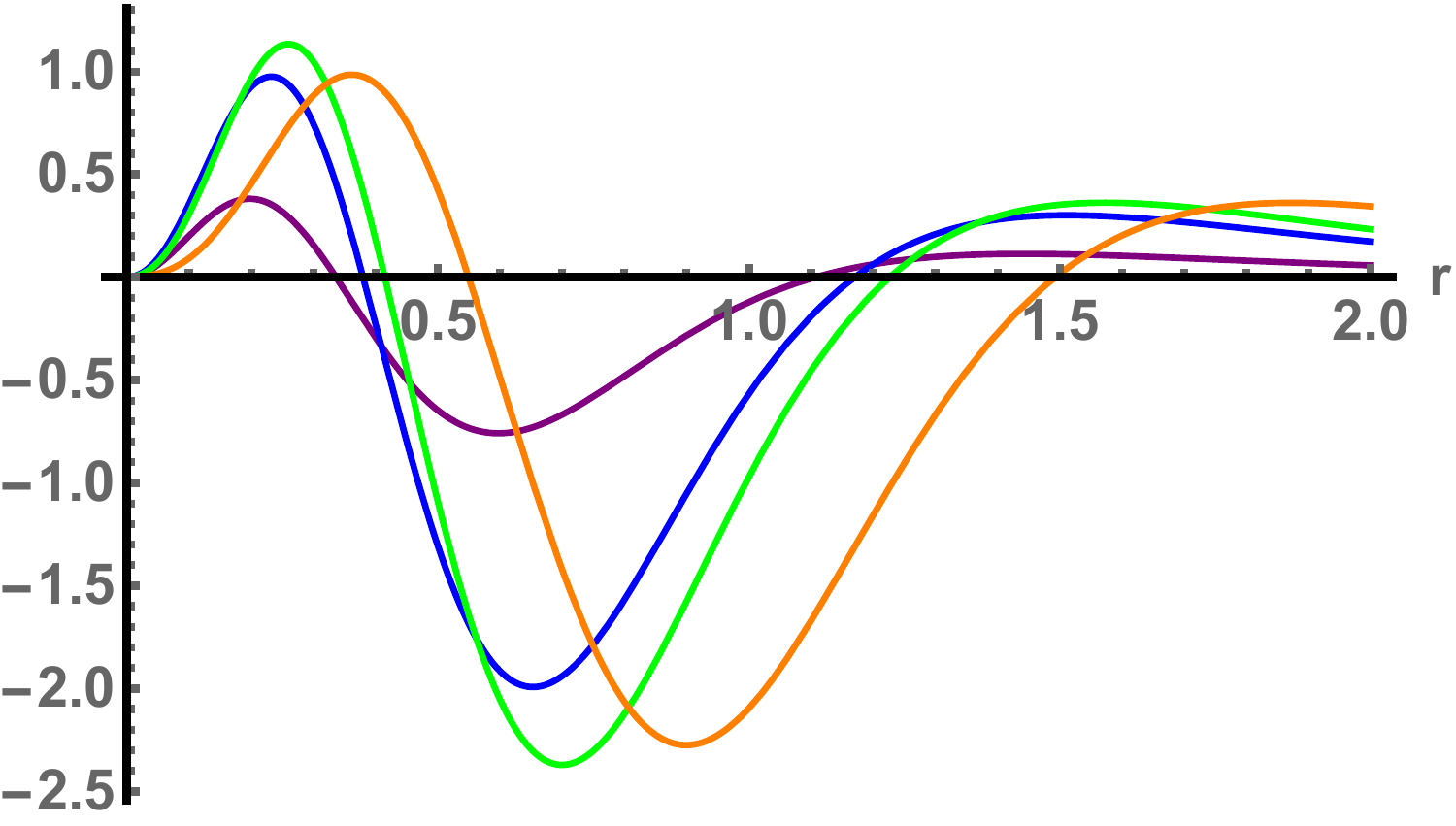}
\end{tabular}
\end{center}
\caption{\small In these plots we depict the quantity in square parenthesis in \eqref{eq:NECD3D5} divided by $Q_f^{3/2}$ and multiplied by $10^M$ as a function of the radial coordinate $r$ for boomerang solutions. The NEC is satisfied whenever this quantity is positive. {\bf{Left}}: Solutions with $Q_f=1$, $n=2$ and varying $m=1,M=5$ (blue), $m=5,M=3$ (red). {\bf{Right}}: Solutions with $n=2$, $m=2,M=4$ and varying $Q_f=1/10$ (purple), $1$ (blue), $2$ (green), and $10$ (orange). We observe that although the NEC is satisfied both in the IR and in the UV, it is violated in the intermediate region even for small values of $Q_f$.}\label{fig:d3d5nec}
\end{figure}

Even though the NEC is satisfied in the IR for both boomerang and Lifshitz solutions, the radial NEC is violated in the UV. The NEC in the anisotropic direction is satisfied everywhere, as shown in Fig.~\ref{fig:d3d5necparallel}, so there are no additional constraints on the solutions from the condition ${\cal R}_\parallel \leq 0$. We also observe that $A''(u)\leq 0$ both in the IR and the UV for boomerang solutions, however, we find by a numerical analysis that $A''(u)>0$ in intermediate regions, as shown in Fig.~\ref{fig:d3d5nec}. 
\color{black}

\begin{figure}[h!]
\begin{center}
\begin{tabular}{cc}
\includegraphics[width=0.45\textwidth]{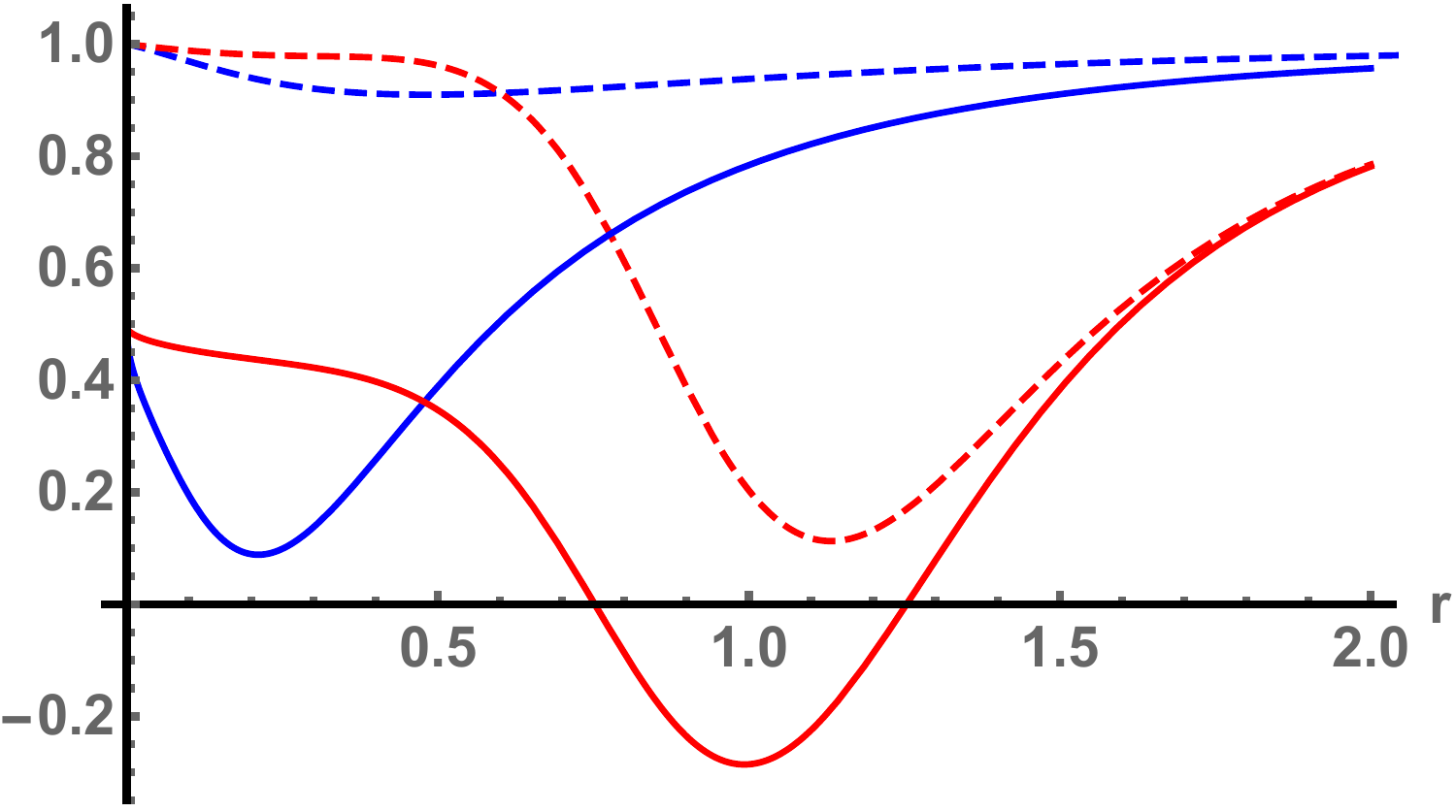} &\includegraphics[width=0.45\textwidth]{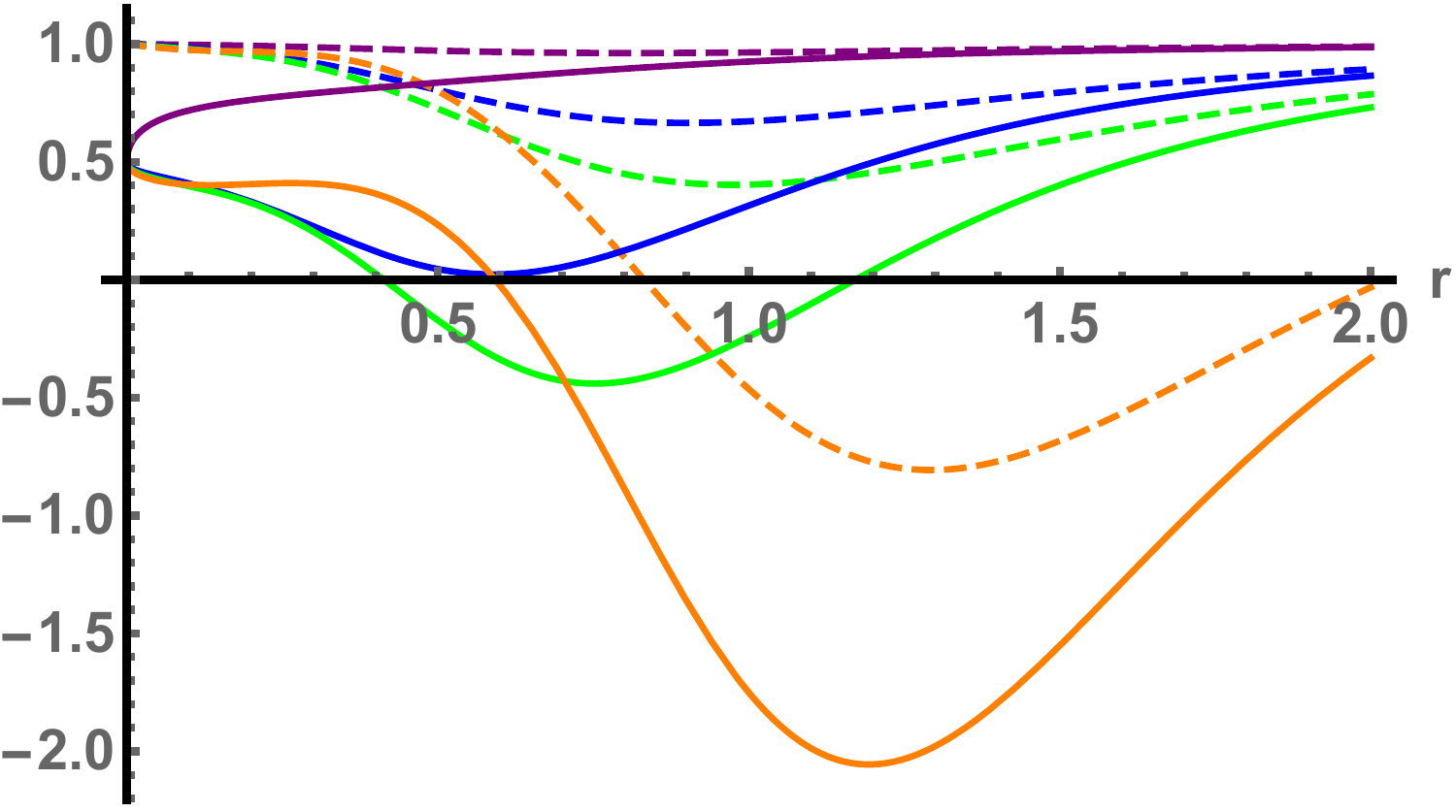}
\end{tabular}
\end{center}
\caption{\small In these plots we depict the quantity in square parenthesis in \eqref{eq:NEC4D3D5} as a function of the radial coordinate $r$ for boomerang solutions. The NEC is satisfied whenever this quantity is positive. {\bf{Left}}: Solutions with $Q_f=1$, $n=1/2$ (solid), $2$ (dashed) and varying $m=1$, (blue), $m=5$ (red). {\bf{Right}}: Solutions with $n=1/2$ (solid), $2$ (dashed), $m=2$ and varying $Q_f=1/10$ (purple), $1$ (blue), $2$ (green), and $10$ (orange). The NEC is satisfied throughout all values of the radial coordinate as long as $Q_f$ is not too large.}\label{fig:d3d5nec4d}
\end{figure}

Finally, we can further reduce to four dimensions along the directions of the D3-D5 intersection and test the NEC in the reduced geometry. We can transform the metric \eqref{eq:d3d5met4d} in domain wall form \eqref{eq:metdw} through the change of radial coordinate
\be
 \frac{du}{dr}=h^{3/4}r^3 e^{-(f+\phi)/2}\equiv \partial_r u \ .
\ee
The warp factor in the domain wall metric is
\be
 A=\frac{1}{4}\log h+\frac{1}{2}(f-\phi)+2\log r \ .
\ee
The NEC in four dimensions is equivalent to the condition \cite{Freedman:1999gp}
\be
 A''(u)\leq 0 \ .
\ee
The radial derivative of the warp factor is, after using the BPS equations \eqref{eq:d3d5BPSeq},
\be
 A'(u)=\frac{1}{4\partial_r u}\left[\frac{h'}{h}+2(f'-\phi')+\frac{8}{r} \right]=\frac{1}{4}h^{-3/4}r^{-4} e^{(f+\phi)/2}\left[6r^2e^{-2f}+4 -\frac{Q_c e^{-2f}}{r^2h}-2e^{-f+\frac{3}{2}\phi} p \right] \ .
\ee
The second derivative of $A(u)$ gives
\bea
 A''(u) & = & -\frac{3}{4F^2(r\partial_r u)^2}\left[9+4F\left(\frac{5}{3}F-1 \right)+\frac{7}{12}H^2 -H\left(3+2F \right)\right. \nonumber\\
        &   & \left. +(H-2-4F)FP+\frac{1}{3}F^2P\left(5P+\frac{2 r p'}{p}\right)\right] \ . \label{eq:NEC4D3D5}
\eea
Using the UV expansions of the solutions,
\be
 A''(u) =  -\frac{3r^2e^{-4f}}{4(\partial_r u)^2} \left[1+{\cal O}\left(\frac{1}{r^4}\right)\right] \leq 0 \ , \ r\to\infty
\ee 
On the other hand, in the IR $r\to 0$:
\begin{itemize}
\item Boomerang flows $n>1$ ($n\neq 5$)
\be
 A''(u) = -\frac{3r^2e^{-4f}}{4(\partial_r u)^2} \left[1+{\cal O}(r^{n-1})  \right] \leq 0 \ .
\ee 
\item Lifshitz flows $1>n\geq 1/3$
\be
A''(u)\approx -\frac{3r^2e^{-4f}}{4(\partial_r u)^2} \left[ \frac{12 n(n+2)}{(n+5)^2}\right] \leq 0 \ .
\ee
\end{itemize}
In this case the NEC $A''(u)\leq 0$ is satisfied for any $n$ in the IR. It turns out that in this case the NEC is also satisfied for intermediate values of the radial coordinate, for a not too large value of $Q_f$ that depends on $(n,m)$ as illustrated in Fig.~\ref{fig:d3d5nec4d}.

\subsection{Holographic $c$-functions}

\begin{itemize}

\item {\bf Domain wall $c$-function and null congruences.}  Since the NEC changes sign in five dimensions, the $c$-functions are non-monotonic. 

In principle one could define quantities with a radial derivative proportional to $A''$, but this also changes sign along the radial direction. For boomerang flows this happens only at intermediate values of the radial coordinate, while for the solutions that become Lifshitz in the IR, since  $A''\geq 0$, the associated $c$-function would be increasing towards the IR, instead of decreasing.
\color{black}
In four dimensions the situation changes, the NEC is satisfied as long as $Q_f$ is not too large and the $c$-functions are then monotonic.

\item {\bf Anisotropic domain wall functions.} The derivative of the $c$-function defined in \eqref{eq:cdw1} is (for $D=4$)
\bea
\partial_u c_{dw,an1}(u) & = & \frac{8}{F^2(r\partial_r u)^2}\left[1+\frac{2}{3}F\left(\frac{7}{6}F-1 \right)+\frac{5}{72}H^2-\frac{1}{9}H\left(3+2F \right)\right. \nonumber\\
& & \left.+\frac{1}{16}F^2P^2+\frac{1}{16}(H-4F)FP  -\frac{1}{32}FP\left(1-\frac{r p'}{p}F \right)\right] \ .\label{eq:NECan1}
\eea

The derivative of the $c$-function defined in \eqref{eq:cdw2} is (for $D=4$)
\bea
 \partial_u c_{dw,an2}(u) & = & \frac{8e^{-\frac{\phi}{3}}}{F^2(r\partial_r u)^2}\left[1+\frac{2}{3}F\left(\frac{7}{6}F-1 \right)+\frac{5}{72}H^2-\frac{1}{9}H\left(3+2F \right)\right.\\
 & & \left.+\frac{5}{72}F^2P^2+\frac{1}{18}(H-4F)FP +\frac{1}{24}FP\left(1+ \frac{r p'}{p}F\right)\right] \ .\label{eq:NECan2}
\eea
Both are similar to the NEC with different coefficients for the terms depending on $Q_f$.

Let us start with the behavior in the IR, $r\to 0$,
\begin{itemize}
\item Boomerang flows $n>1$ ($n\neq 5$)
\bea
 \partial_u c_{dw,an1}(u) & \simeq & \frac{8}{F^2(r\partial_r u)^2}\left(\frac{n+1}{32}\frac{Q_f}{w_{nm}} r^{n-1}\right) \geq 0 \\
 \partial_u c_{dw,an2}(u) & \simeq & \frac{8e^{-\frac{\phi}{3}}}{F^2(r\partial_r u)^2}\left(\frac{n+1}{24}\frac{Q_f}{w_{nm}} r^{n-1}\right) \geq 0 \ .
\eea
Both proposals are monotonically increasing in the IR region for any $n>1$. 
\item Lifshitz flows $1>n\geq 1/3$
\bea
 \partial_u c_{dw,an1}(u) & \simeq & \frac{8}{F^2(r\partial_r u)^2}\left(-\frac{(1-n)^3 (7n+11)}{(n+1)(n+5)}\frac{w_{nm}}{Q_f}r^{1-n} \right) \leq 0 \\
 \partial_u c_{dw,an2}(u) & \simeq & \frac{8e^{-\frac{\phi}{3}}}{F^2(r\partial_r u)^2}\left(\frac{3}{n+5}-\frac{1}{2}\right) \geq 0 \ .
\eea
In this case the $c_{dw,an2}$ is monotonically increasing in the IR, while $c_{dw,an1}$ is decreasing.
\end{itemize}

However, using the UV expansions of the solutions, for $r\to\infty$, we find
\bea
 \partial_u c_{dw,an1}(u) & \simeq & \frac{8}{F^2(r\partial_r u)^2}\left(-\frac{Q_f}{8r^4} \right) \leq 0 \\
 \partial_u c_{dw,an2}(u) & \simeq & \frac{8e^{-\frac{\phi}{3}}}{F^2(r\partial_r u)^2}\left(-\frac{Q_f}{12r^4}  \right) \leq 0 \ .
\eea
Both proposals for holographic $c$-functions are monotonically {\em decreasing} towards the UV, in contrast to the holographic $c$-functions determined by the NEC.
 
From this behavior we conclude that although $c_{dw,an2}$ would be a good candidate for a holographic $c$-function in the IR in both the boomerang and Lifshitz solutions, it is not monotonic. $c_{dw,an1}$ would be a good candidate only for the boomerang solutions but also fails to be monotonic due to wrong behavior in the UV region.

\item {\bf Refraction index in the radial direction.} We find the refraction index either from \eqref{eq:d3d5met10d} or \eqref{eq:d3d5met5d}, 
\be
 {\bm n}_r=r e^{-f} h^{1/2} \ .
\ee
The derivative along the radial direction gives
\be
 \frac{d}{dr} {\bm n}_r={\bm n}_r\left[ \frac{1}{r}-f'+\frac{1}{2}\frac{h'}{h}\right] \ .
\ee
By using the BPS equations \eqref{eq:d3d5BPSeq} we find
\be\label{eq:refindD3D5}
 \frac{1}{r}-f'+\frac{1}{2}\frac{h'}{h}=-\frac{1}{2r F}\left[ 6+H-4F+2FP\right] \ .
\ee
Using the UV expansions of the solutions,
\be
 \frac{d}{dr} {\bm n}_r\simeq -6\frac{{\bm n}_r}{2r F}\leq 0 \ , \ r\to\infty \ .
\ee
On the other hand, in the IR ($r\to 0$):
\begin{itemize}
\item Boomerang flows $n>1$ ($n\neq 5$)
\be
 \frac{d}{dr} {\bm n}_r\simeq -6\frac{{\bm n}_r}{2r F}\leq 0 \ .
\ee
\item Lifshitz flows $1>n\geq 1/3$
\be
 \frac{d}{dr} {\bm n}_r\simeq -\frac{36}{n+5}\frac{{\bm n}_r}{2r F}\leq 0 \ .
\ee
\end{itemize}
The refraction index is decreasing in both the UV and IR limits. In order for it not to be monotonically decreasing it would be necessary that there are at least two critical points where $\frac{d}{dx}{\bm n}_r=0$ at intermediate values of the radial coordinate. This again seems highly unnatural and indeed our non-exhaustive numerical analysis seems to confirm that the refraction index is monotonically decreasing, see Fig.~\ref{fig:d3d5nr}.

\begin{figure}[h!]
\begin{center}
\begin{tabular}{cc}
\includegraphics[width=0.45\textwidth]{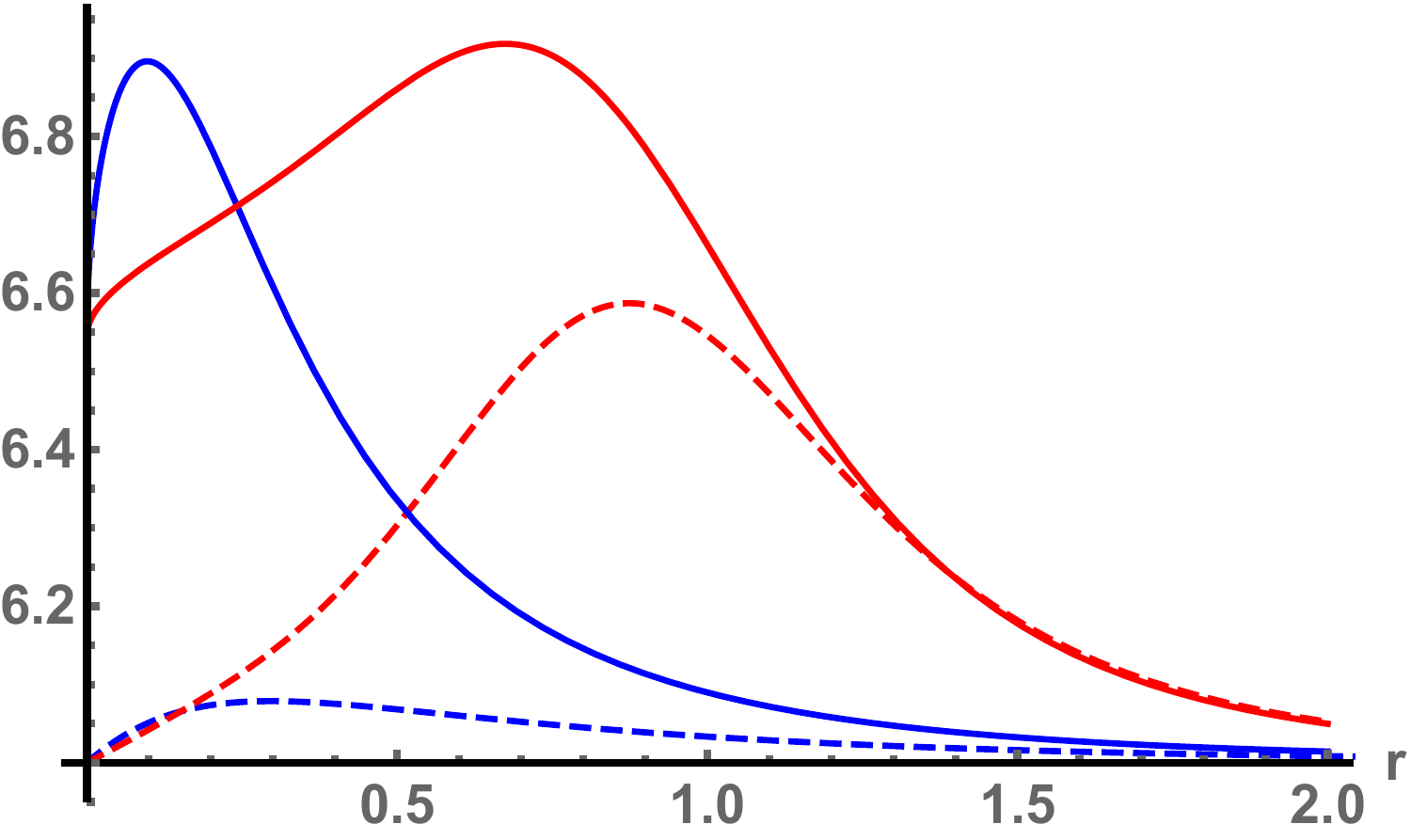}  & \includegraphics[width=0.45\textwidth]{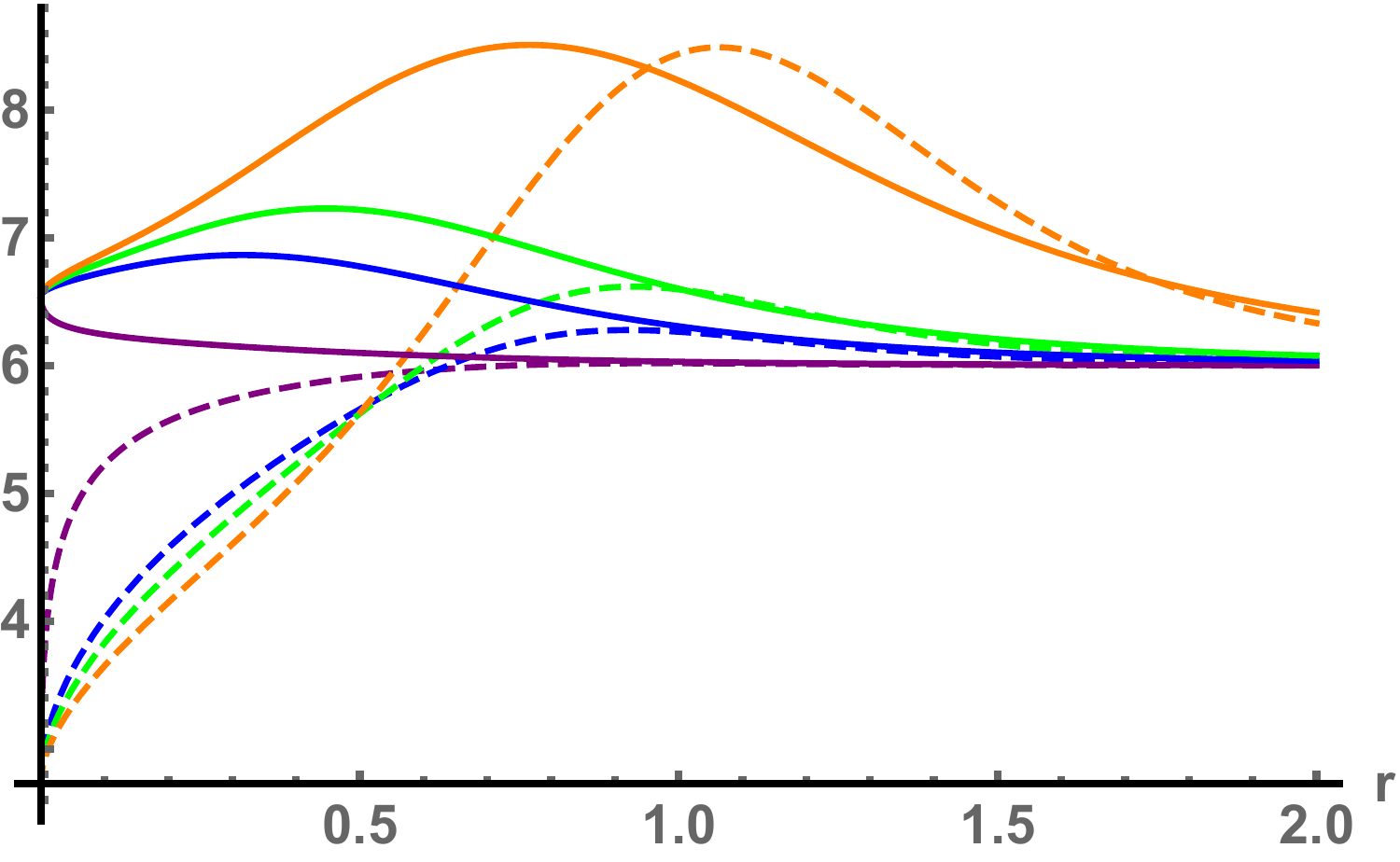}
\end{tabular}
\end{center}
\caption{\small In these plots we depict the quantity in square parenthesis in \eqref{eq:refindD3D5} as a function of the radial coordinate $r$. {\bf{Left}}: Solutions with $Q_f=1$, and varying $(n,m)= (1/2,1)$ (solid blue), $(1/2,5)$ (solid red), $(2,1)$ (dashed blue), and $(2,5)$ (dashed red). {\bf{Right}}: Solutions with $(n,m)=(2,2)$ and varying $Q_f=1/10$ (purple), $1$ (blue), $2$ (green), and $10$ (orange). Since the sign remains positive the refraction index is monotonic for these solutions.}\label{fig:d3d5nr}
\end{figure}

\item {\bf Entanglement entropy.} Since $A''\leq 0$ is satisfied in the UV, it could had been possible that the $c$-function associated to the entanglement entropy for a strip  was monotonic in some cases. However, the analysis made in \cite{Hoyos:2020zeg} found that this is not the case, the entanglement $c$-function turns out to be non-monotonic.
\color{black}

\end{itemize}

\section{Discussion}\label{sec:discussion}

In this paper we constructed several sturdy supergravity solutions by flavor deformations of well-known $(2+1)$- and $(3+1)$-dimensional dual quantum field theories at strong coupling. Our interpolating classical supergravity solutions map to non-trivial renormalization group flows and we were particularly interested in the following question: ``Can one construct quantities in the gravity dual that are monotonic and can be associated with the number of degrees of freedom?'' We showed that each and every holographic $c$-function fails to be monotonic as evaluated on our solutions in general, but nevertheless have the potential to faithfully capture the number of degrees of freedom in the dual field theory. We also discussed the role played by the null energy conditions in leading to such a controversial statement. In this section we will discuss further ramifications and the potential loopholes in our arguments.

Let us first summarize which of isotropic backgrounds respect the null energy condition in Table~\ref{table:NEC}.
\begin{table}[h!]
\begin{center}
	\begin{tabularx}{0.8\textwidth}{cYYY} \toprule
		 & UV & IR  & Intermediate energy \\
		\midrule
		ABJMf ($N_f/N_c\ll 1$) & always & $n<2(\sqrt 2-1)$ &  always  \\
		D2-D6  & always & always           & $N_f<N_f^{\text{crit}}$  \\
		D3-D7  & never  & always           &  $r<r_\text{crit}$   \\
		reduced D3-D5 & always & always & $N_f<N_f^{\text{crit}}$ \\
		\bottomrule
	\end{tabularx}
	\caption{Testing if the Null Energy Condition is satisfied in the dimensionally reduced theory, after ``integrating'' out the internal compact directions. The first column corresponds to the asymptotically large holographic radial coordinate while the second column to the small radial positions, dual to ultraviolet and infrared energy scales in the corresponding QFTs, respectively. The last column corresponds to a generic radial position in the bulk, {\emph{i.e.}}, intermediate energy scales in the field theory.}\label{table:NEC}
\end{center}
\end{table}
In particular,
\begin{itemize}
 \item NEC is violated when the ambient spacetime is (3+1)-dimensional: D3-D7 and the defect D3-D5.
 \item NEC {\emph{can}} be satisfied in both genuinely (2+1)-dimensional quantum field theories: ABJMf and D2-D6, as well as in the effective (2+1)-dimensional theory living at the D3-D5 intersection.
 \item None of the proposed anisotropic holographic c-theorems hold for D3-D5; though see additional comments below.
\end{itemize}

All the cases we studied are supersymmetric, causal, and void of curvature singularities. They are thus all good candidates for holographic duals if one is willing to consider violations of the NEC and/or the possible absence of a $c$-theorem. Let us hence draw lessons of physics interest of our results and ponder about their meaning for the holographic duality:
\begin{enumerate}
 \item A non-monotonic brane distribution does not translate necessarily into violations of the NEC. It could be that distributions different than the ones we have considered here preserve the NEC. It would then be interesting to study the constraints the NEC imposes on brane distributions and find whether they can be satisfied even if the distribution is not monotonic.
 \item Whenever the NEC is violated the usual holographic $c$-theorems fail. It could be either that we have not yet identified the correct holographic $c$-function, or that there is a genuine failure of the $c$-theorem in the dual field theory. It should be noted that some quantities like the refraction index are monotonic along the RG flow, however their relation to the counting of degrees of freedom is unclear. 
\item A failure of the $c$-theorem in the field theory dual could be tested through the entanglement entropy in the $(2+1)$-dimensional cases. The $(3+1)$-dimensional $c$-theorem is harder to test, as it requires computing the $2\to 2$ scattering amplitude of the effective dilaton. If it turns out that the $c$-theorem is indeed not satisfied this raises the question of the underlying reason in the field theory dual, in other words what a non-monotonic brane distribution implies for the dual.
 \item Assuming the $c$-theorem in the field theory dual is indeed not satisfied when the NEC is violated, there is a further question of whether similar violations might happen when quantum corrections to a classical background saturating the NEC start to be taken into account. Demanding that such violations are absent could impose further constraints on the possible gravitational theories at the quantum level, ruling out some even if they do not have issues at the classical level.
 \item In the anisotropic configurations there is no $c$-theorem as such, but it is not unreasonable to assume that the non-monotonic behavior has a similar physical origin as in the isotropic cases. In particular, for boomerang flows even if there is no $c$-theorem, there are isotropic fixed points on both ends of the RG flow. If the usual $c$-functions give a counting of the number of degrees of freedom, then one expects that at least some weaker version of the $c$-theorem holds. According to the analysis in \cite{Hoyos:2020zeg}, this turns out to be true for a $c$-function obtained from the entanglement entropy of a strip with sides separated along one of the directions parallel to the D5-branes. By reducing to the effective (2+1)-dimensional theory along the isotropic directions we were able to find a monotonic $c$-function in a large family of configurations. Whether this is a feature peculiar to the D3-D5 background or can be generalized to other anisotropic geometries, like the ones recently found in \cite{Guarino:2021kyp,Arav:2021gra} is an open question.
 \end{enumerate}

Supergravity solutions similar to the ones constructed here were obtained in \cite{Conde:2011aa,Nunez:2010sf} to study the Higgsing and Seiberg dualities of cascading theories and their relation to the tumbling phenomena in theories of extended technicolor. These supergravity solutions are generated by means of a U-duality rotation of a background with wrapped color D5-branes and sources of  massive flavor D5-branes. After the rotation, a background with D3-branes is generated, which is the gravity dual of a ${\cal N=1}$ supersymmetric field theory. This solution displays a cascade of Seiberg dualities in the far UV, similar to the Klebanov-Strassler backgrounds, and a cascade of Higgsings (sequences of spontaneous symmetry breaking steps) at intermediate scales. Crucially, in these solutions, a non-monotonic flavor profile in the seed solution is needed, in order to have a sensible field theory interpretation  for the D3-brane flux in the rotated background. The precise behavior of the profile at the UV is determined by imposing consistency with the field theory interpretation or, as happens in our case, positivity of the brane energy density. It would be interesting to explore the possibility that the solutions found here could serve as seeds for new supergravity backgrounds obtained by U-duality rotations and thereupon investigate the various positivity bounds.

\addcontentsline{toc}{section}{Acknowledgments}
\paragraph{Acknowledgments}
We would like to thank Francesco Bigazzi, Anton Faedo, Adolfo Guarino, Esko Keski-Vakkuri, and Juan F. Pedraza for comments and discussions. C.~H. is partially supported by the Spanish Ministerio de Ciencia, Innovaci\'on y Universidades through the grant PGC2018-096894-B-100. N.~J. is supported in part by the Academy of Finland grant no. 1322307. J.~M.~P. is supported in part by the Academy of Finland grant no. 1322307 and by a Royal Society Research Fellow Enhancement Award. A.~V.~R. is funded by the Spanish grant FPA2017-84436-P, by Xunta de Galicia-Conseller\'\i a de Educaci\'on  (ED431C-2017/07). This work has received financial support from Xunta de Galicia (Centro singular de investigación de Galicia accreditation 2019-2022), by European Union ERDF, and by the “María de Maeztu” Units of Excellence program MDM-2016-0692 and the Spanish Research State Agency. 
 
\appendix

\section{ABJMf}\label{app:ABJM}

\subsection{Preliminary analysis}\label{app:ABJMprel}

Let us start by introducing the basis of vielbeins that will be suitable to describe the geometry. Our solution will be a warped $AdS_4$ times a $\mathbb{CP}^3$ manifold. The $\mathbb{CP}^3$ will be written as a $S^2$ bundle over a $S^4$, with the fibration constructed using the self-dual $SU(2)$ instanton on the $S^4$. Consider a system of invariant forms of $SU(2)$ satisfying $d\omega_i=\frac{1}{2}\epsilon_{ijk}\omega^j \wedge \omega^k$. With it we construct:
\begin{eqnarray}
 E_1 & = & d\theta+\frac{\xi^2}{1+\xi^2} (\sin \varphi \omega^1-\cos \varphi \omega^2) \nonumber\\
 E_2 & = & \sin \theta (d\varphi-\frac{\xi^2}{1+\xi^2} \omega^3)+\frac{\xi^2}{1+\xi^2} \cos \theta (\cos \varphi \omega^1+\sin \varphi \omega^2)
\end{eqnarray}
which will describe the fibration. The range of the angular coordinates is: $0\leq \theta <\pi, 0\leq \varphi <2\pi $, while that for the non-compact coordinate is $\xi\in[0,\infty)$.  Let us now define suitable vielbeins for the $S^4$:
\begin{eqnarray}
\mathcal{S}_\xi & = & \frac{2}{1+\xi^2}d\xi \nonumber \\
S_1 & = & \sin \varphi \omega^1-\cos \varphi \omega^2 \nonumber \\
S_2 & = & \sin \theta \omega^3-\cos \theta (\cos \varphi \omega^1+\sin \varphi \omega^2) \nonumber \\
S_3 & = & -\cos \theta \omega^3-\sin \theta (\cos \varphi \omega^1+\sin \varphi \omega^2) \ .
\end{eqnarray}
Let us further define: $\mathcal{S}_a=\frac{\xi}{1+\xi^2} S_a$ for $a=1, 2, 3$. Let us also define a `master' function:
\be
 W\equiv \frac{4}{k}h^{\frac{1}{4}}e^{2f-g-\phi}r \ .
\ee
Then the integration of the BPS differential system boils down to solving for the following differential equation for $W$ \cite{Bea:2013jxa}:
\be\label{eq:Wmasterequation}
 W''+4\eta'+(W'+4\eta)\bigg( \frac{W'+10\eta}{3 W}-\frac{W'+4\eta+6}{r(W'+4\eta)}\bigg)=0 \ .
\ee
Given the solution for $W$ on can construct the full geometry:
\bea
 e^g & = & \frac{r}{W^{\frac{1}{3}}} \exp \bigg[ \frac{2}{3} \int^r d\xi \frac{\eta(\xi)}{W(\xi)} \bigg] \\
 e^f & = & \sqrt{\frac{3r}{W'+4 \eta}} W^{\frac{1}{6}} \exp \bigg[ \frac{2}{3}\int^r d\xi\frac{\eta(\xi)}{W(\xi)} \bigg] \\
 h & = & 4\pi^2 \frac{N_c}{k}e^{-g} (W'+4 \eta) \bigg[ \int^\infty_r d\xi \frac{\xi e^{-3g(\xi)}}{W(\xi)^2} +\beta \bigg] \ ,
\eea
where if we impose $h \rightarrow 0$ as $r \rightarrow \infty$ then $\beta=0$. The dilaton is given by:
\be
 e^\phi=\frac{12}{k} \frac{rh^{\frac{1}{4}}}{W^{\frac{1}{3}} (W'+4\eta)} \exp \bigg[\frac{2}{3} \int^r d\xi\frac{\eta(\xi)}{W(\xi)} \bigg] \ .
\ee

\subsection{Perturbative expansion in small number of flavors}\label{app:ABJMper}

To get solutions representing flavor, one should look for solutions in which $\eta$ goes to a constant in the UV ($x \rightarrow \infty$). In the absence of flavor altogether, $\eta=1$, $W=2r$ is a solution to (\ref{eq:Wmasterequation}) which leads to $AdS_4\times \mathbb{CP}^3$. We are interested in those solutions for which we asymptotically have $\eta\to 1$, {\emph{i.e.}}, only asymptotically anti de Sitter solutions for the background geometry. We therefore demand that the profile goes to zero in the UV, in other words, $\eta\to 1$ fast enough. We will require $T_{00}>0$. This constrains the additive piece to $\eta$ to decay not faster than $r^{-2}$. To solve the master equation, we will consider small deviations from $\eta=1$, and solve perturbatively in the $\hat\epsilon$ deformation. Let us consider the following profile:
\be
 \eta=1+\hat\epsilon p =1+\hat\epsilon \frac{r^n}{(1+r^m)^{\frac{n+2}{m}}} \ .
\ee
Accordingly, we will adopt the ansatz:
\be
 W(r)=W_0(r)+\hat\epsilon W_1(r)+\hat\epsilon^2W_2(r)+\hat\epsilon^3W_3(r)+\ldots 
\ee
with $W_0=2r$. By expanding the master equation to first order in $\hat\epsilon$, we find
\be
 r^2W''_1+2rW'_1-6W_1+2r^{1+n}(1+r^m)^{-\frac{2+m+n}{n}}(7+2n+3r^m)=0 \ .
\ee
There is a simple analytic solution 
\bea
W_1 & =& \frac{c_1+c_2 r^5}{r^3}+r^{1+n} (1+r^m)^{-\frac{2+n}{m}} +\frac{(2+n)r^{1+n}(1+r^m)^{-\frac{2+n}{m}}}{5(4+n)} F[1,\frac{2}{m},\frac{4+m+n}{m},-r^m] \nonumber \\
& & -\frac{6(2+n)r^{1+n}}{5(n-1)}F[\frac{n-1}{m},\frac{2+m+n}{m},\frac{m+n-1}{m},-r^m] \ .
\eea

By imposing regularity at $r\rightarrow 0$ fixes $c_1=0$. The coefficient $c_2$ can be determined by imposing that the UV behavior is still $W\sim 2r$ at leading order. Expanding for large values of $x$ we get the leading order contributions:
\be
 W(r\rightarrow \infty)=\bigg(c_2 -\frac{6}{5}\frac{\Gamma(\frac{3+m}{m})\Gamma(\frac{n-1}{m})}{\Gamma(\frac{2+n}{m})}\bigg)r^2+ \mathcal{O}(r^{-1})
\ee
which clearly fixes $c_2$:
\be
 c_2=\frac{6}{5}\frac{\Gamma(\frac{3+m}{m})\Gamma(\frac{n-1}{m})}{\Gamma(\frac{2+n}{m})} \ .
\ee
Therefore, the solution at first order in $\hat\epsilon$ reads:
\bea
 W & \approx & 2r+ \hat\epsilon \bigg[ \frac{6}{5}\frac{\Gamma(\frac{3+m}{m})\Gamma(\frac{n-1}{m})}{\Gamma(\frac{2+n}{m})}r^2 +\frac{(2+n)r^{1+n}(1+r^m)^{-\frac{2+n}{m}}}{5(4+n)} F[1,\frac{2}{m},\frac{4+m+n}{m},-r^m] \nonumber\\
   &   & +r^{1+n} (1+r^m)^{-\frac{2+n}{m}}-\frac{6(2+n)r^{1+n}}{5(n-1)}F[\frac{n-1}{m},\frac{2+m+n}{m},\frac{m+n-1}{m},-r^m] \bigg] \ .
\eea
We now proceed to construct the geometry to first order in $\hat\epsilon$:
\be
 \int dr \frac{\eta}{W}=\int dr \frac{1+\hat\epsilon {}{p}}{W_0+\hat\epsilon W_1} \approx \int dr \bigg[ \frac{1}{W_0}+\hat\epsilon \big( \frac{{}{p}}{W_0}-\frac{W_1}{W_0^2} \big) \bigg]\equiv I_0+\hat\epsilon I_1\ .
\ee
With the previous result, we immediately get:
\be
 e^g \approx \frac{r}{(W_0+\hat\epsilon W_1)^{\frac{1}{3}}}\exp [\frac{2}{3}(I_0+\hat\epsilon I_1)] \approx \frac{r}{({W}_0)^{\frac{1}{3}}}e^{\frac{2}{3}I_0} \bigg[1+\hat\epsilon \bigg( \frac{2}{3}I_1-\frac{W_1}{3W_0} \bigg)\bigg]\equiv \frac{r}{2^{\frac{1}{3}}}(1+\hat\epsilon g_1)
\ee
and
\bea
 e^f & \approx & \sqrt{\frac{3r}{W'_0+4+\hat\epsilon (W'_1+4{}{p})}}(W_0+\hat\epsilon W_1)^\frac{1}{6}\exp[\frac{2}{3}(I_0+\hat\epsilon I_1)] \nonumber\\
 & \approx & \sqrt{\frac{3r}{W'_0+4}}W^{\frac{1}{6}}_0e^{\frac{2}{3}I_0}\bigg[1+\hat\epsilon \bigg(-\frac{1}{2}\frac{W'_1+4{}{p}}{W'_0+4}+\frac{W_1}{6W_0}+\frac{2}{3}I_1\bigg)\bigg] \nonumber\\
 & \equiv & \frac{r}{2^{\frac{1}{3}}}(1+\hat\epsilon f_1) \ .
\eea 
Similarly for the warp factor:
\bea
 h & \approx & \frac{4\pi^2 N_c}{k}\frac{2^{\frac{1}{3}}}{r}(W'_0+4)\bigg[1+\hat\epsilon \bigg(-g_1+\frac{W'_1+4{}{p}}{W'_0+4}\bigg)\bigg] \int_r^\infty d\xi \frac{2}{\xi^2W_0^2}\bigg[1+\hat\epsilon (-3g_1-2\frac{W_1}{W_0})\bigg] \nonumber\\
  & \approx & \frac{4\pi^2N_c 2^{\frac{1}{3}}}{kr^4}\bigg[1+\hat\epsilon \bigg(-g_1+\frac{W'_1+4{}{p}}{W'_0+4}+6r^3 \int_r^{\infty} \frac{d\xi}{2\xi^4}\big(-3g_1-2\frac{W_1}{W_0} \big) \bigg) \bigg] \nonumber \\
  & \equiv & h_0(1+\hat\epsilon h_1) \ ,
\eea
where $h_0=\frac{4\pi^2N_c 2^{\frac{1}{3}}}{kr^4}$. We can finally compute the dilaton:
\bea
 e^\phi & \approx & \frac{12r}{k}\frac{(h_0(1+\hat\epsilon h_1))^{\frac{1}{4}}}{(W_0+\hat\epsilon W_1)^{\frac{1}{3}}(W'_0+4+\hat\epsilon (W'_1+4{}{p}))}e^{\frac{2}{3}(I_0+\hat\epsilon I_1)} \nonumber\\
 & \approx & \frac{2rh_0^{\frac{1}{4}}}{kW_0^{\frac{1}{3}}}e^{\frac{2}{3}I_0}\bigg[1+\hat\epsilon \bigg(\frac{h_1}{4}-\frac{W_1}{3W_0}-\frac{W'_1+4{}{p}}{W'_0+4}+\frac{2}{3}I_1\bigg)\bigg] \nonumber\\
 & \approx & \frac{2^\frac{5}{4}N_c^{\frac{1}{4}}}{k^\frac{5}{4}}\sqrt{\pi} \bigg[1+\hat\epsilon \bigg(\frac{h_1}{4}-\frac{W_1}{3W_0}-\frac{W'_1+4{}{p}}{W'_0+4}+\frac{2}{3}I_1\bigg)\bigg] \nonumber\\
 & \equiv & \frac{2^\frac{5}{4}N_c^\frac{1}{4}}{k^{\frac{5}{4}}}\sqrt{\pi}[1+\hat\epsilon \phi_1] \ ,
\eea
where we have defined the first order corrections by:
\bea
 g_1 & = & \frac{2I_1}{3}-\frac{W_1}{3W_0} \label{eq:ABJMg1}\\
 f_1 & = & -\frac{1}{2}\frac{W'_1+4{}{p}}{W'_0+4}+\frac{W_1}{6W_0}+\frac{2I_1}{3} \label{eq:ABJMf1}\\
 h_1 & = & -g_1+\frac{W'_1+4{}{p}}{W'_0+4}+6r^3\int_r^{\infty}\frac{d\xi}{2\xi^4}(-3g_1-2\frac{W_1}{W_0}) \label{eq:ABJMh1}\\
\phi_1 & = & \frac{h_1}{4}-\frac{W_1}{3W_0}-\frac{W'_1+4{}{p}}{W'_0+4}+\frac{2I_1}{3} \label{eq:ABJMphi1} \ .
\eea

\subsection{Asymptotic expansions}\label{app:ABJMexp}
Let us now study the geometry in both asymptotic regions. In the IR ($r \rightarrow 0$) we find for the master function
\be
 W = 2r+\hat\epsilon \bigg[\frac{6r^2}{5}\frac{\Gamma(\frac{3+m}{m})\Gamma(\frac{n-1}{m})}{\Gamma(\frac{2+n}{m})}+\frac{2(7+2n)}{(1-n)(4+n)}r^{1+n}+\ldots \bigg] \ ,
\ee
which leads to
\bea
g_1 & = & -\frac{2r}{5}\frac{\Gamma(\frac{3+m}{m})\Gamma(\frac{n-1}{m})}{\Gamma(\frac{2+n}{m})}+\frac{(1+n(4+n))}{(n-1)n(4+n)}r^n+\ldots \\
f_1 & = & -\frac{3r}{10}\frac{\Gamma(\frac{3+m}{m})\Gamma(\frac{n-1}{m})}{\Gamma(\frac{2+n}{m})}+\frac{(2+n(6+n))}{2(n-1)n(4+n)}r^n+\ldots  \\
h_1 & = & \frac{4r}{5}\frac{\Gamma(\frac{3+m}{m})\Gamma(\frac{n-1}{m})}{\Gamma(\frac{2+n}{m})}+\frac{2(6+10n-3n^2-n^3)}{(n-3)(n-1)n(4+n)}r^n+\ldots \\
\phi_1 & = & -\frac{3r}{5}\frac{\Gamma(\frac{3+m}{m})\Gamma(\frac{n-1}{m})}{\Gamma(\frac{2+n}{m})}+\frac{3(-14+n+n^2)}{2(12-13n+n^3)}r^n+\ldots \ .
\eea
In the UV ($r \rightarrow \infty$) we can instead expand the master function as follows
\bea
W & = & 2r+\hat\epsilon \bigg[r^{-1}+\frac{2(2+n)(3-2m)}{m(3+m)(m-2)}r^{-1-m}+\frac{\Gamma(\frac{m-2}{m})\Gamma(\frac{4+n}{m})}{5\Gamma(\frac{2+n}{m})}r^{-3} \\
 & & + \frac{2(2+n)\Gamma(-\frac{2}{m})\Gamma(\frac{4+n}{m})}{5m^2\Gamma(\frac{2+n}{m})}r^{-3-m}+\ldots \bigg] \ , 
\eea
which yields
\bea
 g_1 & = &  -\frac{1}{4r^2}+\frac{(2+n)(m^2-3)r^{-2-m}}{m(m+2)(m^2+m-6)}-\frac{1}{5r^4}\frac{\Gamma(\frac{3+m}{m})\Gamma(\frac{n-1}{m})}{\Gamma(\frac{2+n}{m})}+\ldots \\ 
 f_1 & = &  -\frac{1}{4r^2}+\frac{2}{5r^4}\frac{\Gamma(\frac{3+m}{m})\Gamma(\frac{n-1}{m})}{\Gamma(\frac{2+n}{m})}+\frac{(2+n)(-6-2m+m^2)r^{-2-m}}{2m(2+m)(m^2+m-6)}+\ldots \\
 h_1 & = &  \frac{3}{5r^2}-\frac{2(-18+m(m-2)(5+m))(2+n)}{m(m-2)(3+m)(5+m)} r^{-2-m}+\ldots  \\
 \phi_1 & = & -\frac{3}{5r^2}+\frac{2}{5r^4}\frac{\Gamma(\frac{3+m}{m})\Gamma(\frac{n-1}{m})}{\Gamma(\frac{2+n}{m})}+\frac{3(-12+m(3+m))(2+n)}{2(m-2)m(m+3)(5+m)}r^{-2-m}+\ldots  \ .
\eea

\subsection{Dimensionally reduced effective action}\label{app:ABJMred}

In this subsection we detail how to obtain the dimensionally reduced geometry. Let us take the reduction Ansatz:
\be
 ds^2=e^{2\gamma}g_{pq}\,dz^p dx^q
 +e^{-\frac{2}{3}\gamma+2\lambda}ds^2_{i}+e^{-\frac{2}{3} \gamma-4 \lambda}ds^2_{ab} \ ,
\ee
where $g_{pq}$ is a four-dimensional metric and the scalars $\gamma$ and $\lambda$ depend on the 4d coordinates $z^{p}=(x^0, x^1, x^2, r)$. Moreover,  $ds^2_i=\mathcal{S}_{\xi}^2+\sum_{i=1}^{3}\mathcal{S}_i^2$, and $ds_{ab}^2=\sum_{i=1}^2E_i^2$. In order to construct the four-dimensional reduced action for an arbitrary profile $\eta(r)$, we proceed as in \cite{Hoyos:2020zeg,Jokela:2019tsb} and obtain
\bea
  S & = & \frac{1}{2\kappa^2_{4}} \int d^4x\sqrt{-g}\Bigg[R-\frac{1}{2}(\partial \phi)^2-\frac{8}{3}(\partial \gamma)^2-12 (\partial \lambda)^2-V-V^{branes}\Bigg] \ ,
\eea
where we have introduced the potentials:
\bea
 V & = & 2e^{\frac{8}{3}\gamma-8 \lambda}-12e^{\frac{8}{3} \gamma-2 \lambda}-2e^{\frac{8}{3}\gamma+4\lambda}+\frac{k^2}{8}e^{\frac{3}{2}\phi+\frac{10}{3}\gamma}(e^{8\lambda}+2\eta^2e^{-4\lambda})+\frac{9N^2\pi^4}{2}e^{-\frac{\phi}{2}+6\gamma} \nonumber \\
 V^{branes} & = & k e^{\frac{3}{4}\phi+3\gamma} (-2+2\eta+e^{-\frac{4}{3}\gamma-2\lambda}\nabla_v \eta) \nonumber \ , 
\eea
and where $\nabla_v \eta$ is the directional derivative of the profile $\eta$ along the unit vector $v$ that points along the radial direction $r$:
\be
\nabla_v \eta=v^n\partial_n\eta=\frac{1}{\sqrt{g_{rr}}}\partial_r \eta \ ,
\ee
and $v^n=\frac{1}{\sqrt{g_{rr}}}\delta^n_r$.

  The map between the scalar fields in the reduced action and the background functions reads
\bea
 e^\lambda & = & e^{\frac{1}{3}(f-g)} \\
 e^\gamma & = & h^{-\frac{3}{4}}e^{-2f-g+\frac{3}{4}\phi} \\
 g_{\mu\nu} & = & \eta_{\mu\nu}he^{2g+4f-2\phi} \\
 g_{rr} & = &\frac{h^2}{r^2}e^{4f+4g-2\phi} \ .
\eea

Let us now compute the equations of motion. Let us group the scalars as $(\psi_i)\equiv (\phi,\gamma,\lambda)$ and denote $(\alpha_{\psi_i}) = (\alpha_\phi,\alpha_\gamma,\alpha_\lambda)\equiv (1,\frac{3}{16},\frac{1}{24})$. Then we have the following equations of motion for the scalars
\be
 \Box \psi_i =  \alpha_{\psi_i} \partial_{\psi_i} (V+V^{branes})  \ ,
\ee
and for the metric we have
\be
R_{\mu \nu}-\frac{1}{2}g_{\mu\nu}R-\sum_{i}\frac{1}{2\alpha_{\psi_i}}[\partial_\mu \psi_i \partial_\nu \psi_i-\frac{1}{2}g_{\mu \nu} (\partial \psi_i)^2]+\frac{1}{2}g_{\mu \nu}V-T^{branes}_{\mu \nu}  = 0 \ , 
\ee
where we have denoted
\bea
 T^{branes}_{\mu\nu} & = & -\frac{1}{2}g_{\mu \nu}V^{branes} \\
 T^{branes}_{rr} & = & -k g_{rr}e^{\frac{3}{4}\phi+3\gamma}(\eta-1) \ .
\eea

\section{D2-D6}\label{app:D2D6}

\subsection{Preliminary analysis}\label{app:D2D6prel}
In our D2-D6 backgrounds the internal manifold is a generic compact six-dimensional nearly-K\"ahler manifold and the two-form $J$ is related to its almost complex structure. The expression of $J$ depends on the particular case we consider. In the case of the six-sphere the explicit form of $J$ was given in Appendix B of \cite{Faedo:2015ula}.  In this Appendix we write down $J$ for another example, namely the case in which the six-dimensional manifold is the principal orbit of a Ricci flat cone with $G_2$ holonomy which is topologically ${\mathbb S}^4\times {\mathbb R}^3$ \cite{BryantSalamon,Gibbons:1989er}.  Let us consider a system of invariant forms of $SU(2)$ satisfying $d\omega_i=\frac{1}{2}\epsilon_{ijk}\omega^j \wedge \omega^k$. With it we construct the following set of one-forms
\bea
 e_1 & = & \frac{1}{2}[d\theta+\frac{\xi^2}{1+\xi^2} (\sin \varphi \omega^1-\cos \varphi \omega^2)] \\
 e_2 & = & \frac{1}{2}[\sin \theta (d\varphi-\frac{\xi^2}{1+\xi^2} \omega^3)+\frac{\xi^2}{1+\xi^2} \cos \theta (\cos \varphi \omega^1+\sin \varphi \omega^2)] \\
 e_3 & = & \frac{\sqrt{2}}{1+\xi^2}d\xi  \\
 e_4 & = & \frac{1}{\sqrt{2}}\frac{\xi}{1+\xi^2}[-\cos \theta \omega^3-\sin \theta (\cos \varphi \omega^1+\sin \varphi \omega^2)]\\
 e_5 & = & \frac{1}{\sqrt{2}}\frac{\xi}{1+\xi^2}[\sin \varphi \omega^1-\cos \varphi \omega^2]  \\
 e_6 & = & \frac{1}{\sqrt{2}}\frac{\xi}{1+\xi^2}[\sin \theta \omega^3-\cos \theta (\cos \varphi \omega^1+\sin \varphi \omega^2) ] \ .
\eea
The range of the coordinates  $\theta$, $\varphi$ and $\xi$ is: $0\leq \theta <\pi, 0\leq \varphi <2\pi $, $0\leq \xi <\infty$.
These forms constitute a natural system of vielbeins for the principal orbits of the $G_2$ cone constructed from the bundle of self-dual two-forms over ${\mathbb S}^4$. Defining $ds_6^2=\sum_{i=1}^6\,(e^i)^2$,  the metric of the $G_2$ cone is $ds^2_7=dr^2+r^2\, ds_6^2$. Moreover, for this internal manifold the two-form $J$ is:
\be
 J=e^1\wedge e^2+e^3\wedge e^4+e^5\wedge e^6 \ .
\ee
The system of BPS equations we wish to solve for is
\bea
 \chi' & = & Q_f \frac{p}{r^2}e^{2\chi} \\
  h' & = & -Q_c\frac{e^{-2\chi}}{r^6}-3Q_f \frac{p}{r^2}e^{2\chi}h \ .
\eea
The solution for $\chi$ is readily available
\be
 e^{-2\chi}  =  1+2\int_r^{\infty}dz \frac{Q_f p(z)}{z^2} = 1+\frac{2Q_f}{5r^5}F[\frac{5}{m},\frac{4+n}{m},\frac{5+m}{m},-r^{-m}] \ .
\ee
The warp factor can be written in terms of $\chi$:
\be
 h  =  e^{-3\chi}\int_r^{\infty}dy \frac{Q_c e^{\chi(y)}}{y^6} \ ,
\ee
but we have not been able to find a closed form expression for this integral and hence need to resort to numerics in general. We will continue with asymptotic analysis.


\subsection{Asymptotic expansions}\label{app:D2D6exp}

Let us analyze the asymptotic behavior as the holographic radial coordinate takes values close to the Poincar\'e horizon or to the boundary. Let us spell out the asymptotic expressions for the functions entering the metric. In the IR region ($r\to 0$) we have
\bea
 h  & = &  \frac{Q_c}{5 c_n r^5} \left(1-\frac{2 Q_f}{c_n}r^{n-1}\left(\frac{1}{n-1}-\frac{n+4}{m(m+n-1)}r^m \right) \right)+\ldots \\
 \chi & = &  \sqrt{c_n}\left(1- \frac{ Q_f}{c_n}r^{n-1}\left(\frac{1}{n-1}+\frac{n+4}{m(m+n-1)}r^m \right)\right)+\ldots \ ,
\eea
where
\be
 c_n=1+Q_f\frac{2\Gamma\left(\frac{m+5}{m} \right)\Gamma\left(\frac{n-1}{m} \right)}{\Gamma\left(\frac{n+4}{m} \right)} \ .
\ee
In the UV region ($r\to \infty$) we find:
\bea
 h & = &   \frac{Q_c}{5 r^5} \left(1-\frac{Q_f}{r^{5+m}}\frac{(n+4)(3m+25)}{m(m+5)(m+10)}\right)+\ldots \\
 \chi & = &  1- \frac{ Q_f}{r^{5+m}}\frac{n+4}{m(m+5)}+\ldots \ .
\eea

\subsection{Dimensionally reduced effective action}\label{app:D2D6red}

Let us take the reduction Ansatz:
\be
 ds^2=e^{\gamma}ds^2_4+e^{-\frac{\gamma}{3}}ds^2_{6}
\ee
with:
\be
 ds^2_4=g_{pq}dz^q dz^q\ , 
\ee
where $z^p$ are the coordinates $(x^0, x^1, x^2, r)$. 
This yields the following reduced action for an arbitrary profile $p(r)$
\bea
S & = & \frac{1}{2\kappa^2_{4}} \int d^4x\sqrt{-g}\Bigg[R-\frac{1}{2}(\partial \phi)^2-\frac{2}{3}(\partial \gamma)^2-V-V^{branes} \Bigg] \ ,
\eea
where we have introduced the potentials
\bea
 V & = & -30e^{\frac{4}{3}\gamma}+\frac{Q_c^2}{2}e^{3\gamma-\frac{\phi}{2}}+\frac{3}{2}Q_f^2p^2e^{\frac{3}{2}\phi+\frac{5}{3}\gamma} \\ 
 V^{branes} & = & 3Q_f e^{\frac{3}{4}\phi+\frac{3}{2}\gamma}(4p+e^{-\frac{2}{3}\gamma}\nabla_v p)\ , 
\eea
where $\nabla_v p$ is the directional derivative along the radial unit vector $v$.
\be
 \nabla_v p=v^n\partial_n p=\frac{1}{\sqrt{g_{rr}}}\partial_r p \ ,
\ee
where $v^n=\frac{1}{\sqrt{g_{rr}}}\delta^n_r$. 
 
We have the following relations for the metric and scalar fields in terms of the other background fields:
\bea
 e^\gamma & = & h^{-\frac{3}{2}}r^{-6}e^{\frac{3}{2}\phi-6\chi}\\
 g_{\mu\nu} & = & \eta_{\mu \nu}hr^6e^{-2\phi+6\chi} \\
 g_{rr} & = & h^2r^6e^{-2\phi+8\chi} \ .
\eea
Let us define $(\psi_i)\equiv (\phi,\gamma)$ and denote $(\alpha_{\psi_i}) \equiv (\alpha_\phi,\alpha_\gamma)\equiv (1,\frac{3}{4})$. All this leads to the following equations of motion for the scalars
\be
 \Box \psi_i= \alpha_{\psi_i} \partial_{\psi_i} (V+V^{branes}) \ .
\ee
For the metric we find
\be
 R_{\mu \nu}-\frac{1}{2}g_{\mu\nu}R-\sum_{i}\frac{1}{2\alpha_{\psi_i}}[\partial_\mu \psi_i \partial_\nu \psi_i-\frac{1}{2}g_{\mu \nu} (\partial \psi_i)^2]+\frac{1}{2}g_{\mu \nu}V-T^{branes}_{\mu \nu}=0 \ ,
\ee
where have introduced the brane sources
\bea
 T^{branes}_{\mu\nu} & = & -\frac{1}{2}g_{\mu \nu}V^{branes} \\
 T^{branes}_{rr} & = & -6Q_f p g_{rr} e^{\frac{3}{4}\phi+\frac{3}{2}\gamma} \ .
\eea

\section{D3-D7}\label{app:D3D7}

\subsection{Preliminary analysis}\label{app:D3D7prel}

Let us start by introducing a suitable system of vielbeins to present the five-sphere  ${\mathbb S}^5$ as the Sasaki-Einstein manifold obtained as a $U(1)$ bundle over ${\mathbb C}{\mathbb P}^2$. Using the coordinates $\chi  \in [0,\pi]$ and $\tau \in [0,2\pi]$, and a system of invariant forms of $SU(2)$ which satisfy $d\omega^i=\frac{1}{2}\epsilon^{ijk}\omega^j \wedge \omega^k$ we define
\bea
 e^1 & = & \frac{1}{2}\cos \frac{\chi}{2}\omega^1\\
 e^2 & = & \frac{1}{2}\cos \frac{\chi}{2}\omega^2\\
 e^3 & = & \frac{1}{2}\cos \frac{\chi}{2}\sin \frac{\chi}{2}\omega^3\\
 e^4 & = & \frac{1}{2}d\chi
\eea
as well as the one-form
\be
 A=\frac{1}{2}\cos^2\frac{\chi}{2}\omega^3 \ .
\ee
We can now  construct the metric of the K\"ahler-Einstein base as:
\be
 ds^2_{{\mathbb C}{\mathbb P}^2}=\sum_{i=1}^4 (e^i)^2
\ee
from which we construct the Sasaki-Einstein manifold after adding the fiber
\be\label{S5_U1_bundle}
 ds^2_{{\mathbb S}^5}=ds^2_{{\mathbb C}{\mathbb P}^2}+(d\tau+A)^2\ .
\ee
We will now outline the integration of the system of BPS equations \ref{eq:d3d7BPSeq}. We start by integrating the equation of motion for the dilaton:
\be
 e^{-\phi}=e^{-\phi_0}-Q_f \int_0^r dr' \frac{p(r')}{r'} \ , 
\ee
where $e^{\phi_0}=e^{\phi(r=0)}$. To integrate the equations for $f$ and $g$, we will combine them in a single variable $w=e^{2f-2g}$. If we now denote $F(r)=e^{\phi_0-\phi}$, then $w$ satisfies the following equation:
\be
 w'=\frac{6}{r}(1-w)w +\frac{F'}{F}w
\ee
which can be readily integrated, giving the solution:
\be
 w=\frac{r^6F}{C_1-6\int^r_0dr'F'(r')r'^5} \ .
\ee
Let us define $I\equiv 6\int_0^r dr'  F'(r')r'^5 $ and $\hat{\eta} \equiv Q_f e^\phi \int_0^r dr' r'^5 p(r')$. After using the BPS equation for the dilaton, one can verify that $I=-6e^{\phi_0} \int^r_0r'^5e^{-\phi}dr'=-e^{\phi_0-\phi}[r^6+\hat{\eta}]$ where we have used integration by parts. Therefore, we get:
\be 
 w=(1+r^{-6}(C_1 e^{\phi-\phi_0}+\hat{\eta}))^{-1} \ .
\ee
Furthermore we find
\be
 g'=\frac{w}{r}=\frac{r^5}{C_1 e^{\phi-\phi_0}+\hat{\eta}+r^6} \ .
\ee 
The following relations hold: $\hat{\eta}'=\phi'(\hat{\eta}+r^6)$, $6r^5=(C_1 e^{\phi-\phi_0}+\hat{\eta}+r^6)'-\phi'(C_1 e^{\phi-\phi_0}+\hat{\eta}+r^6)$. Therefore:
\be 
 g'=\frac{1}{6}\frac{(C_1 e^{\phi-\phi_0}+\hat{\eta}+r^6)'}{C_1 e^{\phi-\phi_0}+\hat{\eta}+r^6}-\frac{1}{6} \phi'
\ee
which also can be immediately integrated. After straightforward manipulations we get:
\bea
 e^g & = & \beta e^{-\frac{\phi}{6}}r (1+r^{-6} (C_1 e^{\phi-\phi_0}+\hat{\eta}))^{\frac{1}{6}} \nonumber \\
 e^f & = & \beta e^{-\frac{\phi}{6}}r(1+r^{-6}(C_1 e^{\phi-\phi_0}+\hat{\eta}))^{-\frac{1}{3}} \nonumber \\
 h & = & \int_r^\infty dr' Q_c \frac{e^{-4g}}{r'} \ ,
\eea
where $\beta$ is an integration constant, and the range for the holographic coordinate is $r \in [0,\infty)$. 

Let us now use the profile:
\be
 p=\frac{r^n}{(1+r^m)^{\frac{n+4}{m}}} 
\ee
which is guaranteed to give a positive $T_{00}$ in the UV ($r \rightarrow \infty$). Let us define $W_1\equiv F(\frac{n}{m},\frac{4+n}{m},\frac{m+n}{m},-r^m)$ and $W_2 \equiv F(\frac{4+n}{m},\frac{6+n}{m},\frac{6+m+n}{m},-r^m)$. The BPS equations can then be straightforwardly integrated, leading to 
\bea
 e^{-\phi} & = & e^{-\phi_0}(1-\frac{r^n}{n}Q_f e^{\phi_0}W_1) \\
 \hat{\eta} & = & \frac{r^{6+n}}{6+n}\frac{Q_f e^{\phi_0}W_2}{1-\frac{r^n}{n}Q_f e^{\phi_0}W_1} \ .
\eea
We will constrain the value of the integration constant $C_1$ by imposing a regularity condition in the IR region ($r \rightarrow 0$). One can see that, at $r\rightarrow 0$:
\bea
 e^g & \approx & \beta C_1^\frac{1}{6}e^{-\frac{\phi_0}{6}} \\
 e^f & \approx & r^3 \beta C_1^{-\frac{1}{3}}e^{-\frac{\phi_0}{6}} \ .
\eea
Therefore, the fiber and the base of the Sasaki-Einstein manifold shrink at a different pace, which forces us to impose $C_1=0$ to avoid the collapse of the cycle. With this  we arrive at the final expressions for the functions $f,g$
\bea
 e^g & = & \beta e^{-\frac{\phi_0}{6}}r \bigg[1 +r^n Q_f e^{\phi_0} \big( \frac{W_2}{6+n}-\frac{W_1}{n}\big)\bigg]^\frac{1}{6} \\
 e^f & = & \beta e^{-\frac{\phi_0}{6}}r \bigg[1-\frac{r^n}{n}Q_f e^{\phi_0}W_1 \bigg]^{\frac{1}{6}} \bigg[1+\frac{r^n}{6+n} \frac{Q_f e^{\phi_0} W_2}{1-\frac{r^n}{n}Q_f e^{\phi_0}W_1}\bigg]^{-\frac{1}{3}} \ .
\eea

\subsection{Asymptotic expansions}\label{app:D3D7exp}

We will now spell out the asymptotic expressions for the metric functions. Close to the IR region ($r \rightarrow 0$) we have
\bea
 e^f & \approx & \beta r e^{-\frac{\phi_0}{6}} \big[ 1-Q_f \frac{e^{\phi_0}(2+n)}{2n (6+n)}r^n +Q_f \frac{e^{\phi_0}(4+n)(2+m+n)}{2m (m+n)(6+m+n)}r^{n+m}  \big]  \nonumber\\
 e^g & \approx & \beta r e^{-\frac{\phi_0}{6}}  \big[  1-Q_f \frac{e^{\phi_0}}{n(6+n)}r^n +Q_f \frac{e^{\phi_0}(4+n)}{m(m+n)(6+m+n)}r^{n+m}  \big] \nonumber\\
 e^\phi & \approx & e^{\phi_0} \big[ 1+Q_f \frac{e^{\phi_0}}{n}r^n-Q_f \frac{e^{\phi_0}(4+n)}{m(m+n)}r^{n+m} \big]  \nonumber\\
 h & \approx & \frac{Q_c}{4 \beta^4 r^4}e^{\frac{2}{3} \phi_0} \big[ 1+16 \frac{Q_f e^{\phi_0} r^{n}}{n(n+6)(4-n)} -16 \frac{(4+n)Q_f e^{\phi_0} r^{n+m} }{m(m+n)(6+m+n)(4-n-m)}  \big] \ .\nonumber
\eea
Close to the UV region ($r\rightarrow \infty$), we have
\bea
 e^f & \approx & \beta r \gamma^{\frac{1}{6}}e^{-\frac{\phi}{6}} \nonumber\\
 e^g & \approx & \beta r \gamma^{\frac{1}{6}}e^{-\frac{\phi}{6}}  \nonumber\\
 e^\phi & \approx & \frac{e^{\phi_0}}{\gamma}  \big[ 1-Q_f \frac{e^{\phi_0}}{\gamma}\frac{r^{-4}}{4}+Q_f \frac{e^{\phi_0}}{\gamma}\frac{4+n}{m(m+4)}r^{-4-m}-Q_f \frac{e^{\phi_0}}{\gamma} \frac{(4+n)(4+m+n)}{4m^2(2+m)}r^{-4-2m} \big]  \nonumber\\
 h  & \approx & \frac{Q_c}{4 \beta^4 r^4}\frac{e^{\frac{2}{3} \phi_0}}{\gamma^{\frac{2}{3}}} \ ,\nonumber
\eea
where $\gamma=1-Q_f e^{\phi_0}\frac{\Gamma[\frac{4}{m}]\Gamma[\frac{m+n}{m}]}{n \Gamma[\frac{4+n}{m}]}$.

\subsection{Dimensionally reduced effective action}\label{app:D3D7red}
Let us take the following reduction Ansatz:
\be
 ds^2=e^{\frac{10}{3} \gamma}g_{pq}dz^p dz^q+e^{-2(\gamma+\lambda)}ds^2_{{\mathbb C}{\mathbb P}^2}+e^{2(4\lambda-\gamma)}(d\tau+A)^2 \ ,
\ee
where $g_{pq}$ is the 5d metric.

The reduced action for the profile $p(r)$ reads 
\be
 S=\frac{1}{2 \kappa_{5}^2}  \int d^5x \sqrt{-g_5} \big[ R_5-\frac{40}{3} (\partial \gamma)^2-20(\partial \lambda)^2-\frac{1}{2}(\partial \phi)^2 -V-V^{branes}\big] \ ,
\ee
where we have introduced the potentials
\bea
 V & = & 4 e^{\frac{16}{3} \gamma+12 \lambda}-24 e^{2\lambda+\frac{16}{3}\gamma}+\frac{1}{2} Q_f^2 p^2e^{2 \phi-8\lambda+\frac{16}{3} \gamma}+\frac{1}{2} Q_c^2 e^{\frac{40}{3} \gamma} \\
 V^{branes} & = & 4 Q_f e^{\frac{16}{3} \gamma+\phi+2 \lambda}(p+\frac{1}{4}e^{-\frac{8}{3} \gamma-6 \lambda} \nabla_v p) \ ,
\eea
and where $\nabla_v p=v^n \partial_n p=\frac{1}{\sqrt{g_{rr}}}\partial_r p$, with $v^n=\frac{1}{\sqrt{g_{rr}}}\delta^n_r$.

By defining $(\psi_i)\equiv (\phi,\gamma,\lambda)$ and denoting $\alpha_{\psi_i} = (\alpha_\phi,\alpha_\gamma,\alpha_\lambda)\equiv (1,\frac{3}{80},\frac{1}{40})$ the equations of motion are cast in the form for the scalars
\be
 \Box \psi_i  =   \alpha_{\psi_i} \partial_{\psi_i} (V+V^{branes}) \ ,
\ee
and for the metric
\be 
 R_{\mu \nu}-\frac{1}{2}g_{\mu \nu}R-\sum_i \frac{1}{2 \alpha_{\psi_i}}[\partial_\mu \psi_i \partial_\nu \psi_i-\frac{1}{2}g_{\mu \nu} (\partial \psi_i)^2]+\frac{1}{2}g_{\mu \nu}V-T^{branes}_{\mu \nu}  =  0 \ ,
\ee
supplemented with the sources
\bea
 T^{branes}_{\mu \nu} & = & -\frac{1}{2}g_{\mu \nu} V^{branes} \nonumber \\
 T^{branes}_{rr} & = & -g_{rr} 2 Q_f e^{\frac{16}{3}\gamma+\phi+2 \lambda}p \ .
\eea

\section{D3-D5}\label{app:D3D5}

\subsection{Preliminary analysis}\label{app:D3D5prel}

Let us represent the five-sphere ${\mathbb S}^5$ as a $U(1)$ bundle over ${\mathbb C}{\mathbb P}^2$ (see (\ref{eq:d3d7BPSeq})). The ten-dimensional backreacted metric can then be written as
\bea
 ds^2_{10} &  = & h^{-{1\over 2}}\,\big[-(dx^0)^2+(dx^1)^2+(dx^2)^2\,+\,e^{-2\phi}\,(dx^3)^2\big]\nonumber\\
          &    & +h^{{1\over 2}}\,\Big[r^2 e^{-2f}\,dr^2\,+\,r^2\,ds^2_{{\mathbb C}{\mathbb P}^2}\,+\,e^{2f}\,(d\tau+A)^2\Big]\ .
\eea
Here $A$ is a one-form on ${\mathbb C}{\mathbb P}^2$ inherent to the non-trivial $U(1)$ bundle.  The BPS equations can be combined into a single second-order equation for a master function $W(r)$, in terms of which $f$ and $\phi$ are given by
\be
 e^{2f}\,=\,{6\,r^2\,W\over 6\,W\,+\,r\,{dW\over dr}} \ \ , \ \ e^{-\phi}\,=\,W\,+\,{1\over 6}\,r\,{d W\over dr}\ .
\ee
The warp factor $h$ can be written in terms of the following integral
\be
 h = Q_c\,e^{-\phi(r)}\,\int_{r}^{\infty}{d\bar r\over \bar r^5\,W(\bar r)}\ ,
\ee
The second-order differential equation satisfied by the master function $W$ is
\be
 {d\over dr}\Big(r\,{d W\over dr}\Big)\,+\,6\,{d W\over dr} = -{6\,Q_f\,p\over r^2\,\sqrt{W}}\ .
\ee
 The solution for the profiles we study is
\bea
 W(r) & = & 1\,+\,Q_f\,\,\Bigg[{1\over 4 r^4}\,F\Big({4\over m}, {3+n\over m}; {4+m\over m};- r^{-m}\Big)\nonumber \\
          &   & +{r^{n-1}\over 5+n} F\Big({5+n\over m}, {3+n\over m}; {5+m+n\over m};-  r^m\Big)\Bigg]\ .
\eea
The brane distribution can be written conveniently as 
\be
 p=\sqrt{W}\frac{r^n}{(1+r^m)^{\frac{n+3}{m}}} \ .
\ee

\subsection{Asymptotic expansions}\label{app:D3D5exp}

The expansion in the UV region of the geometry is
\be
 W = 1+\frac{3Q_f }{4r^4}+\ldots \ , \ r\to\infty \ .
\ee
This yields a decreasing density:
\be
 p = \frac{1}{r^3}\left(1+\frac{3Q_f }{8r^4}-\frac{n+3}{mr^m}+\ldots \right) \ . 
\ee
The expansions of the dilaton and the warp factors are
\be
 e^{-\phi}\approx 1+\frac{Q_f }{4 r^4} \ , \ e^{2 f}\approx r^2\left( 1+\frac{Q_f }{2  r^4}\right) \ , \  h\approx \frac{Q_c}{4r^4} \ .
\ee

The behavior in the IR $r \to 0$ depends on the profile of the D5-brane density, in particular, on the value of the exponent $n$. We can distinguish two cases depending on whether $n>1$ or $n<1$, with a limiting case $n=1$ between the two. The master function has the following IR expansions, depending on the value of $n$,
\be
W\approx  \left\{  \begin{array}{ll}
 w_{n,m}+\frac{6  Q_f }{(n+5)(1-n)} r^{n-1} & , \ n>1 \\
 -Q_f \log (r) & , \ n=1 \\ 
 \frac{6 Q_f }{(n+5)(1-n)}  r^{n-1} & , \ n<1 \ , \end{array} \right. 
\ee
where
\be
 w_{n,m}\,=\,1+{\Gamma\Big({4\over m}\Big)\,\Gamma\Big({n-1\over m}\Big)\over m\,\Gamma\Big({3+n\over m}\Big)}\,Q_f \ .
\ee
From these expressions one can infer the expansion for the D5-brane density
\be
 p \approx \left\{\begin{array}{ll}
     \sqrt{w_{n,m}}  r^n  & , \ n>1 \\
     \sqrt{\frac{6 Q_f }{n+5}}  \, r(-\log( r))^{1/2} & , \ n=1 \\
      \sqrt{\frac{6 Q_f }{(n+5)(1-n)}} r^{\frac{3n-1}{2}} & , \ n<1 \ .\end{array}\right. 
\ee
For the dilaton and metric functions one finds, for $n>1$,
\bea
 e^{-\phi} & \approx & w_{n,m}-\frac{ Q_f }{n-1} r^{n-1}\approx w_{n,m}\nonumber\\
 e^{2f} & \approx & r^2\left( 1+\frac{ Q_f }{w_{n,m}\,(n+5)} r^{n-1}\right)\approx r^2 \nonumber\\
 h & = & \frac{Q_c}{4r^4}\left(1+{\cal O}( r^{n-1})\right) \ .
\eea
When $n<1$ the anisotropy along the spatial direction transverse to the D5-branes survives in the IR and the geometry becomes a scaling solution, a Lifshitz-type. The expansion of the dilaton and warp factors of the metric is ($r\to 0$):
\bea
 e^{-\phi} & \approx &  \frac{Q_f }{1-n}r^{n-1}\nonumber\\
 e^{2f} & \approx &  \frac{6}{n+5}r^2\nonumber\\
 h & = & \frac{n+5}{6(n+3)}\frac{ Q_c}{r^4}\left(1+{\cal O}( r^{1-n})\right) \ .
\eea

\subsection{Dimensionally reduced effective action in 5d}\label{app:D3D5red}

Let us consider the following reduction Ansatz:
\be\label{10d_5d_metric_ansatz}
 ds_{10}^2\,=\,e^{{10\over 3}\,\gamma}\,g_{pq}\,dz^p\,dz^{q}\,+\,e^{-2(\gamma+\lambda)}\,ds^2_{{\mathbb C}{\mathbb P}^2}\,+\, e^{2(4\lambda-\gamma)}\,(d\tau+A)^2\ , 
\ee
where $g_{pq}$ is the 5d metric. The reduced action reads: 
\bea
 S_{eff} & = & {1\over 2\,\kappa_{5}^2}\,\int d^5z\, \sqrt{-g_5}\Big[R_5-{40\over 3}(\partial\gamma)^2-20(\partial\lambda)^2-{1\over 2}(\partial\phi)^2-{1\over 2}e^{4\gamma+4\lambda+\phi}( {\cal F}_1)^2-U_{scalars}\Big]\nonumber \\
& &- {V_5\over 2\,\kappa_{10}^2}\,\int d^5z\, \sqrt{-\hat g_4}\,U_{branes}\,+\,S_{WZ} \ ,
\label{effective_5d_action}
\eea
where we have introduced the potentials
\bea
U_{scalars} & = & 4\,e^{{16\over 3}\,\gamma+12 \lambda}-24\,e^{{16\over 3}\gamma+2\lambda} +{Q_c^2\over 2}\,e^{{40\over 3}\gamma}  \\ \label{5d_potential}
U_{branes} & = & 6\,Q_f\,e^{{\phi\over 2}\,-\,2\lambda\,+{14\gamma\over 3}}\,\Big(p+{e^{4\lambda-{8\gamma\over 3}}\over 3}\,\nabla_v\,p\Big)\ , \label{U_branes_5d}
\eea
and where $\nabla_v p=v^n \partial_n p=\frac{1}{\sqrt{g_{rr}}}\partial_r p$, with $v^n=\frac{1}{\sqrt{g_{rr}}}\delta^n_r$. We have furthermore introduced a Wess-Zumino action, given by
\be
S_{WZ}\,=\,{V_5\over 2\,\kappa_{10}^2}\,\int {\cal C}_3\wedge \Sigma_2\ ,
\ee
where ${\cal C}_3$ is a three-form potential and $\Sigma_2$ is a smearing two-form, which satisfies:
\be
 d\,{\cal F}_1\,=\,\Sigma_2\,\ ,
\ee
with
\be\label{F_1_ansatz}
 {\cal F}_1\,=\,\sqrt{2}\,Q_f\,p \,dx^3\ .
\ee
The smearing form $\Sigma_2$ takes the form:
\be
 \Sigma_2\,=\,\sqrt{2}\,Q_f\,p'\,dr\wedge dx^3\ . \label{Sigma_2}
\ee
The potential $\mathcal{C}_3$ is defined such that
\be
 {\cal F}_4\,\equiv\,e^{4\gamma+4\lambda+\phi}\ast_5{\cal F}_1= d\,{\cal C}_3\,\,=\,\sqrt{2}\,Q_f\,r\,p\,e^{2\phi}\,dr\wedge dx^0\wedge dx^1\wedge dx^2\ , \label{def_C3}
\ee
where $\ast_5$ denotes the Hodge dual of the 5d theory. This definition is motivated by the Maxwell equation (Bianchi identity) for $\mathcal{F}_1$ ($\mathcal{F}_4$). The map between the scalar fields in the reduced action and the background functions reads
\bea
 e^{\lambda} & = & e^{{f\over 5}}\,r^{-\frac{1}{5}} \nonumber \\
 e^{\gamma} & = & h^{-{1\over 4}}\,e^{-{f\over 5}}\,r^{-\frac{4}{5}}\nonumber\\
 g_{r r} & = & h^{{4\over 3}}\,e^{-{4f\over 3}}r^{\frac{14}{3}} \label{inverse_ansatz_relations_5d}
\eea
and
\bea
 g_{x^\mu x^\nu} & = & \eta_{\mu\nu}h^\frac{1}{3}r^\frac{8}{3}e^{\frac{2}{3}f} \nonumber\\
 g_{x^3\,x^3} & = & h^\frac{1}{3}e^{-2\phi}r^\frac{8}{3}e^{\frac{2}{3}f}\ .  \label{g_5d_relations}
\eea
Let us further define $(\psi_i)\equiv (\phi,\gamma,\lambda)$ and denote $(\alpha_{\psi_i}) = (\alpha_\phi,\alpha_\gamma,\alpha_\lambda)\equiv (1,\frac{3}{80},\frac{1}{40})$ casting the equations of motion for the scalars in the form
\be
\Box \psi_i = \alpha_{\psi_i}\,\partial_{\psi_i}\,U_{scalars}+{1\over 2}\,\alpha_{\psi_i}\,\big( {\cal F}_1\big)^2 \partial_{\psi_i}\Big(e^{4\lambda+4\gamma+\phi}\Big)+{\sqrt{-\hat g_4}\over \sqrt{-g_5}}\alpha_{\psi_i}\partial_{\psi_i} U_{branes}
\label{5d_scalar_eoms_compact}
\ee
and for the metric we get
\bea
R_{\mu\nu}-{1\over 2}\,g_{\mu\nu}\,R - \sum_{i}\,{1\over 2\alpha_{\psi_i}}\,[\partial_\mu\,\psi_i\,\partial_\nu\,\psi_i\,-\,{1\over 2}\,g_{\mu\nu}\,(\partial \psi_i)^2] \nonumber \\ \,+\,{1\over 2}\,g_{\mu\nu}\,U_{scalars}                                       -{1\over 2}\, [\big({\cal F}_1\big)_\mu \big({\cal F}_1\big)_\nu\,-\,{1\over 2}\,g_{\mu\nu}\, \big( {\cal F}_1\big)^2] -T_{\mu\nu}^{branes} & = & 0\ , \label{Einstein_eqs_5d}
\eea
where $T_{\mu\nu}^{branes}$ represent the contribution originating from the brane term:
\bea
 T_{x^\mu x^\nu}^{branes} & = & -{Q_f\over \sqrt{g_{rr}}}\,\Big(3\,e^{-7\lambda+{4\gamma\over 3}\,+\,{3\phi\over 2}}\,p\,+\,{e^{-3\lambda-{4\gamma\over 3}\,+\,{3\phi\over 2}}\over \sqrt{g_{rr}}}\,p'\Big)\,\eta_{\mu\nu}\qquad\qquad
(\mu,\nu=0,1,2)   \nonumber \\
 T^{branes}_{rr} & = & -3\,Q_f\,\big(g_{rr}\big)^{{3\over 2}}\,e^{3\lambda+8\gamma+{3\phi\over 2}}\,\,p \nonumber \\
 T^{branes}_{x^3 x^3} & = & 0 \ .
\eea

\subsection{Dimensionally reduced effective action in 4d}\label{app:D3D5reductionto4d}

Let us consider the following reduction Ansatz:
\be\label{10d_5d_metric_ansatz4d}
 ds_{10}^2\,=\,e^{{10\over 3}\,\gamma-\beta}\,g_{pq}\,dz^p\,dz^{q}\,+e^{\frac{10}{3}
 \gamma+2\beta}(dx^3)^2+\,e^{-2(\gamma+\lambda)}\,ds^2_{{\mathbb C}{\mathbb P}^2}\,+\, e^{2(4\lambda-\gamma)}\,(d\tau+A)^2\ , 
\ee
where $g_{pq}$ is the 4d metric. The reduced action reads: 
\be
 S_{eff}  =  {1\over 2\,\kappa_{4}^2}\,\int d^4z \sqrt{-g}[R-\frac{40}{3}(\partial \gamma)^2-20 (\partial \lambda)^2-\frac{3}{2}(\partial \beta)^2-\frac{1}{2}(\partial \phi)^2-V-V^{branes}] \ , \label{effective_5d_action4d}
\ee
where we have introduced the potentials
\bea
  V & = &  -24 e^{\frac{16}{3}\gamma+2\lambda-\beta}+4e^{\frac{16}{3}\gamma+12\lambda-\beta}+\frac{Q_c^2}{2}e^{\frac{40}{3}\gamma-\beta}+Q_f^2p^2e^{\phi+4\gamma+4\lambda-3\beta} \nonumber \\
  V^{branes} & = & 6Q_f e^{\frac{\phi}{2}-2\beta-2\lambda+\frac{14}{3}\gamma}(p+\frac{1}{3}\nabla_v p e^{\beta+4\lambda-\frac{8}{3}\gamma}) \ ,
\eea
and
\be
  \nabla_vp=v^n\partial_n p=\frac{1}{\sqrt{g_{rr}}} p' \ \,, \ \ v^n=\frac{1}{\sqrt{g_{rr}}}\delta^n_r \ .
\ee

Let us further define $(\psi_i)\equiv (\phi,\gamma,\lambda,\beta)$ and denote $(\alpha_{\psi_i}) = (\alpha_\phi,\alpha_\gamma,\alpha_\lambda,\alpha_\beta)\equiv (1,\frac{3}{80},\frac{1}{40},\frac{1}{3})$ casting the equations of motion for the scalars in the form
\be \label{5d_scalar_eoms_compact4d}
 \Box \psi_i= \alpha_{\psi_i} \partial_{\psi_i} (V+V^{branes})\ ,
\ee
and for the metric we get
\be
R_{\mu \nu}-\frac{1}{2}g_{\mu \nu}R - \sum_{i}\frac{1}{2\alpha_{\psi_i}}[\partial_\mu {\psi_i} \partial_\nu \psi_i-\frac{1}{2}g_{\mu \nu}(\partial \psi_i)^2]+\frac{1}{2}g_{\mu \nu}V-T^{branes}_{\mu \nu} =  0
\label{Einstein_eqs_5d4d} \ ,
\ee 
where $T_{\mu\nu}^{branes}$ represent the contribution originating from the brane term:
\bea
 T_{x^\mu x^\nu}^{branes} & = & -\frac{1}{2}g_{\mu\nu} V^{branes}\qquad\qquad
(\mu,\nu=0,1,2)   \nonumber \\
 T^{branes}_{rr} & = & -3\,Q_f\ g_{rr} e^{-2\beta+\frac{14}{3}\gamma-2\lambda+\frac{\phi}{2}}p .
\eea

\bibliographystyle{JHEP}
\bibliography{refs}

\end{document}